\documentclass[aps,prd,amsmath,amssymb,10pt,letterpaper,balancelastpage,showpacs,superscriptaddress,nofootinbib,floatfix,notitlepage,longbibliography]{revtex4-2}

\usepackage[utf8]{inputenc} % Handles UTF-8 encoding
\usepackage[T1]{fontenc}    % Use T1 encoding for proper font rendering

\usepackage{comment}
\usepackage{graphics}
\usepackage{bm}
\usepackage{color}
\usepackage{xcolor}
\usepackage{amsmath}
\usepackage{amssymb}
\usepackage{mathrsfs}
\usepackage{graphicx}
\usepackage{subcaption}
\usepackage{accents}
\usepackage{adjustbox}
\usepackage{float}%allow forced image/table placement
\usepackage{physics}

\usepackage[
    starfontserif % comment for sans glyphs
    ]{starfont}

\DeclareSymbolFont{starfontsym}{OT1}{sts}{m}{n}
\DeclareMathSymbol{\mathTerra}{\mathord}{starfontsym}{76}

\newcommand{\Rtrap}{R_{\text{trap}}}

\newcommand{\specialColor}{black}

\newcommand{\Bhat}{\hat{B}}

\newcommand{\Ahat}{\hat{A}'}
\newcommand{\ws}{\omega_{\mathTerra}}

\newcommand{\tA}{\theta_{A}}
\newcommand{\Gr}{\Gamma_{0,\text{random}}}
\newcommand{\Gf}{\Gamma_{0,\text{fixed}}}

\newcommand{\gc}{\gamma_c}
\newcommand{\gz}{\gamma_z}

\newcommand{\tobs}{t_{\text{obs}}}
\newcommand{\tTotal}{t_{\text{total}}}
\newcommand{\lambdaDB}{\lambda_{\text{dB}}}

\newcommand{\eq}[1]{eq.~(#1)}
\newcommand{\Eq}[1]{Eq.~(#1)}

\newcommand{\Aunitvec}{\mathbf{\hat{A}'}}

\newcommand{\me}{m_e}
\newcommand{\wcp}{\omega_{c'}}
\newcommand{\wm}{\omega_{m}}

\newcommand{\ic}{i_c}
\newcommand{\fc}{f_c}

\newcommand{\wc}{\omega_{c}}
\newcommand{\wz}{\omega_{z}}
\newcommand{\Amu}{A_{\mu} }
\newcommand{\wfi}{\omega_{fi}}

\newcommand{\nc}{n_c}
\newcommand{\Dw}{\Delta \omega_{A'}}

\newcommand{\rhoDM}{\rho_{\text{DM}}}

\newcommand{\Gammafree}{\Gamma_{c, \text{free}}}
\newcommand{\Gammacavity}{\Gamma_{c, \text{cavity}}}
\newcommand{\wdr}{\omega_d}
\newcommand{\m}{m_{A'}}

\newcommand{\JEM}{J_{\text{EM}}}

\newcommand{\Eobs}{\Ebold^{\text{obs}}}
\newcommand{\Eobsm}{E^{\text{obs}}}

\newcommand{\Jeff}{\Jbold_{\text{eff}}}

\newcommand{\tcoherence}{t_{\text{coherence}}}

\newcommand{\Btwomax}{B_{2,\text{max}}}
\newcommand{\BtwomaxTilde}{\tilde{B}_{2,\text{max}}}
\newcommand{\tave}{t_{\text{ave}}}

\newcommand{\SNR}{\text{SNR}}

\newcommand{\tauceff}{\tau_{c,\text{eff}}}

\newcommand{\Psignal}{P_{\text{signal}}}
\newcommand{\rhosignal}{\rho_{\text{signal}}}
\newcommand{\Pabsorb}{P_{\text{absorb}}}
\newcommand{\Acavity}{A_{\text{cavity}}}

\newcommand{\ncmax}{n_{c,\text{max}}}
\newcommand{\Rbot}{R_{\text{bot}}}

\newcommand{\tSNR}{t_{\text{SNR}}}
\newcommand{\tsignal}{t_{\text{signal}}}
\newcommand{\tdet}{t_\text{det}}
\newcommand{\gdet}{\gamma_{\text{det}}}
\newcommand{\Qdet}{Q_{\text{det}}}
\newcommand{\QDM}{Q_{\text{DM}}}
\newcommand{\zmax}{z_{\text{max}}}
\newcommand{\Nsw}{N_{\text{switch}}}
\newcommand{\Qdeteff}{Q_\text{det,eff}}
\newcommand{\Reff}{\mathscr{R}_{\text{eff}}}

\newcommand{\Bext}{B_{\text{ext}}}
\newcommand{\Bextbold}{\mathbf{B}_{\text{ext}}}
\newcommand{\g}{g_{a\gamma \gamma}}
\newcommand{\Cagg}{C_{a\gamma \gamma}}
\newcommand{\Aa}{A_a}
\newcommand{\Aabold}{\mathbf{A}_a}

\newcommand{\ETE}{E^{\text{TE}}}
\newcommand{\ETM}{E^{\text{TM}}}
\newcommand{\ETMr}{E^{\text{TM}}_{r,0np}}
\newcommand{\ETMrs}{E^{\text{TM}*}_{r,0np}}

\newcommand{\ETMros}{E^{\text{TM}*}_{r,01p}}
\newcommand{\ETMt}{E^{\text{TM}}_{\theta',0np}}
\newcommand{\ETMts}{E^{\text{TM}*}_{\theta',0np}}

\newcommand{\ETMtos}{E^{\text{TM}*}_{\theta',01p}}

\newcommand{\Pon}{P^0_n(\cos\theta')}

\newcommand{\Jhat}{\hat{J}}
\newcommand{\Jhatn}{\hat{J_n}}
\newcommand{\Jhato}{\hat{J_1}}

\newcommand{\kTMnp}{k^{\text{TM}}_{0np}}
\newcommand{\kTMop}{k^{\text{TM}}_{01p}}
\newcommand{\up}{u'_{1p}}
\newcommand{\ksphere}{\kappa_{\text{sphere}}}
\newcommand{\kcylinder}{\kappa_{\text{cylinder}}}
\newcommand{\Gcylinder}{G_{pq}^{\text{cylinder}}}
\newcommand{\Gcylinderstar}{G_{p_*q_*}^{\text{cylinder}}}
\newcommand{\GcylinderNstar}{G_{N_*}^{\text{cylinder}}}
\newcommand{\Gsphere}{G_{p}^{\text{sphere}}}
\newcommand{\Gspherestar}{G_{p_*}^{\text{sphere}}}

\newcommand{\pstar}{{p_*}}
\newcommand{\wpstar}{\omega_{\pstar}}
\newcommand{\wpq}{\omega_{1pq}}

\newcommand{\wNstar}{\omega_{N_*}}
\newcommand{\wN}{\omega_{N}}

\newcommand{\Lag}{\mathcal{L}}

\newcommand{\Prob}{\mathscr{P}}
\newcommand{\adag}{a^{\dag}}

\newcommand{\del}{\bold{\nabla}}

\newcommand{\Afree}{\Abold'_{\text{free}}}

\newcommand{\Efree}{\Eobs_{\text{free}}}

\newcommand{\Ecavity}{\Eobs_{\text{cavity}}}

\newcommand{\Ecavityxm}{\Eobsm_{\text{cavity},x}}
\newcommand{\Efreexm}{\Eobsm_{\text{free},x}}

\newcommand{\Hfree}{H_{c,\text{free}}(t)}
\newcommand{\kappaBREAD}{\kappa_{\text{BREAD}}}

\newcommand{\psibold}{\boldsymbol{\psi}}

\newcommand{\Abold}{\mathbf{A}}
\newcommand{\Bbold}{\mathbf{B}}

\newcommand{\Ebold}{\mathbf{E}}

\newcommand{\Jbold}{\mathbf{J}}

\newcommand{\kbold}{\mathbf{k}}

\newcommand{\pbold}{\mathbf{p}}

\newcommand{\rbold}{\mathbf{r}}

\newcommand{\vbold}{\mathbf{v}}

\newcommand{\xbold}{\mathbf{x}}

\usepackage{hyperref}
\hypersetup{
    colorlinks,
    citecolor=black,
    filecolor=black,
    linkcolor=black,
    urlcolor=black
}

\begin{document}
\title{Highly Excited Electron Cyclotron for QCD Axion and Dark-Photon Detection}
%\date{October 7, 2024}
\author{Xing Fan}
\affiliation{Center for Fundamental Physics, Department of Physics and Astronomy, Northwestern University, Evanston, Illinois 60208, USA}
\affiliation{Department of Physics, Harvard University, Cambridge, Massachusetts 02138, USA}
\author{Gerald Gabrielse}
\affiliation{Center for Fundamental Physics, Department of Physics and Astronomy, Northwestern University, Evanston, Illinois 60208, USA}
\author{Peter W.~Graham} 
\affiliation{Stanford Institute for Theoretical Physics, Department of Physics, Stanford University, Stanford, California
94305, USA}
\affiliation{Kavli Institute for Particle Astrophysics \& Cosmology, Department of Physics, Stanford University, Stanford, California 94305, USA}
\author{Harikrishnan Ramani} 
\affiliation{Stanford Institute for Theoretical Physics, Department of Physics, Stanford University, Stanford, California 94305, USA}
\affiliation{Department of Physics and Astronomy, University of Delaware, Newark, Delaware 19716, USA}
\author{Samuel S. Y. Wong}
\email{samswong@stanford.edu}
\affiliation{Stanford Institute for Theoretical Physics, Department of Physics, Stanford University, Stanford, California 94305, USA}
\author{Yawen Xiao}
\email{ywxiao@stanford.edu}
\affiliation{Stanford Institute for Theoretical Physics, Department of Physics, Stanford University, Stanford, California 94305, USA}

\begin{abstract} %note: in revtex4-2 class, abstract must be before maketitle

We propose using highly excited cyclotron states of a trapped electron to detect meV axion and dark photon dark matter, marking a significant improvement over our previous proposal and demonstration [Phys.~Rev.~Lett.~129, 261801]. When the axion mass matches the cyclotron frequency $\omega_c$, the cyclotron state is resonantly excited, with a transition probability proportional to its initial quantum number, $n_c$. The sensitivity is enhanced by taking $n_c \sim 10^6 \left( \frac{0.1~\text{meV}}{\omega_c} \right)^2$.
By optimizing key experimental parameters, we minimize the required averaging time for cyclotron detection to $t_{\text{ave}} \sim 10^{-6} $ seconds, permitting detection of such a highly excited state before its decay. An open-endcap trap design enables the external photon signal to be directed into the trap, rendering our background-free detector compatible with large focusing cavities, such as the BREAD proposal, while capitalizing on their strong magnetic fields. Furthermore, the axion conversion rate can be coherently enhanced by incorporating layers of dielectrics with alternating refractive indices within the cavity. Collectively, these optimizations enable us to probe the QCD axion parameter space from 0.1 meV to 2.3 meV (25--560 GHz), covering a substantial portion of the predicted post-inflationary QCD axion mass range. This sensitivity corresponds to probing the kinetic mixing parameter of the dark photon down to $\epsilon \approx 2 \times 10^{-16}$.
\end{abstract}

\maketitle

\tableofcontents

\section{Introduction}
The particle nature of dark matter is one of the major open problems in fundamental physics~\cite{DMDiscovery1933,DMRotationCurve1980,DMGravitationalLensing2006,DMBulletCollision2000,DM_WMAPGalaxyCenter2007,Planck2018-1}. Since the dark matter mass is unknown, dark matter candidates can be categorized by their mass and spin. Ultralight bosons are a broad class of dark matter candidates for which the dark matter has a mass much lighter than 1 eV and has a macroscopically large occupation number, manifesting as a coherent classical wave~\cite{FuzzyCDM2000,DarkPhotonMisalignmentMechanism2011,ATheoryOfDarkMatter2009,WISPyDarkMatter2012,ReviewScalarFieldDarkMatter,UltraLightScalarCosmologicalDarkMatter}. 
In most models, no symmetry forbids their coupling to Standard Model photons, giving rise to the exciting prospect of detecting them in the form of precision measurements of electromagnetic waves~\cite{DPLimitsReview2021,HuntForDarkPhoton2020,ADMXSideCar2018,ADMXTechnicalDesignReview2021,CAPP2020_7ueV,CAPP2020_13ueV,CAPP2021_10ueV,Chiles:2021gxk,Hochberg:2021yud,SinglePhotonReview2011}. In this paper, we propose a new detection scheme for such dark matter models in the $0.1-2.3$ meV mass range using a trapped electron in a highly excited state, as a vastly improved version of our previous proposal \cite{Fan_2022}. Related ion trap technologies have also been proposed to search for millicharged particles \cite{MillichargeIonTrap}, and superconducting qubits have been proposed for lower frequencies \cite{TransmonQubit,transmonAxion}.

Among various kinds of ultralight bosons, the most well-motivated model is the QCD axion \cite{AxionSearchExp,AxionThyReview} that could potentially explain dark matter and also solve the strong CP problem \cite{StrongCP1,StrongCP2}. The QCD axion is a pseudo-Goldstone boson associated with the breaking of the hypothetical global $U(1)_{\text{PQ}}$ symmetry at an unknown high energy scale $f_a$ \cite{WeinbergAxion,HarmlessAxion}. Due to its coupling with QCD, the axion has a mass of $m_a \approx 0.6 ~\text{meV} \left(\frac{10^{10} ~\text{GeV}}{f_a} \right)$ \cite{DMRadioGUT}. Its mixing with pions generates a photon coupling given by the Lagrangian \cite{DMRadioGUT}
\begin{align}
\Lag &\supset -\frac{\g}{4}a F_{\mu\nu}\tilde{F}^{\mu\nu}+\frac{1}{2}m_a a^2 \\
&=\g a \Ebold \cdot \Bbold +\frac{1}{2}m_a a^2 ~,
\end{align}
where $a$ is the axion field and the axion-photon coupling constant $\g=\Cagg \alpha/(2\pi f_a)$ is proportional to its mass $m_a$, due to their mutual dependence on $f_a$. $\Cagg$ is an $\mathcal{O}(1)$ model-dependent parameter \cite{DMRadioGUT}; for KSVZ model, $\Cagg \sim - 1.92$ \cite{KSVZ1,KSVZ2}; for DFSZ model, $\Cagg \sim 0.75$ \cite{HarmlessAxion, DFSZ2}. More generally, axion-like-particles (ALPs) \cite{Graham:2013gfa,Hewett:2012ns,Ringwald:2012cu,Ringwald:2012hr,Baker:2011na,Arias:2010bh,Jaeckel:2010ni,Ehret:2010mh,OSQAR:2011xex,Battesti:2010dm,Conlon:2006tq,Arvanitaki:2009fg,Acharya:2010zx,Pospelov:2012mt} are a broader class of dark matter candidates that do not solve the strong CP problem \cite{ALPObs}, so that $m_a$ and $\g$ need not be related. The specified Lagrangian implies that axions can be converted to photons in a background magnetic field.

The post-inflationary QCD axion \cite{InflationaryAxion1,InflationaryAxion2,InflationaryAxion3} is of particular interest. If the $U(1)_{PQ}$ symmetry is broken after inflation, then the observed dark matter relic density predicts a unique axion mass, which was recently estimated to be in the range $m_a \in (0.04~\text{meV},0.18~\text{meV})$~\cite{PostInflationAxionMass}. As we shall see, the proposed new detection scheme probes a large region of axion parameter space and, in the most powerful incarnation, reaches down to the QCD axion line in the $0.1-2.3$~meV mass range, thus covering a significant fraction of this particularly well-motivated region.

Another ultralight bosonic dark matter candidate is the dark photon~\cite{DPTheory1986,Okun:1982xi,Graham:2015rva,Dror:2018pdh,Agrawal:2018vin,LowEnergyFrontierReview2010, Ahmed:2020fhc,GravitationalProductionOfDP2021,Co_2019,Co_2021}, a massive vector boson associated with a dark $U(1)'$ gauge symmetry. This can be considered a minimal extension of the Standard Model, with the dark photon kinetically mixed with the Standard Model photon via the Lagrangian 
\begin{align}
    \mathcal{L}\supset -\frac{1}{4}F'_{\mu\nu}F'^{\mu\nu} + \frac{\epsilon}{2} F^{\mu\nu}F'_{\mu\nu}+\frac{1}{2}m_{A'}^2A'_\mu A'^\mu.
\end{align}
Here, $A'_{\mu}$ and $F'_{\mu \nu}$ denote the dark photon and the dark photon field strength, respectively; $\m$ is the dark photon mass; and $\epsilon$ is the kinetic mixing parameter.
Its phenomenology is similar to that of the axion, but its detection does not require a strong magnetic field since no such field is required for dark-photon-to-photon conversion.

\subsection{Outline and Summary}

In Ref.~\cite{Fan_2022}, we proposed and demonstrated that a single electron suspended in a Penning trap could be a novel quantum detector with the potential for very sensitive searches for ultralight dark matter. A 75-times higher detection sensitivity for dark photons at a frequency of 148 GHz (0.6 meV) was established. Section~\ref{sec: Experiment} reviews the novel method that can be used to search for ground-state-to-first-excited-state transitions of the electron cyclotron motion that are resonantly driven by a dark photon dark matter, when its mass matches the cyclotron frequency. Details are provided in Appendix \ref{sec: transition rate in free space}. The cyclotron state of the electron was monitored in real-time via a quantum non-demolition measurement \cite{QuantumCyclotron}. The electron cyclotron could be so well isolated from Standard Model photons from all other sources that no background was anticipated.  Indeed, the demonstration measurement was completely background-free over a 7-day measurement time. By tuning the trap magnetic field, the cyclotron frequency $\wc$ could be scanned across the $0.1-2.3$ meV mass range. The demonstration measurement was not sensitive to axions because the magnetic field of the Penning trap unavoidably converts axions to an electric field oriented perpendicular to the cyclotron plane---the one field direction that the one-electron detector cannot detect. 
 
This paper proposes the addition of three new methods to enable vastly more sensitive dark photon searches, and to add high sensitivity to the most well-motivated QCD axions with masses between 0.1 to 2.3 meV, a challenging range for current technologies.
\begin{enumerate}
    
\item Section  \ref{sec: electron in a highly excited state} shows how detection sensitivity will be greatly increased by performing rapid measurements on a highly excited cyclotron state.  The large cyclotron quantum number enhances photon absorption, thereby increasing the dark matter signal size. The rapid measurement enables the observation of a dark-matter-induced excitation before the decay of the highly excited state erases the signal.

\item Section \ref{sec:effects of cavity} shows how high axion sensitivity can be achieved by using a large cavity that is optimally designed to convert ultralight dark matter to signal photons, as illustrated in a different context by BREAD \cite{BREAD}.  Our cavity can be designed to efficiently focus  and rotate the polarization of the signal photons for efficient, background-free detection by a 
trapped-electron detector. 

\item Dielectric layers \cite{MultilayerOpticalHaloscopes}, also discussed in Section \ref{sec:effects of cavity},  will be used to further enhance the conversion of ultralight dark matter to signal photons. 

\end{enumerate}

\begin{figure}[htbp!]
    \centering
    \begin{subfigure}{\textwidth}
        \centering
        \includegraphics[width=0.75\textwidth, bb=0 0 850 600]{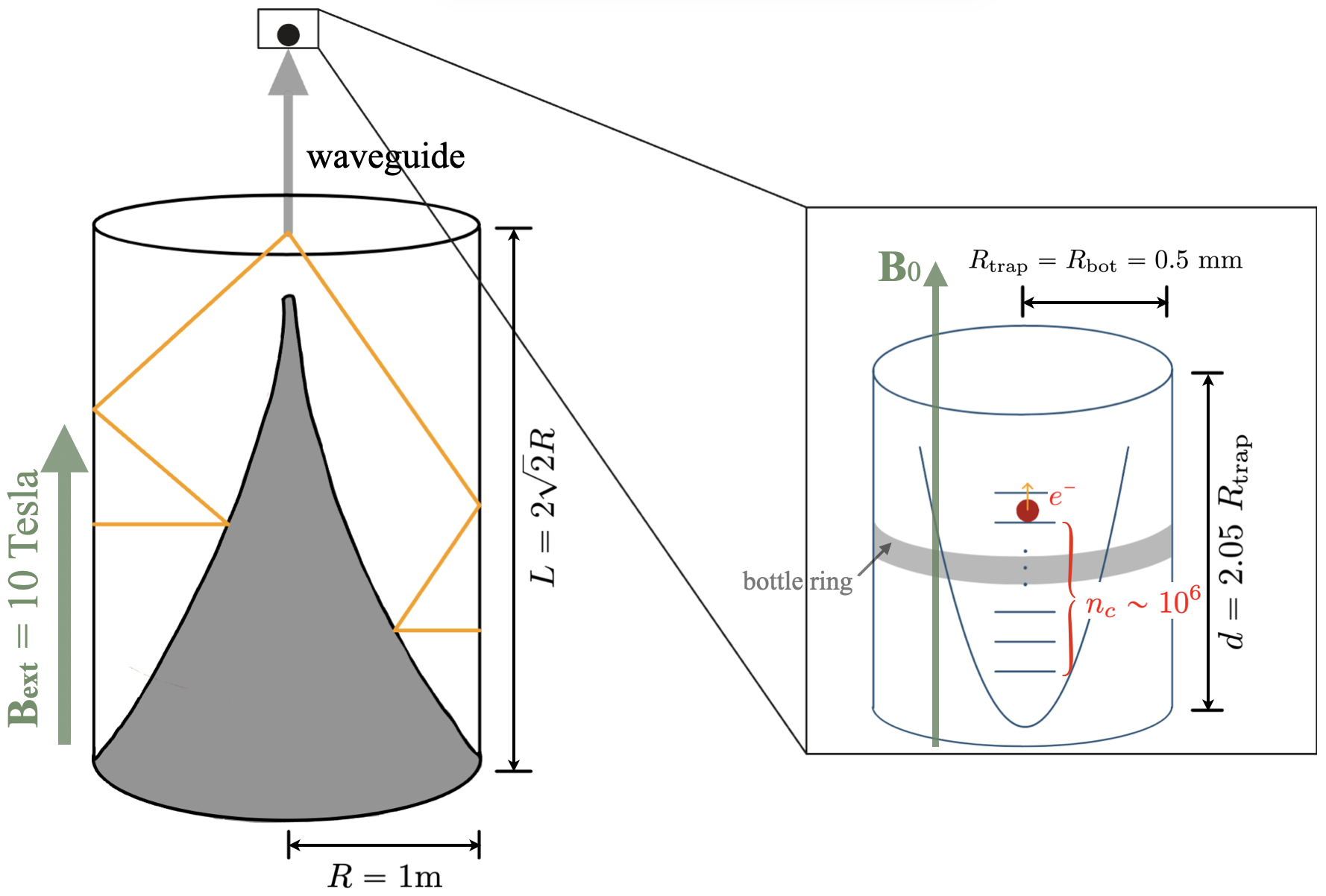}
        \caption{Schematic diagram of the proposed experimental design.}
        \label{fig: BREAD open trap}
    \end{subfigure}
    
    \vspace{1cm} % Adjust the vertical space between the subfigures
    
    \begin{subfigure}{\textwidth}
        \centering
        \includegraphics[width=0.5\textwidth, bb=0 0 800 600]{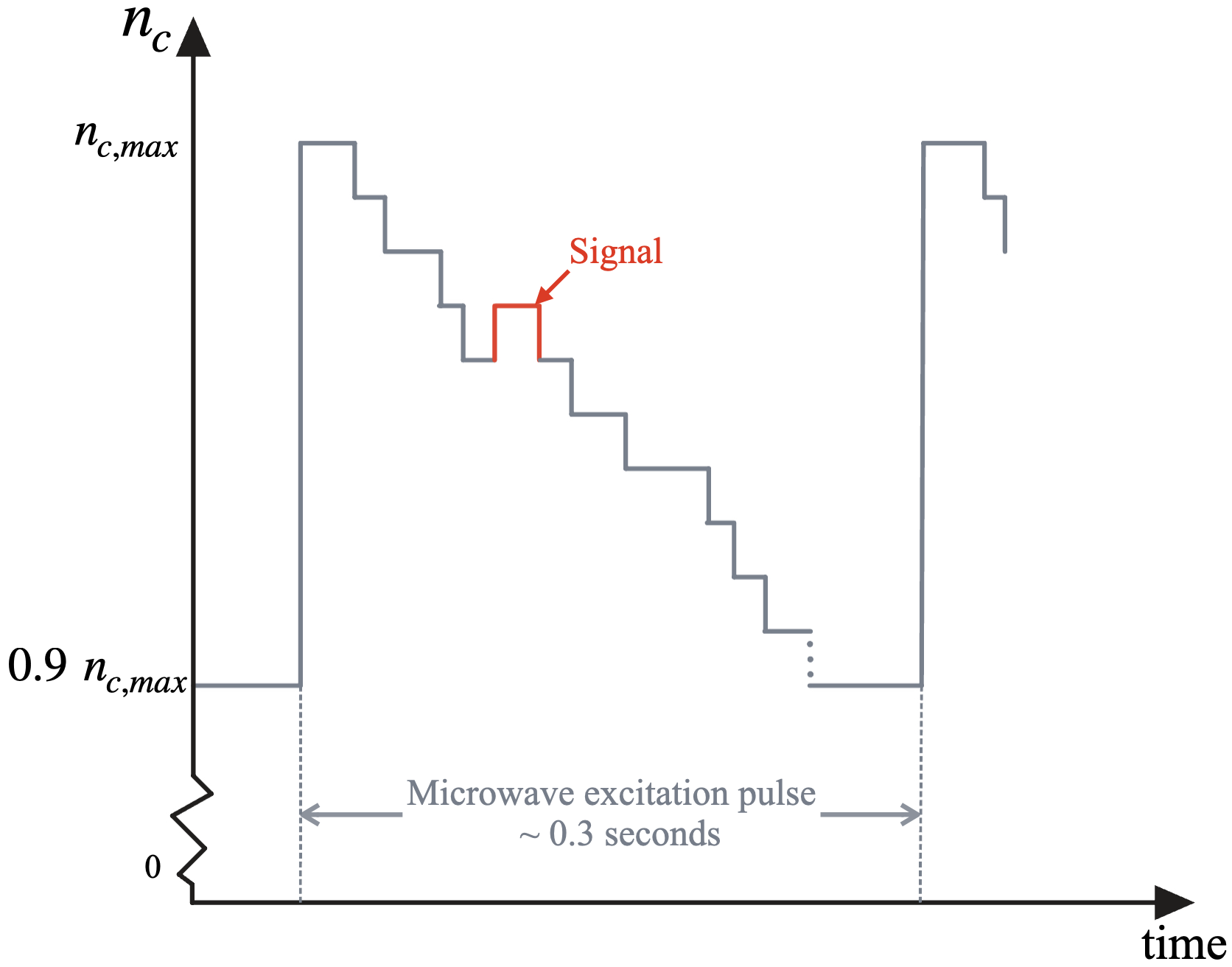}
        \caption{Typical signal form}
        \label{fig: recognize signal}
    \end{subfigure}
    
    \caption{(a) A BREAD-like cylindrical barrel of radius $R=1$~m in a $\Bext=10$~T external magnetic field filled with dielectric layers of alternating refractive indices (not shown) resonantly converts axions into photons (yellow lines) that are subsequently focused into a single point. The polarization of the focused signal is rotated and directed by a waveguide through the opening of an open-endcap Penning trap to a trapped electron, prepared at a highly excited cyclotron state of \textcolor{\specialColor}{$n_c \sim 10^6  \left( \frac{0.1~\text{meV}}{\wc} \right)^2$}. In the figure, we showed the largest value at cyclotron frequency $\wc=0.1$~meV for concreteness.  Figure of BREAD adapted from Ref.~\cite{BREAD}. (b)Although cyclotron jumps can occur in both upward and downward directions, we identify only the upward jumps as our signal to avoid contamination from spontaneous decays. Consequently, the upward jump (shown in red) is recognized as the signal amid numerous decays (shown in gray). To prevent loss of sensitivity, the cyclotron state is never permitted to drop below 90\% of its maximum value by applying a periodic driving pulse with period $\Delta t_\text{pulse} =  0.34\text{ s}\times\left( \frac{0.1~\text{meV}}{\wc} \right)^2$.}
    \label{fig: detection}
\end{figure}

The basic idea of the sensitive measurement incorporating the new methods is represented in a simplified way in Figure~\ref{fig: BREAD open trap}.  A coherent state of the electron's cyclotron oscillator, with an average quantum number of about $10^6\left( \frac{0.1~\text{meV}}{\wc} \right)^2$, is generated using a short driving pulse. Nearly continuous detection of the cyclotron energy will cause the electron to reveal its cyclotron quantum state.  The detection will be provided by monitoring the frequency of the electron's axial oscillation along the magnetic field direction, which is coupled to the cyclotron energy by a strong magnetic bottle gradient.  A series of quantum jumps will be observed as the electron radiates its cyclotron energy into synchrotron radiation one quantum at a time. The signature of dark matter will be any one-quantum increases in cyclotron energy that are caused by either a photon dynamically coupled to a dark photon, or by a photon produced by an axion in a magnetic field.  

Sensitivity projections are presented in Section \ref{sec: projections} and Section \ref{sec: conclusion} as a conclusion.
In Appendix \ref{sec: transition rate in free space}, we provide a rigorous derivation of the sensitivity of our electron cyclotron detection scheme using Rabi flopping and a ``memory-loss model" to account for frequency width. In Appendix \ref{sec: backreaction}, we estimate the backreaction limit, which is a fundamental limit to all dish-type experiments, and compute the photon detection efficiency. In Appendix~\ref{sec:large nc appendix}, we provide details on how to achieve a large cyclotron number. Appendix~\ref{sec: cavity mode appendix} and~\ref{sec:cylindrical focusing} provide details of the cavity modes calculation and focusing. In Appendix \ref{sec:fixed polarization}, we have worked out the consequence of a fixed dark photon polarization scenario for our experiment, which is representative of a class of experiments not addressed in Ref.~\cite{DPLimitsReview2021}: a directionally sensitive quantum sensor with discrete signals. We also provide a table of all required experimental parameters in Appendix~\ref{sec: experimental parameters}.

\begin{figure}[htbp!]
    \centering
    \includegraphics[width=0.6\textwidth]{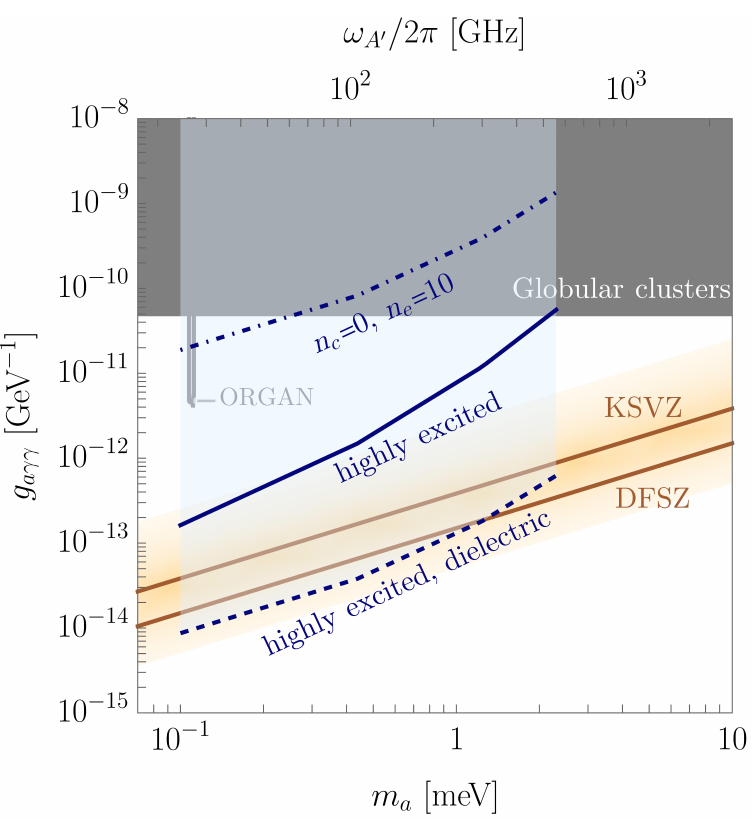}
    %, bb=0 0 370 370
    \caption{Projected sensitivity to axion dark matter. The blue dash-dotted line represents the projected reach incorporating only a conversion cavity of radius $R=1$~m in a $\Bext=10$~T external magnetic field, while retaining the previous detection scheme of starting at the cyclotron ground state $n_c=0$ and trapping $n_e=10$ electrons. The blue solid line incorporates the same conversion cavity but starts from an initial \textcolor{\specialColor}{$n_c \sim 10^6  \left( \frac{0.1~\text{meV}}{\wc} \right)^2$} excited states, which is only possible with one electron, as explained in Section \ref{sec:more electrons}. The blue dashed line additionally includes dielectric layers that enhance axion conversion. The dielectrics are swapped out once a month to make this method compatible with frequency scanning. In all cases, we assume a 1000-day search per decade in frequency. The ``kinks" near 0.5 meV are due to the saturation of the focusing limit. The dark red lines are the prediction of the KSVZ \cite{KSVZ1,KSVZ2} and DFSZ \cite{HarmlessAxion,DFSZ2} axion models. Shaded in gray are nearby bounds from stellar cooling \cite{globularCluster} and the ORGAN experiment \cite{ORGAN}.}
    \label{fig: axion}
\end{figure}

\begin{figure}[htbp!]
    \centering
    \includegraphics[width=0.6\textwidth, bb=0 0 370 370]{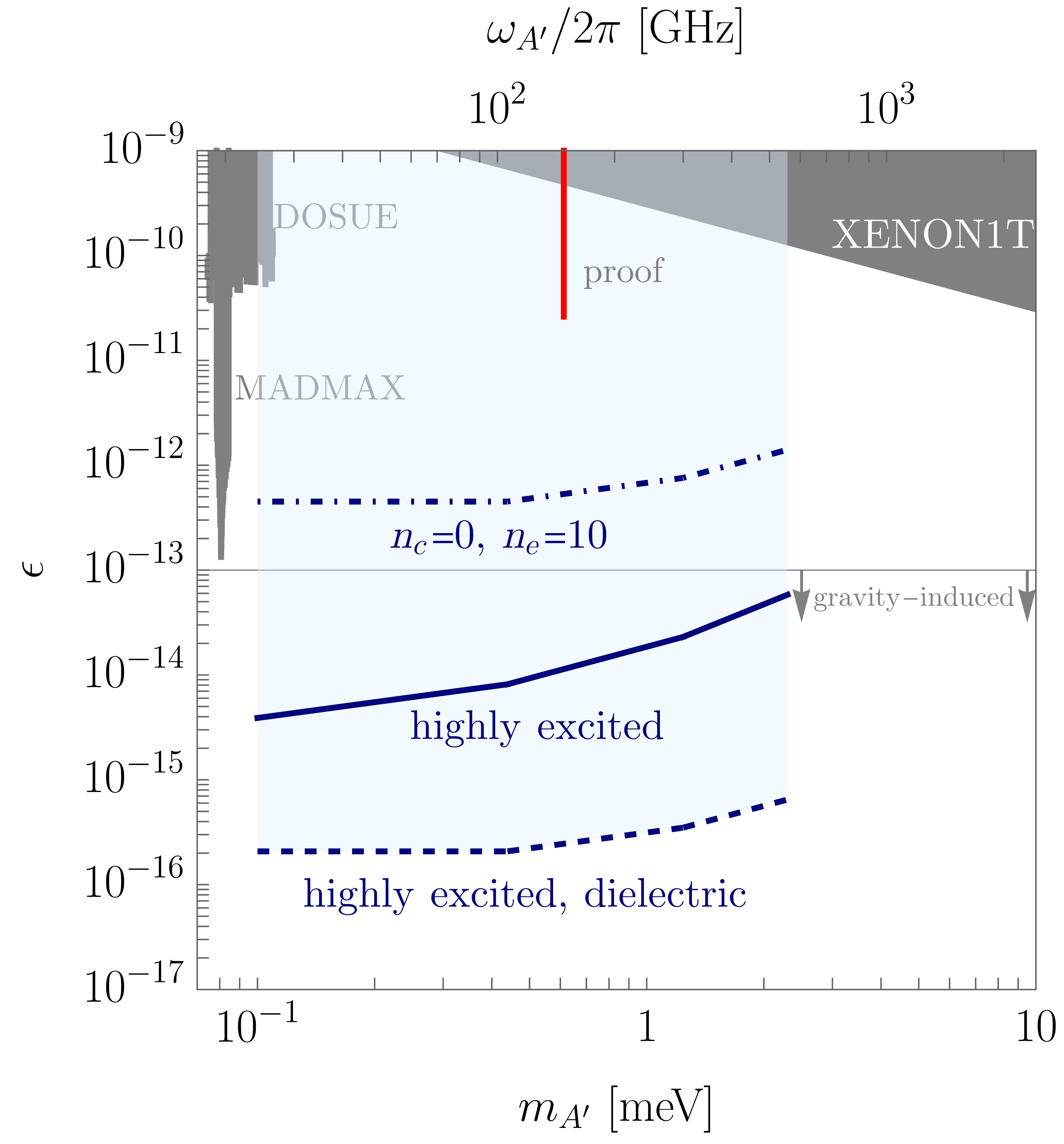}
    \caption{Projected sensitivity to dark photon dark matter. The three blue projection lines are based on the same setup as those in Figure~\ref{fig: axion}. The red vertical line represents the limit set by our proof-of-principle measurement \cite{Fan_2022}. Shaded in gray are nearby bounds from the DOSUE~\cite{DOSUE2022_100ueV}, MADMAX~\cite{MADMAX2024}, and XENON1T~\cite{Xenon1TResultDarkPhoton} experiments. The regions below $\epsilon \sim 10^{-13}$, as indicated by the arrows, may be induced by gravity alone~\cite{PriceTinyKineticMixing}.}
    \label{fig: dark photon}
\end{figure}

\section{Electron Cyclotron as a Dark Matter Detector\label{sec: Experiment}}

\subsection{{Electron Cyclotron Detector}}
\label{sec: 2A}

An electron suspended in a Penning trap was recently used to measure the electron magnetic moment \cite{ElectronMagneticMomentPRL2022} at an unusual precision, approaching 1 part in $10^{13}$, to test the most precise prediction of the Standard Model 
\cite{atomsNewMeasurement2019,atomsTheoryReview2019}.  There is a long history of such measurements \cite{DehmeltMagneticMoment,HarvardMagneticMoment2008,HarvardMagneticMoment2011,ElectronMagneticMomentPRL2022}. The key advance that enabled the great precision of the measurements was the realization of a ``one-electron quantum cyclotron'' \cite{QuantumCyclotron}. The cyclotron motion of one suspended electron was cooled below 100 mK, whereupon the electron occupied only the ground state of this motion.  Sensitive quantum non-demotion (QND) detection was then employed to detect driven,  single quantum transitions to the first excited cyclotron state.  This was the detector that we recently demonstrated could be used to detect meV dark photons with greatly improved sensitivity \cite{Fan_2022}.  

In a Penning trap \cite{Review}, an electron is effectively confined in three dimensions using a combination of a spatially uniform magnetic field $\Bbold = B_0 \hat{z}$ in the axial direction and a static quadrupole potential 
\begin{align}V(\rbold)=\frac{1}{2}m_e\wz^2 \left(z^2-\frac{x^2+y^2}{2}\right).
\end{align}
In this expression we neglect small, higher order anharmonic terms (e.g.\ $x^4$, $y^4$, $z^4$) and the effects of trap electrode imperfections.   
The magnetic field in a symmetric gauge is given by the vector potential $\Abold = \frac{1}{2} \Bbold \times \rbold$. 
In $V$, the single constant $m_e \wz^2 $ is written as the product of the electron mass and the square of an axial angular frequency, for convenience.  The $B$-field produces radial cyclotron motion at a cyclotron frequency modified slightly by $V$. The potential $V(\rbold)$ traps the electron in the axial direction in a harmonic axial motion with axial frequency $\wz$. 

For most of the measurements mentioned above, including the dark photon demonstration measurement, the potential was produced using cylindrical Penning trap electrodes~\cite{CylindricalPenningTrap}, with electrodes on the surface of a cylinder and on flat ends. For this proposal, an open-endcap electrode geometry \cite{OpenTrap} is considered instead, with trap electrodes that are coaxial cylinders.  This choice makes it possible to couple the signal photons from the conversion cavity into the Penning trap within which the electron is centered. 

Neglecting relativistic effects, the Hamiltonian for an electron in an ideal Penning trap is
\begin{align}
    H_0 &= \frac{(\pbold - e\Abold)^2}{2\me} + e V \label{eq:first line H0}.
\end{align}
The axial motion separates naturally from the radial motion.  The axial part of $H_0$ is that of a simple harmonic oscillator with angular axial frequency $\omega_z = \frac{\sqrt{eV_0/m_e}}{D}$,
\begin{align}
 H_z &= \frac{p_z^2}{2 m_e} + \frac{1}{2} \me \omega_z^2 z^2  \\
 &= \hbar \wz \left(\adag_z a_z + \frac{1}{2} \right)~,  
\end{align}
where the effective trap dimension is given by $D^2=\Rtrap^2+\frac{1}{2}d^2$, and $\Rtrap$ and $d$ are the radius and height of the cylindrical trap, respectively.
The second expression uses harmonic oscillator lowering and raising operators, $a_z$ and $a_z^\dagger$, that are familiar functions of $z$ and $p_z$.

The radial motions and Hamiltonian are more complicated because the vector and scalar potentials both depend upon $x$ and $y$.  However, Ref.~\cite{Review} introduced two sets of harmonic oscillator raising and lowering operators that separate the radial Hamiltonian into uncoupled cyclotron and magnetron terms, 
\begin{align}
H_r = \hbar \wcp \left(\adag_c a_c +\frac{1}{2} \right)
- \hbar \wm \left(\adag_m a_m + \frac{1}{2} \right). 
\end{align}
 To an approximation that suffices for our purposes, the angular cyclotron frequency modified by the electrostatic quadrupole, $\wcp \approx \wc$, where $\wc=e B_0/m_e$ is the free-space cyclotron frequency.  Also to an approximation that suffices for our purposes, the magnetron frequency is $\wm \approx \wz^2 /(2\wc)$.  
Both the cyclotron lowering and raising operators ($a_c$ and $\adag_c$) and the magnetron lowering and raising operators ($a_m$ and $\adag_m$) are complicated but manageable functions of $x$, $y$, $p_x$, and $p_y$. The raising and lowering operators for each of the separated motions satisfy the same commutator relations as those for a harmonic oscillator. For example, $[a_c,\adag_c]=1$.

The energy eigenstates of $H_0 = H_z + H_r$ are direct products of oscillator eigenstates, $\ket{n_c,n_z,n_m}$.  
The lowering and raising operators each operate on their respective oscillator states.  For example,
\begin{align}
a_c \ket{n_c,n_z,n_m} &= \sqrt{n_c}  \ket{n_c-1,n_z,n_m}\\  
\adag_c \ket{n_c,n_z,n_m} &= \sqrt{n_c+1}  \ket{n_c+1,n_z,n_m}~.   
\end{align}
The energy eigenvalues are  
\begin{align} 
E_0(n_c,n_z,n_m) &= \left(n_c + \frac{1}{2} \right) \hbar \wc  + \left(n_z + \frac{1}{2} \right) \hbar \wz - \left(n_m + \frac{1}{2} \right) \hbar \wm.
\end{align}
The angular cyclotron frequency $\wc$ is much greater than the angular axial frequency $\wz$, which in turn is much greater than the magnetron frequency, $\wm$.  
For this work, the small difference between the trap-modified cyclotron frequency and the free-space cyclotron frequency is neglected, and the magnetron is ignored.

Consider detecting one-quantum transitions from the cyclotron state $n_c$ to $n_c+1$. The cyclotron energy and state are difficult to detect directly because the cyclotron frequency is in the mm microwave regime. The thousand times lower frequency of the axial oscillation can be much more easily measured.  To couple this frequency to the cyclotron energy, a magnetic ring that encircles the cylindrical trap electrodes is added \cite{DehmeltMagneticBottle,FanBackActionPRL2021} to make a small magnetic gradient, $B_2 z^2 \hat{z}$, for an electron on the trap axis (with $x=0$ and $y=0$).  This adds
\begin{align} \label{eq: Coupling Hamiltonian}
   E_{cz} = \hbar \delta \left(n_c + \frac{1}{2} \right) \left(n_z + \frac{1}{2} \right)
\end{align}
to the energy eigenvalues, with  
\begin{align} \label{eq:deltaexpression}
\delta = \frac{e B_2}{m_e^2 \wz} ~.
\end{align}
The axial frequency then effectively gains a small dependence on the quantum number $\nc$, and this effect is proportional to $\delta$. In effect, whenever the cyclotron state makes a transition from $n_c$ to  $n_c +1$, the axial frequency shifts accordingly
\begin{align}
    \wz \to \wz + \delta ~.
\end{align}
This shift is large enough with current experimental methods so that the quantum state of the cyclotron motion can be unambiguously determined within a fraction of a second.  This way of detecting the cyclotron energy as a shift of the axial frequency is a quantum nondemolition (QND) measurement, insofar as the addition to the Hamiltonian that produces $E_{cz}$ commutes with $H_0$. The important consequence is that repeated measurements of the cyclotron energy do not change the cyclotron quantum state \cite{FanBackActionPRL2021}. 

The above description holds for small cyclotron quantum number $n_c \lesssim 10$. We will address how this is modified when introducing the technique of highly excited states in Section~\ref{sec: electron in a highly excited state}.

The axial frequency is measured by Fourier transforming the oscillating image current generated in the wall of the trap due to the axial motion.  By monitoring this axial frequency, the cyclotron quantum number $n_c$, and hence the quantum state, is continuously monitored.  The ``continuous'' axial detection has an unavoidable averaging time $\tave$ that is needed to distinguish a cyclotron transition signal from Johnson noise. A measured axial frequency shift, $\Delta \wz$, could be caused by some combination of a cyclotron energy shift and noise in the detection electronics. For a given averaging time, we identify a $\Delta \wz$ as a cyclotron excitation if $\Delta \wz$ is larger than $\delta$ by 5 standard deviations of the signal during this averaging time.

A dilution refrigerator keeps the trap electrodes at a temperature that we will assume is $T=10~\text{mK}\approx 10^{-3}~\text{meV}$ so that black body photons do not have enough energy to excite the cyclotron motion, which has an energy threshold of $\hbar \wc = 0.1 \text{--} 2.3$~meV for our search. The theoretical black body excitation rate is \begin{equation}
    \Gamma_\text{BB}=  n_c\gamma_{c,0}\frac{1}{\exp\left(\frac{\omega_c}{T}\right)-1}
    %\simeq n_c\gamma_{c,0} \exp\left(-\frac{\omega_c}{T}\right)
    \approx n_c \times 10^{-51}~\text{s}^{-1},
\end{equation}
where $\gamma_{c,0}={4 \alpha\wc^2}/{(3\me)}\approx$0.3--100~s$^{-1}$ is the ground state cyclotron damping rate in the trap. Even for cyclotron states as high as \textcolor{\specialColor}{$n_c \sim 10^6$} that we will consider in this paper, the Boltzmann suppression makes the background rate negligible.  In the demonstration dark photon measurement \cite{Fan_2022}, we observed no cyclotron excitations for 7 days for a temperature of 30 mK \cite{Fan_2022}. The measurement was thus background-free for 7 days.  

\subsection{Simultaneous Dark Photon and Axion Detection\label{sec:challenges}}

The basic idea of these measurements \cite{Fan_2022} is to search for one-quantum cyclotron transitions caused by dark matter with mass $m$ that is resonant with the cyclotron frequency $\omega_c$ insofar as $mc^2=\hbar \wc$.  A search for dark matter over the largest mass range is done by sweeping the magnetic field to vary the cyclotron frequency over the largest range. In the demonstration apparatus, the magnetic field could likely be reduced so the minimum frequency and energy for the search range is $\wc/(2\pi) = 25~\rm{GHz} \approx 0.1 ~meV/(2\pi\hbar)$. The lowest possible limit depends upon the temperature (to keep the measurement free of background), the relative size of the trap and the cyclotron orbit, and keeping the cyclotron frequency well above the axial frequency.   
In the demonstration apparatus, a maximum frequency $\wc/(2\pi) = 180~\rm{GHz} \approx 0.75 ~meV/(2\pi\hbar)$ would be accessible.  
With a 20 T superconducting solenoid, the maximum could be increased to $\wc/(2\pi) = 560~\rm{GHz} \approx 2.3 ~meV/(2\pi\hbar)$ for the frequency/mass of the dark matter\footnote{{For convenience, we will adopt natural units where $\hbar=c=1$ throughout the remainder of this paper, except in numerical results where we explicitly show the dimensional factors. The conversion between mass and frequency will follow the conventions shown here.  In particular, we will give numbers and label plots by mass in meV and frequency in GHz with this $2 \pi \hbar$ conversion factor between them.  So, for example, a 2.3 meV mass corresponds to a 560 GHz frequency.}}.    

Introducing sensitivity to the axion, in addition to dark photons, requires converting the axion to a photon in a magnetic field away from the Penning trap of the quantum cyclotron. The field converts the axion to a photon polarized along the magnetic field direction.  This polarization must be rotated so as to be perpendicular to the magnetic field within the trap in order to produce cyclotron excitation.  The separated conversion and detection volumes have two additional advantages.  (1) A larger and more efficient conversion ``antenna'' can be used (Section \ref{sec: BREAD}).  (2) The conversion field, and hence the conversion efficiency, can be left fixed at the largest possible value while the field of the one-electron quantum cyclotron is swept over the widest possible range of possible axion masses.

\subsection{Cyclotron Transition Rate}

The rates at which dark photons and axions produce cyclotron excitations are summarized here, as is the constraint that comes from observing no excitation in a given observation time.  The detailed derivation on which this is based is summarized in Appendix \ref{sec: transition rate in free space}.

A dark photon would excite a cyclotron state with quantum number $n_c$ at a rate 
\begin{align}\label{eq:free Gamma final1}
\Gammafree = \frac{\epsilon^2 e^2 \pi (\nc +1)}{2 \me \m} \frac{\rhoDM}{\Delta \omega} \langle \sin^2 \theta \rangle~, 
\end{align}
depending upon a frequency width, $\Delta \omega$. This width is approximately the larger of the cyclotron decay of the state via synchrotron radiation, $\Delta \wc$, and the anticipated width of the dark matter, $\Dw=10^{-6} \m$. $\langle \sin^2 \theta \rangle=\frac{2}{3}$ is the average dark photon polarization. This equation holds true only in free space; the boundary condition of the surrounding conductor gives rise to a cavity factor $\kappa^2$, to be discussed in Section \ref{sec:effects of cavity}:
\begin{align} \label{eq: Gamma cavity dark photon}
\Gammacavity = \kappa^2 \frac{\epsilon^2 e^2 \pi (\nc +1)}{2 \me \m} \frac{\rhoDM}{\Delta \omega} \langle \sin^2 \theta \rangle ~.
\end{align}
%Here, $\kappa^2$ is the cavity factor that captures the boundary condition of the surrounding conductor, to be discussed in Section \ref{sec:effects of cavity}; $\langle \sin^2 \theta \rangle=\frac{2}{3}$ is the average dark photon polarization.

If no excitation is observed over an observation time of $\tobs$, the inequality \begin{align} \label{eq: Gamma cavity inequality}
\Gammacavity < -\frac{1}{\zeta \tobs} \log(1-CL) ~.
\end{align}
expresses a constraint at confidence level $CL=0.9$, where $\zeta$ is the detection efficiency. 

For the axion, we have an analogous formula,
\begin{align} \label{eq: Gamma cavity axion}
  \Gammacavity = \kappa^2  \frac{\g^2 }{m_a^3} \Bext^2 \frac{e^2 \pi (\nc +1) }{2 \me} \frac{\rhoDM}{\Delta \omega} ~,
\end{align}
where $\Bext$ is the external magnetic field that converts axions to photons. The same inequality pertains to the confidence limit.

\section{Electron in a Highly Excited State\label{sec: electron in a highly excited state}}

Our earlier proposal and demonstration was sensitive to any dark matter that was able to excite the cyclotron ground state of one trapped electron into its first excited state. QND detection revealed no background transitions at all for at least days.  With no background, the sensitivity was high enough to realize a 75-times higher dark photon sensitivity at 148 GHz. 

The cyclotron excitation rate in \eq{\ref{eq: Gamma cavity dark photon}} increases linearly as $n_c$ when the electron cyclotron detector is a highly excited cyclotron state rather than its ground state. The higher the cyclotron state, however, the more rapid the decay of the state.  This section explores what is required to realize an enormous sensitivity gain without introducing background.

\subsection{Excited State and Averaging Time\label{sec: excited states}}

The main obstacle to realizing the enhanced sensitivity of a highly excited state is that the decay rate increases with increasing $n_c$. The electric dipole approximation is valid because the largest electron wave function we consider has a size of $0.2$ mm (\eq{\ref{eq: zmax}}), which is still much smaller than the shortest relevant wavelength of a signal photon, $\lambda_{A'} =1.2$ mm.  The selection rules allow for decay only from $n_c$ to $n_c-1$. 
 The free-space decay time for a cyclotron energy eigenstate with quantum number $n_c$, 
\begin{align} \label{eq: tau c}
\tau_{c}= \frac{1}{n_c} \frac{3 m_e c^2} {4\alpha (\hbar \wc)^2} = 3.4 \text{ s} \left( \frac{1}{n_c} \right) \left( \frac{0.1 \text{ meV}}{\hbar \wc}\right)^2= 3.4 \text{ s} \left( \frac{1}{n_c} \right)\left( \frac{24 \text{ GHz}}{\wc/2\pi}\right)^2 ~,
\end{align}
is given by Fermi's Golden Rule.   
The 0.1 s decay time\footnote{Much longer decay times pertained for the demonstration experiment.  There, the lifetime of excited cyclotron states was dramatically increased by the microwave cavity deliberately formed by the trap electrodes \cite{CylindricalPenningTrap,InhibitionLetter}. When all resonant cavity modes had frequencies far from resonance with the TE and TM modes of the trap cavity, this cavity inhibited spontaneous emission by a factor of 200 or more compared to free space \cite{HarvardMagneticMoment2011}. For the measurement in the open Penning trap proposed here, however, inhibited spontaneous emission is not a major factor.} of the first excited state in a 6 T magnetic field decreases inversely with $n_c$. The lifetime also falls inversely as $B^2$ if the field is increased because of the $\wc^2$ dependence.  This will be important when the magnetic field is swept to vary the dark matter energy to which the one-electron detector is sensitive to.  Reducing the resonant frequency of the detector by a factor of 10 increases the excited state lifetime by a factor of 100. 

To avoid missing the signal, the averaging time $\tave$ required to detect a one-quantum transition must be shorter than the cyclotron lifetime $\tau_c$. Therefore, the shortest averaging time we can achieve sets the highest possible $n_c$. However, $\tave$ must be at least as large as the following three time scales.
\begin{enumerate}
    \item $t_{\text{signal}}$, the minimum time it takes a signal to form;
    \item $\tdet$, the time it takes the detector to read out the signal;
    \item $t_{\text{SNR}}$, the integration time needed to achieve a desired signal-to-noise ratio of SNR=5 over Johnson noise.
\end{enumerate}
Thus, we require
\begin{equation}
    \tau_c \geq \tave \geq \max \left( \tsignal, \tSNR, \tdet \right) ~.
\end{equation}
We determine these timescales next to ascertain the largest feasible $n_c$. 

\subsubsection{Signal Formation Time \label{sec: signal formation time}}
When a jump occurs in the cyclotron mode, it takes some time for this change to be reflected in the electron's axial motion and the detected axial image current. The axial oscillation is driven to an amplitude that can be detected, $z_\text{max}$, by applying an RF drive to one of the endcap electrodes. An axial frequency shift of $\delta$ signals a one-quantum cyclotron transition.  The uncertainty principle requires a detection time of at least $1/\delta$ to detect this shift.  This signal settling time, 
\begin{align}
    \tsignal = \frac{1}{\delta} ~,
\end{align}
can be realized using the beat between the axial signal and a reference close to $\omega_z$, as detailed in Appendix \ref{sec: detection time}.  

Achieving the smallest possible signal settling time requires the largest possible magnetic bottle gradient strength, $B_2$, to couple the cyclotron and axial motions, and the lowest possible axial frequency $\wz$, because
\begin{align}  \label{eq: delta and B2}
\delta = \frac{e B_2}{m_e^2 \wz} ~.  
\end{align}
The large $B_2$ that is desired increases in proportion to the saturation magnetization 
of a ferromagnetic ring of radius $\Rbot$ that encircles the trap, and the smaller possible $\Rbot$ is thus desired. For a cobalt-iron bottle ring with the proposed 0.5~mm radius, the saturation is reached at $B=0.1$~T.
The maximum value is then
\begin{align} \label{eq: B2 max}
\Btwomax = 3 \times 10^6 \text{ T}/\text{m}^2 \left( \frac{0.3~\text{mm}}{\Rbot}\right)^2 \equiv \frac{\BtwomaxTilde}{\Rbot^2} ~.
\end{align}

The axial oscillation frequency is given by the expression
\begin{align}
    \wz=\sqrt{\frac{e V_0}{\me D^2}} ~,
\end{align}
where the effective trap dimension is given by $D^2=\Rtrap^2+\frac{1}{2}d^2$.  The trap height is fixed at $d \approx 2 \Rtrap$ due to an anharmonicity constraint \cite{OpenTrap}. This specific geometry ensures that a signal photon is not significantly suppressed when reaching the center of the trap, where the electron is located, even if the longest photon wavelength considered, $\lambda_{A'} = 2\pi/\wc= 12$~mm exceeds the trap aperture~\cite{SubwavelengthAperture,BetheDiffractionSmallHole}. Consequently, the minimum achievable trap and bottle sizes are only constrained by current fabrication limitations, allowing for radii as small as $\Rtrap =\Rbot=0.5$~mm, with the encircling magnetic bottle ring integrated onto the surface of the open trap (see Figure~\ref{fig: BREAD open trap}). 

Furthermore, the voltage $V_0$ can be easily varied between 0 and 100 V, enabling a wide range of possible values for $\wz$. While, in principle, $\wz$ can be minimized to reduce $\tsignal$, it should be noted that the detection time, to be discussed next, exhibits an inverse dependence on $\wz$, ultimately leading us to select $\wz \sim 2\pi \times 10$ MHz as the optimal operating frequency.

Using the maximum achievable $B_2$ for the selected small bottle size and axial frequency, we obtain a remarkably short signal formation time of:
\begin{align}
    \tsignal = \frac{1}{\delta}=2.9 \times 10^{-6} ~\text{s} \left(\frac{\wz/2\pi}{10~\text{MHz}}\right)\left(\frac{\Rbot}{0.5~\text{mm}}\right)^2 ~.
\end{align}

\subsubsection{Detection Time}

The detector used to monitor the axial image current is a resonant RLC circuit~\cite{DehmeltWalls1968} with a bandwidth defined as
\begin{align}
\gdet=\frac{\wz}{\Qdet}  ~,
\end{align}
where the maximum detector quality factor is empirically determined as\footnote{This formula assumes a factor of 2 improvement than currently available by using superconducting wire made of Nb-Ti, since it is in a high magnetic field.}
\begin{align}
\Qdet = 2.8\times 10^5 \left(\frac{1~\text{MHz}}{\wz/2\pi} \right)~.
\end{align}

The resonant circuit requires a time equal to the inverse of the bandwidth to fully build up a detectable signal:
\begin{align}
\tdet= \frac{1}{\gdet} = \frac{\Qdet}{\wz} = 4.46 \times 10^{-4} ~\text{s} \left(\frac{10~\text{MHz}}{\wz/2\pi} \right)^2~.
\end{align}
As a result, while lowering $\wz$ can decrease $\tsignal$, it cannot be done indefinitely, since it will simultaneously increase $\tdet$. Moreover, for an axial frequency shift $\delta$ to be detectable, it must remain smaller than the detector bandwidth, imposing an additional constraint\footnote{This condition can be relaxed if we use more than one drive.},
\begin{align}\label{eq: tsignal lower limit }
    \delta &\leq \gdet \implies \tsignal \geq \tdet ~,
\end{align}
which must be satisfied by the chosen value of $\wz$.

However, the constraint in \eq{\ref{eq: tsignal lower limit }} can also be satisfied by using a smaller effective quality factor, $\Qdeteff < \Qdet$. This adjustment can be achieved by analyzing the resonator's output outside its bandwidth without altering the physical quality factor $\Qdet$~\cite{195pagesDMRadioReview}. Let $q$ denote the ratio between $\Qdeteff$ and $\Qdet$,
\begin{align}
    \Qdeteff \equiv q \times \Qdet~,
\end{align}
which can be varied during data analysis. Under these conditions, the detection time is reduced to:
\begin{align} \label{eq:tave gamma det}
\tdet(q)= \frac{1}{\gamma_{\text{det,eff}}} = \frac{Q_{\text{det,eff}}}{\wz} = 2.72 \times 10^{-6} ~\text{s} \left(\frac{q}{0.0061}\right)\left(\frac{10~\text{MHz}}{\wz/2\pi} \right)^2~.
\end{align}
This approach allows flexibility in optimizing the detection time without compromising the overall quality factor of the system.
While, in principle, we can arbitrarily decrease $q$ to lower the detection time, the SNR time has the opposite dependence on $q$, as we shall soon see, leading us to choose the above optimal value.

\subsubsection{SNR Time}
Another limit on the averaging time is set by $\tSNR$, the duration required to achieve the minimum signal-to-noise ratio above Johnson noise, chosen here as SNR = 5. This value can be estimated from the ratio of the signal energy to the noise energy:
\begin{align}
\SNR^2 &= \frac{E_\text{signal}}{E_{\text{noise}}}\\
&= \frac{\Psignal \tSNR}{T_R/2} \\
&= \frac{E_z \gamma_z \tSNR}{T_R/2} ~, \label{eq: SNR equation}
\end{align}
where the noise energy is defined as half of the temperature of the SQUID detector’s resistor, $T_R = 0.01$ K. The signal power is determined by the dissipation of the axial motion, $\Psignal=E_z \gamma_z$, with $E_z = \frac{1}{2}m_e \wz^2 \zmax^2$ representing the energy of the axial motion and $\gz$ denoting the axial damping rate.
This derivation is consistent\footnote{There appears to be a discrepancy by a factor of $\left(\frac{\delta}{\gz/2}\right)^2$. This factor arises because the amplitude of the electron’s Lorentzian response is proportional to $\zmax 
~ \propto ~ \min\left(\frac{\delta}{\gz/2}, 1\right)$. Ref.~\cite{DehmeltMeasurementTime} addresses the regime $\delta \ll \gamma_z$, while our work focuses on the opposite limit, $\delta \gg \gamma_z$.} with the result obtained in Ref.~\cite{DehmeltMeasurementTime}.

The axial damping rate is given by the expression \cite{ThesisXingFan}
\begin{align}
\gamma_z=\frac{1}{m_e} \left(\frac{e d_1}{d} \right)^2 \Reff,
\end{align}
where $d$ is the height of the trap, $d_1=0.9$ is the image charge parameter \cite{CylindricalPenningTrap,OpenTrap}, and the effective resistance $\Reff$ is modeled using the empirical formula~\cite{ThesisXingFan}
\begin{align}
\Reff=q\times 6000 ~\text{M}\Omega \left( \frac{1 ~\text{MHz}}{\wz/2\pi} \right)^2 ~.
\end{align}

Substituting these expressions into \eq{\ref{eq: SNR equation}}, the averaging time required to achieve the desired SNR is calculated as
\begin{align} \label{eq: tave noise}
    \tSNR &= \frac{\SNR^2 T_R}{2 E_z \gz} \\
    &= \frac{\SNR^2 T_R}{ q\times6000 ~\text{M}\Omega \left( 2\pi ~\text{MHz} \right)^2  } \left(\frac{d}{e d_1 \zmax} \right)^2 \\
    &= 2.85 \times 10^{-6} ~\text{s} ~ \left(\frac{0.0061}{q}\right)\left(\frac{\Rtrap}{2.5~\zmax}\right)^2 ~, \label{eq: tSNR last step}
\end{align}
where we have used $d = 2\Rtrap$. The maximum axial amplitude,
\begin{align} \label{eq: zmax}
\zmax=\frac{\Rtrap}{2.5}=0.2~\text{mm}~,
\end{align}
is set by the axial anharmonicity constraint discussed in Appendix \ref{Minimum Averaging Time}.

\subsubsection{Maximum Cyclotron Number}
We optimized the three time scales ($\tsignal$, $\tdet$, and $\tSNR$) with respect to the independent variables $q$ and $\wz$, subject to both the axial anharmonicity constraint and the band width constraint \eq{\ref{eq: tsignal lower limit }}. The detailed calculations are provided in Appendix \ref{Minimum Averaging Time}. The resulting optimal values are $q=0.0061$ and $\wz/2\pi=9.8$ MHz, yielding a lower bound on the averaging time:
\begin{align} \label{eq: lower equal upper time limit}
   \tave = 2.85 \times 10^{-6} ~\text{s}\leq \tau_c = 3.4 \text{ s} \left( \frac{1}{n_c} \right)\left(\frac{24~\text{GHz}}{\wc/2\pi} \right)^2 = 3.4 \text{ s} \left( \frac{1}{n_c} \right) \left( \frac{0.1 \text{ meV}}{\m}\right)^2 ~,
\end{align}
with the corresponding optimal cyclotron number
\begin{align}
    \textcolor{\specialColor}{\ncmax  = 1.2 \times 10^6 \left(\frac{24~\text{GHz}}{\wc/2\pi} \right)^2 = 1.2 \times 10^6 \left(\frac{0.1~\text{meV}}{\m} \right)^2}~.
\end{align}

The large $n_c$ used here violates assumptions underlying the description of magnetic bottle coupling in Section~\ref{sec: 2A}. The Hamiltonian governing the effective axial motion consists of the pure axial motion and the axial-cyclotron bottle coupling~\cite{ThesisXingFan},
\begin{align}
    H_{z,\text{eff}}=H_z + H_{cz} &= \frac{1}{2} m_e \wz^2 z^2 + 2 \mu_B B_2 n_c z^2 \label{eq:Ezeff}\\ 
     &\equiv \frac{1}{2} m_e \omega_{z,\text{eff}}^2 z^2 ~,
\end{align}
where $\mu_B = e/2m_e$ is the electron magnetic dipole moment, the pure axial frequency is set by the potential as $\wz^2 \equiv e V_0/m_e D^2$, and the effective axial frequency is given by
\begin{align}
    \omega_{z,\text{eff}} = \wz \sqrt{1+2 n_c \delta/\wz} ~.
\end{align}
For small $n_c$, the axial energy is dominated by the first term in \eq{\ref{eq:Ezeff}}, allowing us to interpret $\wz \approx \omega_{z,\text{eff}}$ as the physical axial frequency. However, when $n_c$ becomes large, the second term dominates, and $\wz$ merely serves as a parameter, while $\omega_{z,\text{eff}}$ is the effective axial frequency.

To keep $\omega_{z,\text{eff}}$ stable, we implement a voltage feedback mechanism on $V_0$, ensuring that $\omega_{z,\text{eff}}$ remains fixed at $2\pi \times 9.8$~MHz. Consequently, $V_0$ now depends on $n_c$, and for $n_c \gg 10$, it must change sign to cancel the large contribution from $n_c$, resulting in an imaginary $\wz$. The maximum required $|V_0|$ is about 200 V. This change in electric field direction also changes the sign and magnitude of the magnetron mode: $\wm$ becomes negative and much larger. Instead of having a metastable magnetron with a slow increase in quantum number, we have a slowly decreasing magnetron number with a stable ground state, while the lifetime remains unchanged.

Furthermore, when $n_c$ jumps by 1, the effective axial frequency shifts by $\Delta\omega_{z,\text{eff}} = (\wz/\omega_{z,\text{eff}})\delta $, which follows the same functional form as the original shift $\delta$, but with $\wz$ replaced by $\omega_{z,\text{eff}}$. Therefore, once the voltage feedback is implemented and $\omega_{z,\text{eff}}$ is stabilized, we can dispense with these subtleties and continue to use the previous formulas with the understanding that $\wz$ in fact denotes $\omega_{z,\text{eff}}$, and $\delta$ in fact denotes $\Delta\omega_{z,\text{eff}}$. All other relevant properties of the Penning trap system remain unchanged except for $\omega_m$.

We conduct several additional consistency checks in Appendix \ref{sec: consistency checks} to ensure that such a large cyclotron quantum number, $\ncmax$, does not introduce additional complications (e.g., an excessively large radial wave function size). All relevant experimental parameters are summarized in Table~\ref{tab: parameters} in Appendix~\ref{sec: experimental parameters}. Notably, every parameter used in our analysis is either compatible with current technology or anticipated to be achievable in the near future.

\subsubsection{Preparing a High $n_c$ State}
To prepare the highly excited state \(\ket{\ncmax}\), we periodically apply a resonant external cyclotron drive to the trapped electron. As the wave function expands, the resulting frequency shift in the cyclotron mode is approximately 0.045~GHz, which remains insignificant compared to the axial bottle broadening of $\Delta \wc/2\pi = 1.21$ GHz (see~\eq{\ref{eq:line width GHz}}). The required driving power is on the order of $10^{-10}$ mW, which is far lower than the typical 10 mW output of commercially available sources.

Such a classical driving procedure typically produces coherent states rather than pure cyclotron eigenstates. However, the nearly continuous QND measurements of axial frequency rapidly collapse the coherent state into a definite cyclotron number state, $\ket{\nc}$. Mathematically, a coherent state is represented as a superposition of infinitely many energy eigenstates, parameterized by a complex number $\alpha$:
\begin{align}
    \ket{\alpha} = e^{-|\alpha|^2/2} \sum_{n_c=0}^{\infty} \frac{\alpha^{n_c}}{\sqrt{n_c!}} \ket{n_c} ~.
\end{align}
We employ a resonant drive to generate a coherent state with $|\alpha|^2 = \ncmax$. The resulting distribution of cyclotron states follows a Poisson distribution with mean value $|\alpha|^2 = \ncmax$~\cite{GlauberCoherentState}:
\begin{align}
    |\braket{\nc}{\alpha}|^2 = \frac{|\alpha|^{2\nc}}{\nc!} e^{-|\alpha|^2} ~.
\end{align}
Thus, up to a minor quantum fluctuation, this approach yields the desired state, $\ket{\nc}$, where $\nc \approx \ncmax \pm \sqrt{\ncmax}$. Given the known initial value of $\wz$, along with the corresponding frequency shift $\delta$, we can precisely determine the specific cyclotron number of the system.

Since the observation time per frequency bin, $\tobs$ (see next subsection), is significantly longer than a typical cyclotron lifetime, $\tau_c$, the cyclotron state will inevitably undergo numerous decays during the course of the experiment. Selection rules permit only the transitions $\nc \to \nc - 1$, which manifest as a frequency shift $\wz \to \wz - \delta$. This decay-induced frequency shift is clearly distinguishable from the opposite shift $\wz \to \wz + \delta$ that would result from a dark matter interaction (see Figure~\ref{fig: recognize signal}). Therefore, we can continue searching for a dark matter signal uninterrupted, even while observing multiple decays.

To prevent a significant reduction in sensitivity due to cyclotron decay, we periodically reinitialize the state to $\ket{\ncmax}$, a process that takes negligible time. By reapplying the preparation when $n_c$ decreases to 90\% of its initial value, the time interval between successive preparation pulses can be estimated using \eq{\ref{eq: tau c}} as:
\begin{align}\label{eq: pulse time}
    \Delta t_\text{pulse}= \sum_{n_c=0.9\ncmax}^{\ncmax}\tau_c(n_c,\wc) = 0.34 \text{ s} \left(\frac{24~\text{GHz}}{\wc/2\pi} \right)^2= 0.34 \text{ s}  \left( \frac{0.1 \text{ meV}}{\m}\right)^2  ~.
\end{align}
The shortest required pulse time is approximately a millisecond, which is entirely feasible for implementation in our experimental setup.

\subsection{Cyclotron Bandwidth and Scanning\label{section:Q_c}}

The frequency width appearing in the transition rate formula \eq{\ref{eq: Gamma cavity dark photon}} is defined as the larger of two widths: $\Delta \omega = \max(\Dw, \Delta \wc)$, as detailed in Appendix \ref{sec: transition rate in free space}. Here, $\Dw = \frac{\m}{Q_{\rm DM}} \approx 10^{-6} \m$ represents the intrinsic dark matter width, and $\Delta \wc = \frac{\omega_c}{Q_c}$ is the cyclotron line width. The dark matter quality factor, $Q_{\rm DM}$, is determined by its virial velocity, while $Q_c$ is the cyclotron quality factor, which depends on specific properties of the experimental apparatus.

It is crucial to scan the cyclotron frequency across bins of width $\Delta \omega$ without leaving gaps in between. For frequencies outside the bandwidth, a detuning parameter $D > \Delta \wc$ results in a quadratic loss of sensitivity in $D$, as predicted by Rabi’s formula \eq{\ref{eq: Rabi}}. This sensitivity loss cannot be compensated by simply increasing the observation time per bin (which would only provide a linear gain), contrasting with the behavior in a thermal-noise-limited experiment~\cite{195pagesDMRadioReview}.

As we elaborate below, the value of $Q_c$ does not impact the reach of our experiment, provided that the experiment remains background-free. This behavior also contrasts with that of a thermal-noise-limited experiment~\cite{195pagesDMRadioReview}, where operating in the resonant limit is always advantageous.

When $Q_c$ exceeds $Q_{\rm DM}$, the transition rate is governed by $\Delta \omega = \Dw$ and thus becomes independent of $Q_c$. The widely adopted strategy in this resonant regime is to allocate an observation time of approximately $\tobs \approx \tTotal/Q_{\text{DM}}$ per bin to maintain comparable sensitivity over an entire decade of dark matter mass. Conversely, when $Q_c$ falls below $Q_{\text{DM}}$, the experiment enters the broadband regime. Here, the transition rate is determined by $\Delta \omega = \Delta \wc$, resulting in a linear decrease of the transition rate with $Q_c$, as shown in \eq{\ref{eq: Gamma cavity dark photon}}. However, the increased signal width in the broadband regime allows for longer observation times per bin, approximately $\tobs \approx \tTotal/Q_c $. Consequently, these effects cancel out, leaving the experimental reach nearly unchanged\footnote{Note that $Q_c = \wc/\Delta \wc$ varies with $\wc$, while $\QDM = 10^{6}$ is a constant. This difference causes the resonant and broadband regimes to exhibit distinct scaling behaviors in sensitivity to $\epsilon$ (or $\g$) as a function of $\m$ (or $m_a$), differing by a factor of $\sqrt{\m}$.}.

Nonetheless, determining $Q_c$ is still essential for calculating the appropriate value of $\tobs$ for the experiment. In our previous work~\cite{Fan_2022}, the electron cyclotron was characterized as a high-Q resonator with a quality factor of $Q_c = 10^7$. However, this characterization changes significantly due to the need to increase $B_2$ to a much larger value in order to reduce the signal formation time (see Section \ref{sec: signal formation time}).

The cyclotron linewidth, $\Delta \wc$, is given by the sum of two terms: the damping rate $\gamma_c = \tau_c^{-1}$ and a bottle broadening term $\wc \frac{B_2}{B_0} \zmax^2$~\cite{ThesisXingFan}, which captures the cyclotron frequency uncertainty induced by the magnetic gradient $B_2 z^2$. Thus, the effective cyclotron linewidth is given by: 
\begin{align} \label{eq: Delta wc}
\Delta \wc &= \wc \frac{B_2}{B_0} \zmax^2 + \gamma_c\\
&= \frac{e}{\me}\tilde{B}_2 \frac{\zmax^2}{\Rbot^2} + n_c \frac{4\alpha \wc^2}{3 m_e} \\
&= 5\times 10^{-3} ~\text{meV} +  2 \times 10^{-7} ~\text{meV}  \left( \frac{n_c}{10^6} \right)\left( \frac{\wc}{0.1 \text{ meV}}\right)^2\\
\Delta \wc/2\pi&= 1.2 ~\text{GHz} +  4.8 \times 10^{-5} ~\text{GHz}  \left( \frac{n_c}{10^6} \right)\left( \frac{\wc/2\pi}{24 \text{ GHz}}\right)^2 \label{eq:line width GHz}
 ~.
\end{align}
Even for the maximum achievable cyclotron number, $n_c$, the linewidth is dominated by the bottle broadening term.
As a result, the quality factor gets degraded to \textcolor{\specialColor}{$Q_c=\frac{\wc}{\Delta \wc} =20\times\left(\frac{\wc}{0.1~\text{meV}}\right)=20\times\left(\frac{\wc/2\pi}{24~\text{GHz}}\right)$}, which is much wider than the dark matter width $\QDM = 10^{6}$ across the entire frequency range of interest. Consequently, the experiment operates in the broadband regime, where the observation time per frequency bin is determined by:
\begin{align}  \label{eq: observation time formula}
    \tobs &= \tTotal \frac{\Delta\wc}{m_{A',\text{max}}-m_{A',\text{min}}}\\
    &=1000~\text{days}~\frac{5\times 10^{-3}~\text{meV}}{1~\text{meV}-0.1~\text{meV}}\\
    &\approx~5.5~\text{days} ~,
\end{align}
assuming a total observation time of 1000 days per decade in frequency.

\subsection{More Electrons? \label{sec:more electrons}}
If we trap $n_e$ electrons, the cyclotron transition rate scales linearly: $\Gammacavity \to n_e \Gammacavity$. Intuitively, this would suggest an improvement in sensitivity. This reasoning holds true in the low cyclotron number regime, and in our previous work~\cite{Fan_2022}, a configuration with $n_e = 10$ electrons was proposed. However, as we demonstrate below, in the optimized regime explored here, trapping more than a single electron ceases to provide a sensitivity advantage.

We experimentally found that only the average axial image current of all $n_e$ electrons can be measured, and the signal shift, $\delta$, scales as $1/n_e$. If a cyclotron transition occurs for one of the electrons, it becomes impossible to identify which specific electron underwent the transition. This introduces the risk that the signal could be canceled by a subsequent decay of another electron before a detection can be made. As a result, the effective cyclotron lifetime is reduced by a factor of $n_e$, $\tauceff = \tau_c/n_e$.
This reduction also lowers the optimal cyclotron number by a factor of $n_e$ (\eq{\ref{eq: lower equal upper time limit}}).
Since the transition rate is proportional to $n_c n_e$, these two effects cancel out, eliminating any net gain in sensitivity from trapping multiple electrons.

However, during the R\&D phase of the experiment, achieving the optimal parameters may not be feasible in the initial iterations. Under these conditions, trapping multiple electrons could temporarily enhance sensitivity. Nevertheless, it is evident that increasing the number of trapped electrons does not contribute to the ultimate sensitivity in the fully optimized regime.

\section{Electron in a Cavity\label{sec:effects of cavity}}
The cyclotron transition rate, as given in \eq{\ref{eq:free Gamma final1}}, only holds when the electron trap is in free space.  In reality, the electron is surrounded by metallic conducting surfaces, which ground any electric fields parallel to their surfaces. This presence of conductors can significantly modify the expected transition rate. 
Counterintuitively, an appropriately chosen geometry of conducting shield can even enhance our sensitivity to dark matter due to a ``focusing"-like phenomenon. We begin by providing an intuitive physical explanation of this effect before proceeding with a detailed mathematical derivation. 

The concept of using dish antennas to focus dark-photon- and axion-induced electromagnetic radiation was first introduced in Ref.~\cite{DishAntennaProposal-2}. The core idea is as follows: as discussed below, a metallic surface exposed to dark photon or axion dark matter generates Standard Model electromagnetic radiation normal to the surface. In the ray optics approximation, a spherical surface would emit rays that converge at the center, thereby concentrating all the radiated power at that focal point. This effect arises under the assumption that the Standard Model radiation is purely outgoing from the conductor’s surface. Such an assumption is valid when a good absorber is positioned at the center of the spherical surface.

In contrast, our setup involves a single electron located at the center of the cavity, which absorbs only a negligible fraction of the available energy. Therefore, both outgoing and incoming modes (or equivalently, standing wave modes) must be considered to accurately describe the focusing effect in an absorber-less cavity, as opposed to purely outgoing modes. We begin by providing an intuitive explanation based on ray optics in Section~\ref{sec: physical picture}. This is followed by a rigorous derivation of the radiation pattern inside the cavity, without relying on the ray optics approximation, in Section~\ref{sec: cavity factor}, demonstrating that the focusing effect still persists.

\subsection{Ray Optics\label{sec: physical picture}}
\begin{figure}[htpb]
    \hspace*{-0.2in}
    \vspace*{0.5cm}
    \centering
    \includegraphics[width=1\textwidth, bb=0 0 700 250]{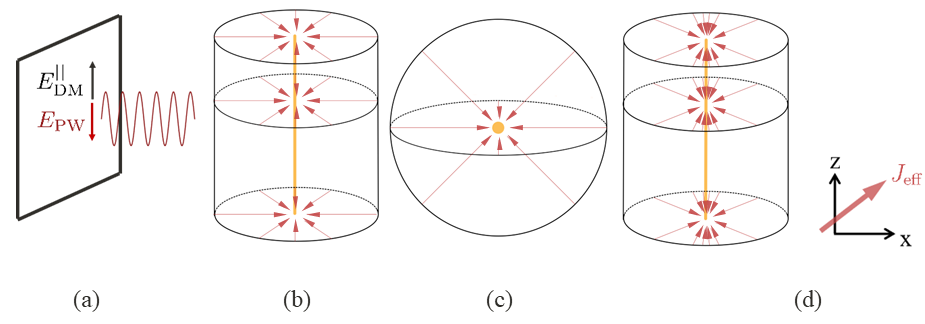}
    \caption{(a) An infinite metal plate sources plane wave. (b) Cylindrical focusing provides a linear enhancement in power. (c) Spherical focusing provides a quadratic enhancement in power. (d) For an effective current pointing in a generic direction, portions of the cavity do not experience a parallel electric field.}
    \label{fig:Focusing}
\end{figure}

We start by estimating the focusing effects with the ray optics intuition here and provide the rigorous justification in the next subsection. Consider an infinite conducting plate immersed in a dark photon dark matter (DPDM) background (everything below is the same for the axion as long as there is an external magnetic field). As long as the electric field sourced by dark photon has a parallel component $E_{DM}^{||}$, the electrons in the conductor will be accelerated and start oscillating. This process must generate radiation in a way that cancels $E_{DM}^{||}$ at the conductor surface in order to satisfy the conductor boundary condition (assuming the wavelength of the dark photon is much larger than the skin depth of the conductor). By planar symmetry, the radiation must be in the form of a plane wave, as shown in Figure~\ref{fig:Focusing}a. In equations, 
\begin{equation}
 E_{\rm DM}^{||} \cos \left(m_{A'} t+\m v x\right) +E_{\rm PW}^{||}\cos\left(\omega t - kx\right) ~.
 \label{eq: matchingA}
\end{equation}
Here $x$ is a space coordinate with $x=0$ defining the metal plate.
Let us start by ignoring the term proportional to $v\approx 10^{-3}$ for now. Requiring that $E^{||}=0$ at $x=0$ gives
\begin{align}
E_{\rm PW}^{||}&=-E_{\rm DM}^{||}=\epsilon \sqrt{\rho_{\rm DM}} \nonumber \\  \omega&=m_{A'} ~.
\end{align} 
So, metal plates generically source $\epsilon$-suppressed plane waves with frequency given by the DPDM mass, $m_{A'}$, as a characteristic signal of dark photon dark matter~\cite{DishAntennaProposal-2}.

Now, suppose we wrap the conducting sheet around to form a cylinder. All radiation sourced by the boundary conductor will go through the central axis, enhancing the energy density there: the incoming power is proportional to the cylinder circumference $2\pi R$, which is concentrated into a region of size the dark photon wavelength $\m^{-1}$; see Figure~\ref{fig:Focusing}b. Therefore, the enhancement is captured by the factor
\begin{align}
\kappa^2 \sim \frac{R}{\m^{-1}}  \qquad \text{(cylindrical focusing)} ~.
\end{align}
Here, we define $\kappa^2$ to be the enhancement in power, such that $\kappa$ is the enhancement in the fields. We just need to multiply our free-space transition rate by a cavity factor $\kappa^2$ to incorporate cavity effects: 
\begin{align} \label{eq: Gamma free to Gamma cavity}
\Gamma_{c, \text{cavity}} &=\kappa^2  \Gammafree ~.
\end{align}

This effect is known as ``focusing." It was first proposed in the context of dish antenna \cite{DishAntennaProposal-2}, and a similar technique was recently proposed by BREAD for a paraboloid geometry~\cite{BREAD}. Cylindrical focusing was employed in our proof-of-principle measurement \cite{Fan_2022}, with the cylindrical trap cavity naturally playing the role of the focusing apparatus, and with an enhancement factor of $\kappa^2 =2.37$ at the chosen frequency.

A spherical geometry would give rise to a more powerful focusing effect. By the same argument as above, the incoming power is proportional to the surface area $4\pi R^2$, giving rise to a quadratic enhancement 
\begin{align}
\kappa^2 \sim \left(\frac{R}{\m^{-1}}\right)^2 \qquad \text{(spherical focusing)}~.    
\end{align}
See Figure~\ref{fig:Focusing}c.
Therefore, in order to maximize the sensitivity to dark photons, we want a large, spherical cavity. In Ref.~\cite{Fan_2022}, we proposed a spherical trap cavity with $\kappa^2 \approx 4000$.

The electron cyclotron is only sensitive to electric fields in the $xy$-plane, perpendicular to the magnetic field. Let the $x$-direction be the dark photon polarization projected onto the plane. Then the portions of the cavity near the intersections with the $x$-axis do not experience a parallel electric field. For example, in the cylinder, not the entire circle contributes to the focusing; see Figure~\ref{fig:Focusing}d. However, this minor effect does not change the scaling of $\kappa^2$.
\subsubsection{Point-Like Nature of Electrons}
We have assumed the electron is point-like, which is valid as long as the dark photon (or axion) Compton wavelength
\begin{align}
\lambda_{A'} = \frac{2\pi}{\m} = 12 ~\text{mm} \left( \frac{0.1 \text{ meV}}{\m} \right)= 12 ~\text{mm} \left( \frac{24 \text{ GHz}}{\wc/2\pi} \right)
\end{align}
 is much larger than the size of the electron wave function. The cyclotron, magnetron, and axial radii are given by \cite{ItanoWineland1982}
\begin{align} \label{eq: cyclotron radius}
r_c &\approx \sqrt{n_c \frac{2}{m_e \wc}} = 3.9 \times 10^{-5} \text{ mm } \left(n_c\right)^{1/2} \left(\frac{24 \text{ GHz}}{\wc/2\pi}\right)^{1/2}= 3.9 \times 10^{-5} \text{ mm } \left(n_c\right)^{1/2} \left(\frac{0.1 \text{ meV}}{\m}\right)^{1/2}  \\
r_m &\approx \sqrt{n_m \frac{2}{m_e \wc}} = 3.9 \times 10^{-5} \text{ mm } \left(n_m\right)^{1/2} \left(\frac{24 \text{ GHz}}{\wc/2\pi}\right)^{1/2}= 3.9 \times 10^{-5} \text{ mm } \left(n_m\right)^{1/2} \left(\frac{0.1 \text{ meV}}{\m}\right)^{1/2}   \label{eq: magnetron radius}\\
r_z &\approx \sqrt{n_z  \frac{2}{\me \wz}}= 6.0 \times 10^{-3} \text{ mm } \left(n_z\right)^{1/2} \left(\frac{1 \text{ MHz}}{\wz/2\pi}\right)^{1/2}  \label{eq: axial radius} ~.
\end{align}
The conditions $r_c, r_m, r_z \ll \lambda_{A'} $ are always satisfied for all relevant quantum numbers.
\subsubsection{Limit on Focuser Size \label{sec: limit on focuser size}}
We cannot arbitrarily increase the size of the cavity to indefinitely enhance focusing. When the radius of the cavity $R$ gets larger than the de Broglie wavelength (i.e. coherence length) of dark photon,
\begin{align}
\lambda_{\text{coherence}} =\lambdaDB= \frac{2\pi}{\m v} = 12 ~\text{m} \left( \frac{0.1 \text{ meV}}{\m} \right)= 12 ~\text{m} \left( \frac{24 \text{ GHz}}{\wc/2\pi} \right)~,
\end{align}
the hitherto-ignored phase $\m v x$ in \eq{\ref{eq: matchingA}} becomes significant and smears out the focal region \cite{FocusingLimit}.  Once this limit is reached, any additional signal power from focusing is spread to a larger volume, but the local power density the electron feels is constant. Therefore, the maximum cavity factor for a spherical cavity is
\begin{align} \label{eq: maximum kappa}
    \kappa_{\text{max,sph}}^2 \sim \left(\frac{2\pi}{v} \right)^2 \approx 4 \times 10^7 ~,
\end{align}
and the maximum cavity radius at which this limit is reached is $R_{\text{max}} = \lambdaDB$. The argument holds for the axion as well.

\subsection{Wave Optics\label{sec: cavity factor}}

In this section, we derive the enhancement factor $\kappa$ using first principles from wave optics, focusing on cylindrical and spherical trap cavities as dark photon converters. As mentioned, since the axion is a scalar with no intrinsic polarization, the electric field produced by its conversion aligns entirely with the direction of the applied magnetic field, rendering it undetectable by our cyclotron sensor, which is sensitive only to fields in the plane perpendicular to this direction. We will address the resolution to this problem using a more complicated focuser geometry in the next subsection.

The key quantity of interest is the observed electric field\footnote{The observed field is defined as the sum of the Standard Model electric field and the $\epsilon$-suppressed dark photon field: $\Eobs = \Ebold+\epsilon \Ebold'$.} $\Ecavityxm(\mathbf{0})$ at the center of the cavity, in the $x$-direction, defined as the projection of the dark photon polarization in the plane perpendicular to the trap’s magnetic field ($\hat{z}$ direction). In a generic cavity, the electric field resulting from the interaction of the dark photon with the cavity needs not align with the $x$-direction. However, this alignment holds for cavities with at least cylindrical symmetry. For such geometries, we define the enhancement factor $\kappa$ in terms of $\Ecavityxm(\mathbf{0})$ and the free-space electric field $\Efreexm$ as:
\begin{align}
\kappa^2 &\equiv \left|\frac{\Ecavityxm(\mathbf{0})}{\Efreexm} \right|^2 ~.
\end{align} 

We begin by solving Maxwell’s equations for the electromagnetic fields inside the cavity. The dark photon induces an effective current~\cite{DMRadio},
\begin{align}
\Jeff=-\epsilon \m^2 A' \Aunitvec ~,
\end{align} that acts as a source term, while the fields themselves must satisfy the cavity’s boundary conditions. Neglecting velocity-suppressed terms and assuming harmonic time dependence of the fields, the general solution can be written as a sum over the cavity’s modes $\Ebold_n(\xbold)$~\cite{DMRadio, DavidHill}:
\begin{equation}
\Ecavity(\xbold) = \sum_n c_n \Ebold_n(\xbold) + \mathcal{O}(v)~,
\end{equation}
where  $n$ is a set of discrete indices labeling the vacuum cavity modes with resonant frequencies  $\omega_n$, and the coefficients $c_n$ are determined by the overlap of the dark photon current with each mode:
\begin{align}
    c_n=\frac{-i \omega}{\omega_n^2 - \omega^2} \frac{\int d^3 x \Ebold_n^*(\xbold) \cdot \Jeff}{\int d^3x |\mathbf{E_n(\xbold)}|^2} ~.
\end{align}
Here, the driving frequency is the dark photon mass $\omega = \m$, and the integral is over the interior of the cavity.

The enhancement factor $\kappa$ becomes:
\begin{align}
\kappa^2 &= \left|  \sum_n \frac{c_n }{i \omega \epsilon A'_x} E_{n,x}(\mathbf{0}) \right|^2 \\
&= \left|  \sum_n \frac{-i \omega}{\omega_n^2 - \omega^2} \frac{\int d^3 x \Ebold_n^*(\xbold) \cdot \left[ -\epsilon \omega^2 (A'_x \hat{x}+A'_z \hat{z})\right]}{\int d^3x |\mathbf{E_n(\xbold)}|^2} \frac{1}{i \omega \epsilon A'_x} E_{n,x}(\mathbf{0}) \right|^2 ~, \label{eq: Ax Az}
\end{align}
where we used $\Efree = \epsilon \Ebold'_{\text{free}}=i\m \epsilon \Afree$. For a cavity with cylindrical symmetry, the dark photon component $A'_z$ cannot generate an $x$-directed field at the origin. Therefore, only the $A'_x$ term contributes:
\begin{align} \label{eq: kappa with cylindrical symmetry}
\kappa^2(\omega) &= \left| \sum_n \frac{\omega^2}{\omega_n^2(1-\frac{2i}{Q_n}) - \omega^2} \underbrace{\frac{\int d^3 x \Ebold_{n}^*(\xbold) \cdot \hat{x}}{\int d^3x |\mathbf{E_{n}(\xbold)}|^2}  \Ebold_{n} (\mathbf{0}) \cdot \hat{x} }_{\equiv G_n} \right|^2 ~.
\end{align}
Roughly speaking, the factor $G_n$ encapsulates the overlap of the cavity mode with the dark photon current, while the factor\footnote{The addition of an imaginary part to the resonant frequencies $\omega_n$, given by the cavity quality factor $Q_n$, renormalizes divergences of $c_n$ by taking into account wall losses \cite{DavidHill}. These quality factors will be suppressed when they are not relevant.}$\frac{\omega^2}{\omega_n^2- \omega^2} $ counts the number of resonances less than $\omega$. For more complex geometries without cylindrical symmetry,  $G_n$  would include contributions from multiple components of \( \Jeff \), making the enhancement $\kappa=\kappa(\theta)$ dependent on the dark photon angle.

The final expression for the cyclotron transition rate in cavity with at least cylindrical symmetry is thus given by:
\begin{align} \label{eq:Gamma cavity}
\Gammacavity = \kappa^2 \frac{\epsilon^2 e^2 \pi  \sin^2 \theta (\nc +1)}{2 \me \m}  \frac{\rhoDM}{\Delta \omega} ~,
\end{align}
where $\kappa^2$ depends on the detailed geometry of the cavity.

We now apply the above general formalism to specific cases of spherical and cylindrical cavities. This will allow us to illustrate how focusing works quantitatively and justify the ray-optics argument given earlier.

As derived in Appendix \ref{sec: spherical cavity}, for a spherical cavity of radius $R$, the cavity factor is expressed as a sum over a discrete index $p$:
\begin{equation} \label{eq: kappa sphere with Gp}
\ksphere^2(\omega) =  \left| \sum_{p=1}^{\infty} \frac{\omega^2}{\omega_p^2 - \omega^2} \Gsphere \right|^2 ~,
\end{equation}
with \( \Gsphere \) defined as:
\begin{align} \label{eq: Gsphere}
\Gsphere
&=- \frac{8}{3}j_1(\up) \left[\frac{u'^5_{1p}}{2 + 2 u'^2_{1p} - 2 u'^4_{1p} + 2 (-1 + u'^2_{1p}) \cos(2 \up) + 
  \up (-4 + u'^2_{1p}) \sin(2 \up)} \right] ~.
\end{align}
Here, $ j_n(x)$ are the spherical Bessel functions, $\Jhatn(x) \equiv x j_n(x)$ are Harrington's spherical Bessel functions, $u'_{np}$ is the $p$-th zero of the derivative $\Jhatn'(x)$, and $\omega_p = u'_{1p}/R$ are the resonant frequencies.

For typical dark photon masses that do not coincide with resonances, the sum is dominated by the nearest mode $\wpstar$ among all the $\omega_p$, giving:
\begin{align}
\ksphere^2(\omega) &\approx \left| \frac{\omega^2}{(\wpstar+\omega) (\wpstar-\omega)} \Gspherestar \right|^2  ~.
\end{align}
We can approximate $\wpstar + \omega \approx 2 \omega$. Since $\omega_p = u'_{1p}/R \approx \omega_{p-1} + \pi/R$ for large $p$, the typical difference is $\wpstar-\omega \approx \frac{\pi}{4R}$. So we have
\begin{align}
\ksphere^2(\omega)
&\approx \left| \frac{2\omega R}{\pi} \Gspherestar \right|^2  ~.
\end{align}
In \eq{\ref{eq: Gsphere}}, the $u'^4_{1p}$ term dominates the denominator for large $p$, so $\Gsphere$ scales as
\begin{align} \label{eq: Gp 4/3}
    |\Gsphere | \approx \frac{4}{3} |j_1(\up) ~ \up| \approx \frac{4}{3} ~,
\end{align}
which is a constant. Finally, we conclude that the cavity factor for sphere scales quadratically:
\begin{align} \label{eq:final sphere kappa estiate}
\ksphere^2(\omega) =  \left| \frac{8\omega R}{3\pi}\right|^2~ \propto ~ (\omega R)^2 =\left(\frac{R}{\m^{-1}}\right)^2 ~,
\end{align}
which matches the ray optics prediction of area-focusing. 

\begin{figure}[htpb]
    \hspace*{-0.2in}
    \vspace*{0.5cm}
    \centering
    \includegraphics[width=0.5\textwidth, bb=0 0 350 250]{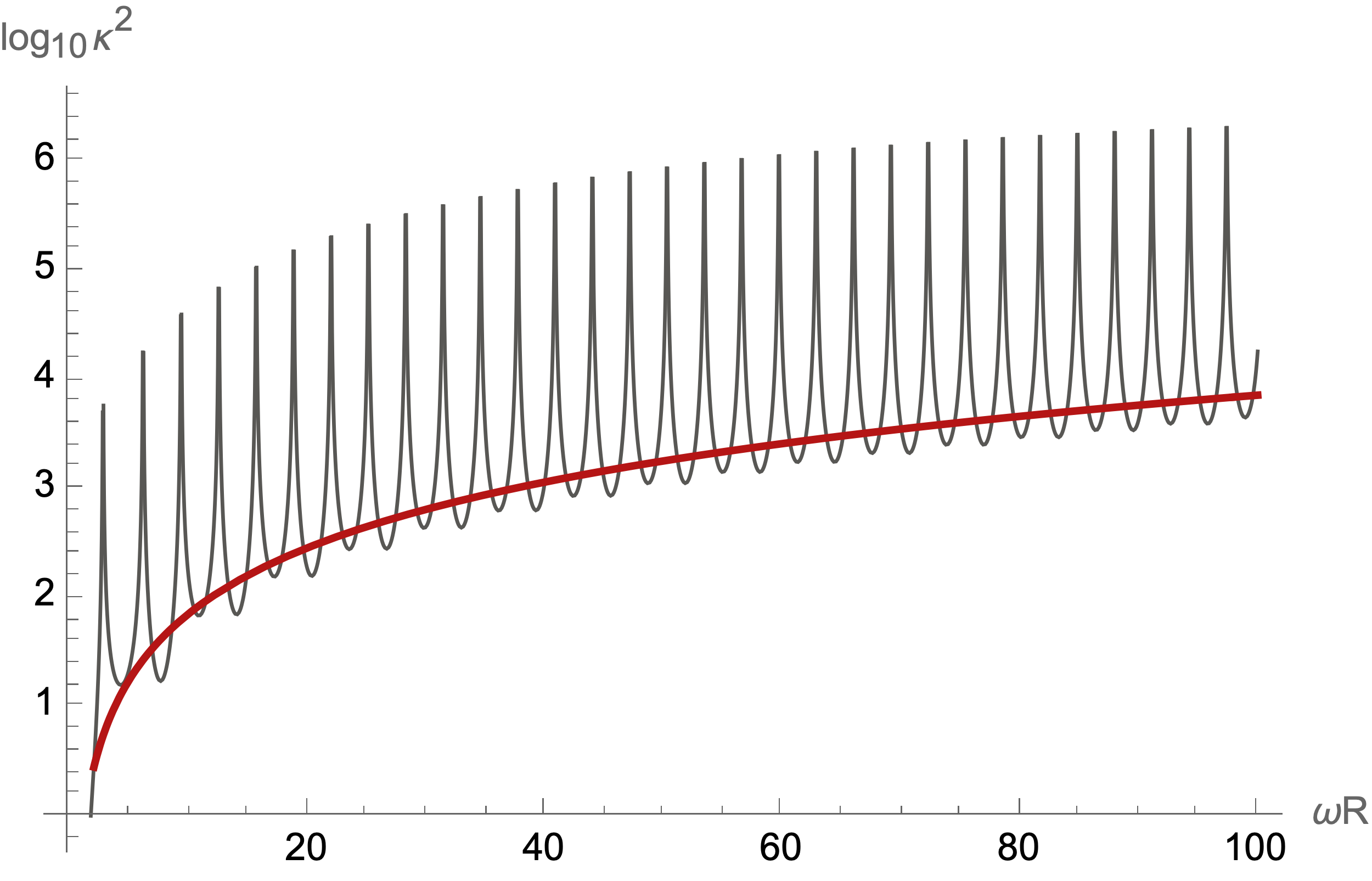}
    \caption{The cavity factor $\ksphere^2$ for a sphere. Exact expression (shown in black; \eq{\ref{eq: Gsphere}}) versus our scaling estimate (shown in red; \eq{\ref{eq:final sphere kappa estiate}}). We have restored the quality factors and set all $Q_p=3000$. }
    \label{fig: kappa sphere}
\end{figure}

In Figure~\ref{fig: kappa sphere}, we show that our estimate \eq{\ref{eq: Gp 4/3}} closely matches the exact expression \eq{\ref{eq: Gsphere}}, up to the cavity modes that we have averaged over. The behavior near the resonance is sensitive to the quality factor of the cavity.

We can also show that a cylindrical cavity has a linear focusing. The calculation is more involved, and we relegate it to Appendix \ref{sec:cylindrical focusing}.

\subsection{Open Trap with Large Conversion Cavity\label{sec: BREAD}}
Having established the validity of the geometric ray optics argument, we now apply it to our experiment.  In our earlier work~\cite{Fan_2022}, we proposed a spherical trap of radius $\Rtrap = 25$ mm, which served as both the trapping and focusing apparatus. This configuration provided a cavity factor of $\kappa^2 \approx (\m \Rtrap)^2 \approx 4000 \left(\frac{\m}{0.5~\text{meV}}\right)^2=4000 \left(\frac{\wc/2\pi}{120~\text{GHz}}\right)^2$. Although significant, this enhancement remains far below the theoretical maximum for a spherical geometry, $\kappa_{\text{max,sph}}^2 \approx 10^7$ (see \eq{\ref{eq: maximum kappa}}). 

The limitation arises because the trap must fit inside the magnetic bottle ring, which requires a small size to generate a large signal (see Section \ref{sec: excited states}). This size constraint conflicts with the need for a larger focusing area. Furthermore, this design is insensitive to axions since the signal photon has an axial polarization. Additionally, using the same magnetic field for both trapping the electron and axion conversion reduces sensitivity at lower frequencies due to the need for scanning the $B_0$ field.

We solve these issues by decoupling the trapping apparatus (radius \(\Rtrap\)) from the focusing apparatus (radius \(R \gg \Rtrap\)). Using an open-endcap Penning trap~\cite{OpenTrap}, which is a hollow cylinder with both ends open, signal photons can enter through one end and interact with the electron positioned at the center. The signal can originate from a separate axion conversion region and be directed into the trap via a waveguide, effectively separating the detection and conversion regions.

For efficient axion conversion and focusing, we adopt a large cylindrical cavity design similar to that proposed by BREAD~\cite{BREAD,PhysRevLett.132.131004}. This setup uses a metal barrel of radius $R = 1$ m and height $L = 2\sqrt{2} R$ within a solenoid providing a strong external magnetic field, $\Bext = 10$ T (see Figure~\ref{fig: BREAD open trap}). Axions are converted into photons at the inner surface and radiate toward the central axis. A parabolic reflector at the center redirects these photons, focusing them to a single point. A waveguide then channels the signal to a photosensor.

The BREAD collaboration plans to use kinetic inductance detectors (KID) and transition-edge sensors (TES) as photosensors. In contrast, we propose using the electron Penning trap, which resolves the polarization issue by designing the waveguide to produce an electric field in the $xy$-plane. This setup also separates the trapping field $B_0$ from the external magnetic field $\Bext$, addressing the scanning issue.

Despite its cylindrical shape, BREAD’s parabolic reflector focuses power from the entire surface area of $A_{\text{BREAD}} = 2\pi R \times L$, resulting in an area focusing factor of $\kappa^2 \sim (\m R)^2$, similar to the spherical focusing discussed earlier. The cavity factor is then:
\begin{align}
    \kappaBREAD^2 &\approx \frac{A_{\text{BREAD,eff}}}{4\pi~\m^{-2}} \\
    &= \frac{\m^2}{4\pi} 2\pi \min(R,\lambdaDB) \times \min(L,\lambdaDB)\\
    &= \frac{1}{2} \m^2 \min\left(R,\frac{2\pi}{\m v}\right) \times \min\left(2\sqrt{2}R,\frac{2\pi}{\m v}\right)\\
    &= \frac{1}{2} \m^2 \min\left[1~\text{m},12.4~\text{m} \left( \frac{0.1~\text{meV}}{\m} \right)\right] \times \min\left[2.8~\text{m},12.4~\text{m} \left( \frac{0.1~\text{meV}}{\m} \right)\right] \label{eq: kappa BREAD}\\
    &= \frac{1}{2} \m^2 \min\left[1~\text{m},12.4~\text{m} \left( \frac{24~\text{GHz}}{\wc/2\pi} \right)\right] \times \min\left[2.8~\text{m},12.4~\text{m} \left( \frac{24~\text{GHz}}{\wc/2\pi} \right)\right] ~, 
\end{align}
where $\lambdaDB = 2\pi / (\m v)$ is the de Broglie wavelength and $v = 10^{-3}$ is the dark matter velocity. At higher frequencies, the focusing limit is saturated (see Section \ref{sec: limit on focuser size}), which causes the kinks in the constraint curves (Figures \ref{fig: axion} and \ref{fig: dark photon}).

Compared to KID/TES sensors, the electron trap has competitive sensitivity and offers two additional advantages. First, it can access lower frequencies ($\m \sim 0.1$ meV), covering parts of the predicted mass range of post-inflationary QCD axions ($m_a \in (0.04~\text{meV},0.18~\text{meV})$)~\cite{PostInflationAxionMass}, while KID/TES are only sensitive to $\m > 0.2$ meV~\cite{BREAD}. Second, the Penning trap is compatible with strong magnetic fields, allowing it to operate directly above the axion-conversion magnet, unlike superconducting KID/TES sensors, which require extensive shielding from high fields.
\subsection{Dielectric Layers \label{sec: Dielectric Layers}}
Ref.~\cite{MultilayerOpticalHaloscopes,garcia2024first,MADMAX2024} demonstrated that axion or dark photon conversion experiments can be enhanced by increasing the conversion volume using a stack of dielectric layers\cite{PhysRevD.88.115002,PhysRevLett.118.091801}, with layer spacing set to match the dark matter Compton wavelength. We point out that this strategy can be integrated with the BREAD setup by adding concentric dielectric cylinders inside the BREAD cavity. It is also compatible with any photosensor and can accommodate frequency scans by periodically switching the dielectric stacks\cite{PhysRevD.105.052010,PhysRevLett.128.231802}. Although a detailed implementation is beyond the scope of this work, we provide a rough estimate of the potential enhancement below, based heavily on Ref.~\cite{MultilayerOpticalHaloscopes,Millar_2017}. 

Suppose we have $N_s$ dielectric stacks, each with $N_l$ layers. Layers within a stack have identical thicknesses, but different stacks can vary slightly to cover a broader frequency range. The total number of layers is limited by the cavity size, $N_s N_l = N_{\max}$. For a single stack, the conversion power $P$ is enhanced by a factor of $N_l^2$ relative to the vacuum power $P_0$:
\begin{align} \label{eq:Nl wrt power}
\frac{P}{P_0} = \frac{1}{2} N_l^2~.
\end{align}
However, the bandwidth decreases by a factor of $1/N_l$, and with $N_s$ neighboring-frequency stacks, the fractional frequency coverage is:
\begin{align}\label{eq: dw/w}
    \frac{\Delta \omega}{\omega} = \frac{N_s}{N_l} ~.
\end{align}
The integrated power across the entire covered frequency range remains constant:
\begin{align}\label{eq: NlNs}
    \frac{P}{P_0} \frac{\Delta \omega}{\omega} = \frac{1}{2} N_l N_s = \frac{1}{2}N_{\max} ~.
\end{align}

However, fractional coverage exceeding $\sim 30\%$ would reduce transparency to propagating signal photons, requiring periodic switching of dielectric stacks with different layer thicknesses to facilitate frequency scan. More frequent switching of the stacks reduces the frequency range per set, thereby allowing for a smaller $N_s$ and a correspondingly larger $N_l$. If we switch stacks $\Nsw = 40$ times over a 1000-day scan (one switch per month) for a decade in frequency, the fractional frequency per set is:
\begin{align}
    \frac{\Delta \omega}{\omega} = 10^{1/\Nsw} -1 \approx \frac{\log 10}{\Nsw} \ll 30\%~.
\end{align}
Substituting this into \eq{\ref{eq: dw/w}} and \eq{\ref{eq: NlNs}}, we find
\begin{align}
    N_l &=\sqrt{\frac{\Nsw}{\log 10} N_{\max}}\label{eq: Nl for 1000}\\
    N_s &=\sqrt{\frac{\log 10}{\Nsw} N_{\max}}\label{eq: Ns for 1000} ~.
\end{align}
The corresponding enhancement in power is:
\begin{align}
\frac{P}{P_0} = \frac{1}{2}\frac{\Nsw}{\log 10} N_{\max} .
\end{align}

Although the ultimate number of layers is limited by the dark matter coherence length, which sets $N_{\max} = v^{-1} N_s =1000 N_s$ for each frequency, where $v \approx 10^{-3}$ is the dark matter velocity, a more stringent constraint arises from fitting the layers within a $R = 1$~m cavity. For instance, at $\m = 0.1$~meV, the Compton wavelength is $\lambda_{A’}  = \frac{2\pi}{\m}= 12$~mm, so at most $\frac{R}{\lambda_{A’}} \approx 83$ layers fit within a $R = 1$~m cavity. Accounting for reduced surface area near the center, we take an effective radius of $R/2$, giving:
\begin{align}
    N_{\max} = \frac{R/2}{2\pi/\wc}~.
\end{align}
Thus, the values of $N_l$ and $N_s$ become:
\begin{align}
    N_l &=\sqrt{\frac{\Nsw}{\log 10} \frac{0.5~\text{m}}{2\pi/\wc}} = 84 \left(\frac{\wc/2\pi}{242~\text{GHz}} \right)^{1/2}= 84 \left(\frac{\m}{1~\text{meV}} \right)^{1/2} \\
    N_s &=\sqrt{\frac{\log 10}{\Nsw} \frac{0.5~\text{m}}{2\pi/\wc}}=5 \left(\frac{\wc/2\pi}{242~\text{GHz}} \right)^{1/2}=5 \left(\frac{\m}{1~\text{meV}} \right)^{1/2}~.
\end{align}
The corresponding enhancement in power is:
\begin{align} \label{eq: dielectric power}
\frac{P}{P_0} = \frac{1}{2}\frac{\Nsw}{\log 10}\frac{0.5~\text{m}}{2\pi/\wc}= 3.5\times 10^3\left(\frac{\wc/2\pi}{242~\text{GHz}} \right)\left(\frac{\Nsw}{40} \right)  = 3.5\times 10^3\left(\frac{\m}{1~\text{meV}} \right)\left(\frac{\Nsw}{40} \right) ~.
\end{align}
The cyclotron transition rate is then enhanced by this factor:
\begin{align} \label{eq: dielectric enhancment}
    \Gammacavity \to \frac{1}{2} N_l^2 \Gammacavity ~.
\end{align}
This enhancement is not exclusive to the electron trap detector and can be applied to other sensors in the BREAD experiment as well.
\section{Projections \label{sec: projections}}
For the dark photon, we derive the constraint on the kinetic mixing parameter, $\epsilon$, using \eq{\ref{eq: Gamma cavity dark photon}} and \eq{\ref{eq: Gamma cavity inequality}} in the broadband regime, where $\Delta \omega = \Delta \wc > \Dw$.  Using \eq{\ref{eq: observation time formula}} for the observation time per bin, the BREAD cavity factor $\kappaBREAD^2$ (\eq{\ref{eq: kappa BREAD}}), the maximum cyclotron number \textcolor{\specialColor}{$ \ncmax = 1.2 \times 10^6 \left(\frac{24~\text{GHz}}{\wc/2\pi} \right)^2 = 1.2 \times 10^6 \left(\frac{0.1~\text{meV}}{\m} \right)^2$}, and adding the dielectric enhancement factor for $N_l$ layers of dielectric per stack (\eq{\ref{eq: dielectric enhancment}}), we get
\begin{align}
  \epsilon^2 = \frac{-\log(1-CL)}{\zeta \tTotal} \frac{m_{A',\text{max}}-m_{A',\text{min}}}{\rhoDM} \frac{2 \me \m}{ e^2 \pi \ncmax \langle \sin^2 \theta \rangle } \frac{1}{ \kappaBREAD^2}\frac{1}{N_l^2/2} ~.
\end{align}

The value of $N_l$ is set by assuming switching dielectric layers $\Nsw=40$ times over $\tTotal=1000$ days, corresponding to roughly once a month (see Section \ref{sec: Dielectric Layers}). Note that we follow the convention of allocating $\tTotal=1000$ days to scan one order of magnitude in frequency but will proportionally spend extra time to cover from 0.1 to 2.3 meV. The detection efficiency\footnote{Here $e$ is Euler's constant, not to be confused with electric charge $e=0.303$ used elsewhere in this equation.} from \eq{\ref{eq: efficiency zeta}} is simply $\zeta = e^{-1}$ since we have saturated the limit $\tave = \tau_c$ and $\delta = 5 \sigma$ for an SNR of 5. Using $\rhoDM = 0.45 \text{ GeV}/\text{cm}^3$ and a confidence level CL=0.9, we can probe down to \textcolor{\specialColor}{$\epsilon \approx 2 \times 10^{-16}$} at $\m=0.1$ meV.

This result is shown in Figure \ref{fig: dark photon}, which also shows the sensitivity projections incorporating only the BREAD experiment and those including BREAD along with excited cyclotron states. The final reach, combining all three designs, is shown in a dotted line as we consider the idea of dielectric layers more futuristic and uncertain in terms of cost. In contrast, the BREAD experiment is already under construction, and our modifications to the electron Penning trap are straightforward and cost-effective using existing technology. The sensitivity gets weaker for $\m \gtrsim 0.5$ meV, as evident by the kinks in the constraint curves, which are due to the saturation of the limit of focusing (see Section \ref{sec: BREAD}). Finally, we note that this setup allows us to meaningfully probe purely gravity-induced kinetic mixing ($\epsilon <10^{-13}$) in the so-called dark matter nightmare scenario \cite{PriceTinyKineticMixing}.

A similar calculation for the axion can be done using \eq{\ref{eq: Gamma cavity axion}}:
\begin{align}
  \g^2 &= \frac{-\log(1-CL)}{\zeta \tTotal} \frac{m_{a,\max}-m_{a,\min}}{\rhoDM} \frac{2 \me m_a^3}{ e^2 \pi \ncmax \Bext^2} \frac{1}{ \kappaBREAD^2}\frac{1}{N_l^2/2} ~.
\end{align}
As shown in Figure \ref{fig: axion}, incorporating all three proposed designs allows us to reach down to
\begin{align}
    \g \approx 9 \times 10^{-15}~\text{GeV}^{-1} \left(\frac{\wc/2\pi}{24 ~\text{GHz}}\right)=9 \times 10^{-15}~\text{GeV}^{-1} \left(\frac{m_a}{0.1 ~\text{meV}}\right)~,
\end{align}
enabling us to probe the QCD axion parameter space. This setup covers the entirety of KSVZ model from 0.1 meV to 2.3 meV and the DFSZ model up to about 1 meV (due to the ``kink" from saturation of focusing limit). This is a particularly well-motivated region of the parameter space because a post-inflationary QCD axion was predicted \cite{PostInflationAxionMass} to have a mass $m_a \in (0.04~\text{meV},0.18~\text{meV})$. Our proposed experiment will be able to explore the mass range from 0.1 meV to 0.18 meV, covering over half of the predicted parameter space.

\section{Conclusion\label{sec: conclusion}}
We have introduced a novel method for detecting QCD axion and dark photon dark matter utilizing highly excited cyclotron states of a trapped electron, significantly improving our previous work \cite{Fan_2022}. By optimizing key experimental parameters, we have reduced the averaging time required to detect a cyclotron excitation to $\tave = 2.9 \times 10^{-6} ~\text{s}$, allowing us to observe the excitation of a cyclotron state with \textcolor{\specialColor}{$n_c = 1.2 \times 10^6 \left(\frac{0.1~\text{meV}}{\m} \right)^2$} before its decay. Since the transition probability scales with the initial quantum number $n_c$, this greatly enhances its sensitivity.

We propose an open-endcap Penning trap design that facilitates the coupling of external photon signals into the trap, leveraging the large focusing power and strong magnetic field of a large conversion cavity (e.g.~the BREAD experiment \cite{BREAD}). Additionally, we propose filling the cavity with dielectric layers of alternating refractive indices to increase the conversion volume. This method is compatible with scanning by periodically switching the dielectric layers, conservatively assumed to be once a month.

These strategies, depicted schematically in Figure~\ref{fig: BREAD open trap}, enable the exploration of the QCD axion parameter space from 0.1 meV to 2.3 meV in a 1000-day-per-decade scan (Figure \ref{fig: axion}), covering a significant portion of the predicted post-inflationary QCD axion mass range $m_a \in (0.04~\text{meV}, 0.18~\text{meV})$ \cite{PostInflationAxionMass}. The experiment also probes the kinetic mixing parameter of the dark photon down to $\epsilon \approx 2 \times 10^{-16}$ (Figure \ref{fig: dark photon}).

To achieve the required averaging time, we need to lower the cyclotron quality factor to \textcolor{\specialColor}{$Q_c = \frac{\wc}{\Delta \wc}= 20 \left( \frac{\m}{0.1~\text{meV}}\right)$}, which is much broader than the dark matter quality factor $\QDM=10^6$. However, this does not diminish sensitivity because it allows for a longer observation time per frequency bin, $\tobs \approx 5.5$ days. We have also shown that when the averaging time is minimized, trapping more than one electron does not improve sensitivity. Our design is advantageous because it is background-free, as demonstrated in Ref.~\cite{Fan_2022}, and B-field compatible, allowing it to operate conveniently above the BREAD magnet.

Moreover, this paper has contributed to the literature by rigorously deriving the focusing effect and justifying the ray-optics approximation in the case of cylindrical and spherical cavity geometries without a perfect absorber. We have also worked out the consequence of a fixed dark photon polarization scenario for our experiment (Appendix \ref{sec:fixed polarization}), which is representative of a class of experiments not addressed by Ref.~\cite{DPLimitsReview2021}: a directionally sensitive quantum sensor (with discrete signals).

Our sensitivity remains well below the ultimate backreaction limit common to all dish-type experiments (Appendix \ref{sec: backreaction}), so there is significant room for improvement. For instance, further lowering the averaging time could allow for an even higher cyclotron state $n_c$. This work thus opens up a ``rapid measurement frontier" in the search for axions. In conclusion, this work provides a clear path to searching the well-motivated yet challenging meV QCD axion parameter space.

\begin{acknowledgments}
We would like to thank David E.~Kaplan and Junwu Huang for discussions and useful suggestions.

This work was supported by the U.S. DOE, Office of Science, National QIS Research Centers, Superconducting Quantum Materials and Systems Center (SQMS) under Contract No.\ DE-AC02-07CH11359. Additional support was provided by NSF Grants No.~PHY-1903756, No.~PHY-2110565, and No.~PHY-2310429; by the John Templeton Foundation Grants No.~61906 and No.~61039; by the Simons Investigator Award No.~824870; by the DOE HEP QuantISED Award No.~100495; by the Gordon and Betty Moore Foundation Grant No.~GBMF7946; and by the Masason Foundation. S.W.~was supported in part by the J.J., L.P., and A.J. Smortchevsky Fellowship and the Clark Fellowship. Y.X.~was supported in part by the Stanford Graduate Fellowship and the Vincent and Lily Woo Fellowship.
\end{acknowledgments}

\appendix
\section{{Cyclotron Transition Rate}\label{sec: transition rate in free space}}
\subsection{Perturbation Hamiltonian by Dark Photon}
The Hamiltonian $H_0$ gets altered by a perturbation due to dark photon dark matter (DPDM). Working in the mass basis, the dark photon Lagrangian contains the term \cite{DMRadio}
\begin{align}
    \Lag \supset -e \JEM^{\mu} (\Amu + \epsilon \Amu') ~.
\end{align}
So, we simply need to replace $\Amu \to \Amu + \epsilon \Amu'$ in the unperturbed Hamiltonian, only keeping terms up to first order in $\epsilon$:
\begin{align} \label{eq: perturbation Hamiltonian}
    H &= \frac{(\pbold - e\Abold - \epsilon e \Abold')^2}{2\me} + e V(\rbold) + \epsilon e V' \\
&= H_0 + \epsilon e \left[ V' -  \frac{(\pbold - e \Abold) \cdot \Abold'}{\me} \right] ~.
\end{align}
We can ignore the ordering of the $\pbold$ and $\Abold'(\xbold,t)$ operators since we will take the zeroth-order approximation of $\Abold'$, which has no position dependence.

Since there is negligible standard model source, $A'_{\mu}$ satisfies the vacuum Proca equation 
\begin{equation}
\left(\partial^2 + \m^2 \right) A'_{\mu} = \epsilon e \JEM^{\mu}=0 ~.
\end{equation}
The most general solution is of the form 
\begin{align}
\Abold'(\xbold,t) = A' \int d^3k ~ \psibold(\kbold) e^{i(\omega t- \kbold \cdot \xbold)} ~,
\end{align} where $\omega = \sqrt{\kbold^2 + \m^2}$ and $A'$ is a constant. In the non-relativistic limit, this simplifies to \cite{DMRadio}
\begin{align} \label{eq: A' with mvx phase}
\Abold'(\xbold,t) &= A' e^{i\m t} \int d^3v ~ \psibold(\vbold) e^{i\m (\frac{1}{2} v^2 t- \vbold \cdot \xbold)} \\
&= A' e^{i\m t} \Aunitvec(\xbold,t) e^{i \varphi(\xbold,t)}
\end{align}
where $\Aunitvec$ is a unit vector. The small but non-zero dark matter velocity $v \sim 10^{-3}$ makes $\Abold'$ an oscillating field with mean frequency $\omega_{A'} = \m$ and a small frequency spread $\Dw \approx \frac{1}{2}\m v^2 \approx  10^{-6} \m$. For now, we will assume $\Abold'$ has a monochromatic frequency; in Appendix \ref{sec:memory loss}, we will take into account $\Dw$ using a memory-loss argument.

The coherence length of $\Abold'$ is large, given by $\lambda_{\text{coherence}} = \frac{2\pi}{\m v}$. The direction $\Aunitvec(\xbold,t)$ has coherence length and time that are at least as large as the phase \cite{DMRadio}. So we can take
\begin{equation} \label{eq: A'}
\Abold'(\xbold,t) 
= A' e^{i\m t} \Aunitvec = A' e^{i\m t} \left[ \sin\theta \hat{x} + \cos\theta \hat{z} \right] ~,
\end{equation}
where $\theta$ is the polar angle with respect to the magnetic field direction $\hat{z}$, and we define the projection of $\Aunitvec$ onto the plane as the $x$-direction without loss of generality.

Charge conservation implies that $A'^{\mu}$ also needs to satisfy the Lorenz gauge-like constraint $\partial_{\mu} A'^{\mu} = 0$. This means $ \partial_t V' = - \del \cdot \Abold'$
which implies that $V' \propto \vbold \cdot \Abold'$. Since $v \sim 10^{-3}$ is small, the $V'$ term can be safely ignored \cite{DMRadio}.

Finally, substituting 
\begin{align}
    \Abold(\rbold) = \frac{1}{2} \Bbold \times \rbold = \frac{B_0}{2} (-y \hat{x} + x \hat{y}) ~ \label{eq: A}
\end{align}
and eq.~(\ref{eq: A'}) into the Hamiltonian, we arrive at
\begin{align}
H
%&= H_0 - \epsilon e \frac{(\pbold - e \Abold) \cdot \Abold'(t)}{\me} \\
&= H_0 - \epsilon e \frac{A'}{\me} \sin(\m t)  
\left[ \sin\theta \left(p_x + e \frac{B_0}{2} y \right) + \cos\theta p_z \right] ~,
\end{align}
where we have taken the imaginary part of the phase to make the Hamiltonian Hermitian. Since $p_z$ can only create and annihilate axial modes ($p_z = i z_0 \me \wz (\adag_z- a_z)$), it cannot drive the transition of cyclotron state. Therefore, for the purpose of calculating cyclotron transitions, we can simply take the $x$-component of the dark photon field $\Abold'_x$, effectively dropping the $p_z$ term:
\begin{align}
\Hfree
&= H_0 - \epsilon e \frac{(\pbold - e \Abold) \cdot \Abold'_x(t)}{\me} \label{eq:Hamiltonian mass basis} \\
&= H_0 - \frac{ \epsilon e A' \sin\theta}{\me} \sin(\m t)
 \left(p_x + e \frac{B_0}{2} y \right) \\
&\equiv H_0 + W(t)\\
&\equiv H_0 + W \sin(\m t) ~, \label{eq:define W}
\end{align}
where we defined the perturbation Hamiltonian $W(t)$ and its time-independent part $W$. We also added subscripts to emphasize this Hamiltonian only applies to the cyclotron mode in free space.

\subsection{Rabi's Formula and Selection Rule}
Next, consider the cyclotron transition with the initial and final states, $\ket{i} =\ket{i_c, i_m, i_z}$ and $\ket{f} = \ket{f_c, i_m, i_z}$. The corresponding Bohr frequency is $\wfi = \wc |\Delta_c|$, where $\Delta_c \equiv \fc - \ic$. If the perturbing frequency matches this Bohr frequency, $\m \approx \wfi$, or, in other words, the detuning parameter $D \equiv \m -\wfi$ is small compared to $\m$, then only this particular transition will be resonantly driven. We can then apply the secular approximation, in which we solve the Schrodinger equation by discarding all non-resonant terms. The resulting transition probability is given by the Rabi's formula \cite{QMCT}:
\begin{equation} \label{eq:Rabi}
\Prob_{if}(t) = \frac{|W_{fi}|^2 }{|W_{fi}|^2+D^2} \sin^2 \left[ \sqrt{|W_{fi}|^2 + D^2} \frac{t}{2} \right] ~, 
\end{equation}
where $W_{fi} \equiv \bra{f} W \ket{i}$ is the transition matrix element, with $W$ defined in eq. (\ref{eq:define W}).

It remains to compute the matrix element
\begin{align}
    W_{fi} = \frac{-\epsilon e A' \sin\theta}{\me} \bra{f}\left( p_x + e \frac{B_0}{2} y \right)  \ket{i} ~.
\end{align}
Writing $p_x$ and $y$ in terms of the cyclotron creation and annihilation operators and dropping the magnetron operators as the magnetron state is assumed to be unchanged, 
\begin{align}
     p_x + e \frac{B_0}{2} y \to i \sqrt{\frac{m_e \wc}{2}} ( \adag_c - a_c) ~,
\end{align} the matrix element evaluates to
\begin{align} 
|W_{fi}|^2 &= \left( \epsilon e A' \sin \theta \right)^2 \frac{\wc}{2\me} \left[ (\ic +1) \delta_{\fc,\ic+1} - \ic \delta_{\fc,\ic-1}\right]\label{eq:Wfi squared} ~.
\end{align}
We see that to leading order in $\epsilon$, a selection rule enforces $|\Delta_c| = 1$ and $\wfi = \wc$. This implies we only need to consider one jump at a time. Since de-excitation is indistinguishable from radiative decay, we henceforth only consider the $n_c \to n_c + 1$ transition. We will also use the symbol $n_c$ instead of $i_c$.

\subsection{Memory-Loss Model of Frequency Width \label{sec:memory loss}}
The Rabi's formula, \eq{\ref{eq:Rabi}}, with $|W_{fi}|^2$ given by \eq{\ref{eq:Wfi squared}}, is the expression for the transition probability of a monochromatic perturbation. We need to modify this result for two reasons: the dark photon has a frequency width $\Dw = 10^{-6} \m$, and the cyclotron also has a line width $\Delta \wc$. In particular, the monochromatic approximation only holds until a time at which the frequency width leads to a phase shift of $2\pi$. This time is known as the coherence time $\tcoherence$. Since we are only sensitive to the overlap of the two sinusoids (the cyclotron and the dark photon), the coherence time is set by the larger of the two widths i.e. the smaller of the two independent coherence times:
\begin{align} \label{eq: coherence time definition}
\tcoherence = \frac{2\pi}{\max \left(\Dw, \Delta \wc \right)} \equiv \frac{2\pi}{\Delta \omega}~.
\end{align}
We will model this frequency width by using Rabi's formula only from $t=0$ to $t= t_{\text{coherence}}$. If the averaging time, the duration over which measurements are averaged, is longer than the coherence time, we assume the sinusoid loses its memory of the phase and starts over discontinuously, but the probabilities add up between successive periods.

For averaging time longer than a coherence time, we can write it as 
\begin{equation}
    \tave \approx N\frac{2\pi}{\Delta \omega}
\end{equation}
where $N \geq 1$ is an integer. Adding up the successive probabilities from $N$ time intervals, we have
\begin{equation}
\Prob_{if}(t) = N \frac{|W_{fi}|^2 }{|W_{fi}|^2+D^2} \sin^2 \left[ \sqrt{|W_{fi}|^2 + D^2} \frac{1}{2} \frac{2\pi}{\Delta \omega} \right] ~.
\end{equation}
The cyclotron transition rate is given by
\begin{align}
\Gammafree &= \frac{d \Prob_{if}(t) }{dt} \\
&= \frac{d \Prob_{if}}{dN} \frac{dN}{dt} \\
&=  \frac{|W_{fi}|^2 }{|W_{fi}|^2+D^2} \sin^2 \left[ \sqrt{|W_{fi}|^2 + D^2} \frac{\pi}{\Delta \omega}\right]  \frac{\Delta \omega}{2\pi} ~\label{eq: Rabi},
\end{align}
where the subscript emphasizes that this formula only applies in free space (i.e. ignoring the cavity).

To cover one order in the dark photon mass range, we need to scan the cyclotron frequency using a frequency bin size\footnote{Note that when $\Delta \omega=\Dw=10^{-6} \m$, the bin size is a function of $\m$, but when $\Delta \omega=\Delta \wc$, it is a constant.} of $\Delta \omega$. Now, define a dimensionless detuning parameter $\tilde{D} \equiv \frac{D}{\Delta \omega} = \frac{\m - \wc}{\Delta \omega}$, where $\wc$ is the center of a given frequency bin. The maximum relevant $\tilde{D}$ is $1/2$, since a dark photon further away than this would be discovered at the neighboring bin. The average ``distance" to dark photon is then $\tilde{D} =1/4$. So we can take the approximation that $\tilde{D} \ll 1$ and $|W_{fi}| \ll \Delta \omega$, since $W_{fi}$ contains a factor of $\epsilon$. Taylor expanding, we get a version of Fermi's golden rule
\begin{align}\label{eq:gammafree}
\Gammafree \approx \frac{|W_{fi}|^2 }{|W_{fi}|^2+D^2} \left[ \sqrt{|W_{fi}|^2 + D^2} \frac{\pi}{\Delta \omega}\right]^2  \frac{\Delta \omega}{2\pi} =  \frac{\pi |W_{fi}|^2 }{2 \Delta \omega} ~.
\end{align}

If the averaging time is shorter than a coherence time $\tave < \tcoherence$, we show in Appendix \ref{sec: coherence time} that a similar calculation implies the transition probability is Zeno-suppressed. In principle, this would hurt the sensitivity and set a limit to this detection method, but it is so far from the leading limit that we will not discuss it here.

Substituting eq.~(\ref{eq:Wfi squared}), we get the cyclotron absorption rate in free space:
\begin{align}
\Gammafree &\approx  \frac{\left( \epsilon e A' \sin \theta \right)^2 (\nc +1)}{2 \me }  \frac{\pi\wc}{2 \Delta \omega} \label{eq: Gamma A square}\\
&\approx \frac{\epsilon^2 e^2 \pi \sin^2 \theta (\nc +1)}{2 \me \m}  \frac{1}{\Delta \omega}\left( \frac{1}{2} A'^2 \m^2 \right)\\
&= \frac{\epsilon^2 e^2 \pi \sin^2 \theta (\nc +1)}{2 \me \m} \frac{\rhoDM}{\Delta \omega} ~, \label{eq:Final Gamma+}
\end{align}
where in the second line, we used the resonance condition $\wc \approx \m$; and in the last line, we used the fact that the dark matter density is $\rhoDM =  \frac{1}{2} A'^2 \m^2$ \cite{DMRadio}. As we will show in Section \ref{sec:effects of cavity}, the effects of a cavity (with at least a cylindrical symmetry) can be incorporated by an overall cavity factor $\kappa^2$:
\begin{align}
\Gamma_{c, \text{cavity}} &=\kappa^2  \Gammafree ~.
\end{align}
The value of the $\sin^2 \theta$ factor in $\Gammafree$ depends on the polarization of $A'$, which, depending on the production mechanism, can be anywhere from fixed to rapidly oscillating in random directions \cite{DPLimitsReview2021}. We will assume the random polarization scenario (discussed in Appendix \ref{sec: limit}) and address the fixed case in Appendix \ref{sec:fixed polarization}.

\subsection{{Sub-Coherence Time Transition Rate} \label{sec: coherence time}}
We continue the calculation of the transition rate from above for the case of $\tave \leq \tcoherence$. Then, we have $N\leq 1$, which is no longer required to be an integer. We can get the transition rate simply by differentiating the Rabi's formula \eq{\ref{eq:Rabi}}
\begin{align}
\Gammafree(t) &= \frac{d \Prob_{if}(t) }{dt} \\
&= \frac{|W_{fi}|^2 }{|W_{fi}|^2+D^2} \sin \left[ \sqrt{|W_{fi}|^2 + D^2} \frac{t}{2}\right]\cos \left[ \sqrt{|W_{fi}|^2 + D^2} \frac{t}{2}\right] \sqrt{|W_{fi}|^2 + D^2}~.
\end{align}
Because for $t\in(0,\tave)$, $ \sqrt{|W_{fi}|^2 + D^2} \frac{1}{2} t \leq N \pi \tilde{D} \lesssim \pi \tilde{D}$, we can still Taylor expand, which gives
\begin{align}\label{eq:gammafree2}
\Gammafree(t) \approx \frac{|W_{fi}|^2 }{|W_{fi}|^2+D^2}  \sqrt{|W_{fi}|^2 + D^2} \frac{t}{2} \sqrt{|W_{fi}|^2 + D^2} =  \frac{t}{2}  |W_{fi}|^2 ~.
\end{align}

Naively substituting $t=\tave= N\frac{2\pi}{\Delta \omega}$ with $N=1$ into this result and comparing with \eq{\ref{eq:gammafree}}, we will see the two limits differ by a factor of $2$.  This is because in the regime $\tave<\frac{2\pi}{\Delta \omega}$, the transition rate cannot be approximated as a constant. Instead, it should be considered as a function that increases linearly with time with an average of
\begin{align}\label{eq:gammafreeave}
\Gamma_{c, \text{free,average}}
= \frac{\Gammafree(0)+\Gammafree(\tave)}{2}= \frac{\pi |W_{fi}|^2 }{2 \Delta \omega} ~,
\end{align}
which agrees with the other approximation at $N=1$.

Notice \eq{\ref{eq:gammafree2}} implies that if the averaging time is shorter than the coherence time, the transition probability is Zeno-suppressed by a factor of $\frac{\tave}{\tcoherence}$. In principle, this would be a limiting factor to our experiment. However, the coherence time is very short for our parameters:
\begin{align}
    \tcoherence= \frac{2\pi}{\Delta \wc} = 8 \times 10^{-10} ~\text{s} ~.
\end{align}
This is much shorter than the leading limit from the SNR time (\eq{\ref{eq: tave noise}}), \textcolor{\specialColor}{$\tSNR \sim 10^{-6}$} s. Furthermore, if the cyclotron width were to be a limit, there is some freedom to adjust $\Delta \wc$ by adjusting $B_2$ (\eq{\ref{eq: Delta wc}}).

Finally, recall in Appendix \ref{sec:memory loss}, we took the coherence time to be the smaller between that of dark matter coherence time and cyclotron coherence time (\eq{\ref{eq: coherence time definition}}):
\begin{align}
    \tcoherence = \min\left({\frac{2\pi}{\Dw},\frac{2\pi}{\Delta\wc}}\right).
\end{align}
When $\Delta \wc > \Dw$, this causes a suppression in cyclotron transition rate $\Gamma_c$ (see \eq{\ref{eq:gammafree}}): $ \Gamma_c \to \frac{\Dw}{\Delta \wc} \Gamma_c$. The discussion of this appendix gives another perspective to this phenomenon: we can interpret the cyclotron motion as a measurement of the dark photon motion, and the suppression of cyclotron transition rate is just a Zeno suppression.

\subsection{Transition Probability \label{sec: limit}}
Write the cyclotron transition rate as
\begin{align} 
\Gammacavity = \Gamma_0 \sin^2 \theta(t)  ~,
\end{align}
where we have isolated the time-dependent polarization angle, $\theta(t)$, which is rapidly oscillating with a dark photon coherence time $\frac{2\pi}{\Dw} \sim 10^{-5}$ s. 

Next, let the transition probability from the cyclotron initial state to the excited state at time $t$ be $P_1(t)$. With a decay rate $\gc$, the probability satisfies the differential equation
\begin{equation}
    \frac{dP_1}{dt} = \Gamma_0 \sin^2 \theta(t)  - \gamma_c P_1 ~,
\end{equation}
until the next averaging time when it is interrupted by measurement. 

Since this polarization is random and unknown, and it is independent of the system, we can take its average value:

\begin{align}
    \sin^2 \theta(t)\to \langle \sin^2\theta(t) \rangle ~.
\end{align}
To find this average, first note that the area element on a unit sphere is proportional to $d(\cos \theta)$. For $\Abold'$ to be pointing in a random direction in the unit sphere, $\cos\theta$ must have a uniform probability density $P(\cos\theta)=\frac{1}{2}$. Then we can take the expected value under this distribution:
\begin{align}
\langle \sin^2\theta \rangle &=  \int_{-1}^{1} d (\cos\theta) \sin^2\theta P(\cos\theta)  = \frac{2}{3} ~. \label{eq:sine square}
\end{align}

Imposing the initial condition $P_1(0)=0$, the solution to the differential equation is
\begin{equation}
    P_1(t) = \frac{(2/3) \Gamma_0}{\gc} \left( 1-\exp(-\gc t) \right) ~.
\end{equation}
This formula only applies within one averaging time, after which the probability gets reset to $P_1=0$. The observation time $\tobs$ available at a given frequency bin satisfies $\tobs \gg \tave$ (see Table \ref{tab: parameters}). And since the measurement time $\tave$ must be shorter than the lifetime of the excited state for the detection to be possible, we take $\gc \tave \ll 1$. Then, the probability of detecting no transition in an observation time $\tobs$ is given by the Poisson formula
\begin{align}
    P_0(\tobs)&=(1-P_1(\tave))^{\tobs/\tave}\\
    &\approx\exp(-\Gammacavity \tobs) ~.
\end{align}

Since it takes an averaging time $t_{\text{ave}}$ to resolve a signal, and a detection threshold of $\Delta \wz > 5 \sigma$ is chosen, an excitation that decays in a time less than $\frac{5 \sigma}{\delta} t_{\text{ave}}$ will not be detected, resulting in a detection efficiency \cite{Fan_2022}
\begin{align} \label{eq: efficiency zeta}
\zeta = \int_{(5 \sigma/\delta) t_{\text{ave}} }^{\infty} dt \frac{1}{\tau_c} e^{-t/\tau_c} = \exp(- \frac{5\sigma}{\delta} \frac{\tave}{\tau_c}) ~.
\end{align}
The effective observation time is thus $\zeta \tobs$. We can then set the 90\% confidence level limit ($CL = 0.9$) on DPDM as

\begin{align}
P_0(\zeta \tobs) = 1-CL,
\end{align}
and the corresponding relation on $\Gammacavity$ becomes  \cite{Fan_2022, LeoParticlePhysicsTextBook}
\begin{align}
\Gammacavity = \kappa^2 \frac{\epsilon^2 e^2 \pi (\nc +1)}{2 \me \m} \frac{\rhoDM}{\Delta \omega} \langle \sin^2 \theta \rangle < -\frac{1}{\zeta \tobs} \log(1-CL) ~ \label{eq: epsilon CL constraint}.
\end{align}
The observation time available per bin $\tobs$ will be discussed in Section \ref{section:Q_c}.

\subsection{Transition Rate for Axion}
To detect axion, we need a strong external magnetic field $\Bext$, which generates an effective vector potential due to axion
\begin{align}
\Aabold &= -\g \frac{\sqrt{2 \rhoDM}}{m_a^2} e^{i m_a t} \Bextbold ~.
\end{align}
To find the sensitivity to axion, we can just make the replacement $\epsilon \mathbf{A'} \to \Aabold$ in the perturbation Hamiltonian \eq{\ref{eq: perturbation Hamiltonian}}. In the free cyclotron transition rate \eq{\ref{eq:Final Gamma+}}, we make the replacement 
\begin{align}
\left(\epsilon \mathbf{A'} \langle \sin \theta \rangle \right)^2 \to \Aa^2
\end{align}
since the axion electric field has a fixed, rather than random, polarization given by the constant $\Bextbold$, which we assume can be engineered to point in the $xy$-plane (see Section \ref{sec: BREAD}). The free transition rate for axion \eq{\ref{eq:Final Gamma+}} then reads
\begin{align} \label{eq: axion Gamma}
\Gammafree 
&=  \frac{\g^2 }{m_a^3} \Bext^2 \frac{e^2 \pi (\nc +1) }{2 \me} \frac{\rhoDM}{\Delta \omega} ~,
\end{align}
where $m_a = \wc$. 
\section{Backreaction: Fundamental Limit to Dish-Type Experiments\label{sec: backreaction}}
In this appendix, we point out that all dish-type experiments for axion or dark photon dark matter search, including our electron-trap proposal as well as other photosensors considered in BREAD \cite{BREAD}, have the same fundamental limit due to backreaction. This is the point at which the detector absorbs all available signal photons. We compute this limit by comparing the signal power with the heating rate of a perfect absorber.

For dark photon dark matter, the focused signal-photon energy density is $\rhosignal= \kappa^2 \epsilon^2 \rhoDM$. The focal volume is set by the dark photon Compton (rather than de-Broglie) wavelength $\lambda_{A'} = \frac{2\pi}{\m}$ since the dark photon is converted to a photon\footnote{This is true even if the detector is much smaller than the volume occupied by the photon, which is the case for our electron wave function. This counter-intuitive fact is a consequence of quantum mechanics, as demonstrated by the familiar atomic (of size $\sim 10^{-10}$ m) absorption of optical photons (of size $\sim 10^{-6}$ m).}. The time scale at which the cavity converts dark photon to photon is also set by the inverse dark photon Compton frequency $T \sim \frac{2\pi}{\m}=\lambda_{A'}$. In the ideal focusing scenario, $\kappa^2$ is the ratio of the cavity area to an area of size the photon wavelength, $\kappa^2 \sim \frac{\Acavity}{4\pi \lambda_{A'}^2}$. Finally, including the dielectric enhancement factor $\frac{1}{2}N_l^2$ [\eq{\ref{eq:Nl wrt power}} and \eq{\ref{eq: dielectric power}}], the available signal power is
\begin{align}
\Psignal \sim \frac{\rhosignal \lambda_{A'}^3}{T}\times \frac{1}{2}N_l^2 = \epsilon^2 \rhoDM \frac{\Acavity}{4\pi}\times \frac{1}{2}N_l^2 ~.
\end{align}

In the limiting scenario, a perfect detector absorbs a single photon in the available observation time per bin, $\tobs \approx 5.5$~days [for $\tTotal =1000$~days; see \eq{\ref{eq: observation time formula}}]. As long as the detector bandwidth is broader than that of dark matter $Q<\QDM=10^6$, which is the case for us and all BREAD sensors, all of the available energy can be absorbed. The absorbed power is then
\begin{align}
\Pabsorb = \frac{\m}{\tobs} ~.
\end{align}
The backreaction limit is reached when $\Psignal = \Pabsorb$.

%, where $Q$ is the detector quality factor (analogous to the cyclotron quality factor $Q_c$ above; not to be confused with the quality factor of our detection circuit $\Qdet$).

%Notice that, unlike the situation for our experiment, the dependence on $Q$ is not canceled here. Recall from Section \ref{section:Q_c} that as we lower $Q$, the transition rate decreases while the available observation time increases and the two effects cancel out to the first order. For a perfect detector, however, as $Q$ gets lowered, other parts of the experiment can be improved to keep the transition rate at its maximum value, by definition of a perfect detector. So, decreasing $Q$ strictly improves the sensitivity, and a strictly backreaction-limited detector should be a broadband detector ($Q=1$).

Putting these together, the backreaction limit for dark photon is
\begin{align}
\epsilon &\approx \left(\frac{8\pi \m}{\tobs \rhoDM \Acavity N
_l^2} \right)^{1/2} \\
&\sim 10^{-18} \left(\frac{40}{N_\text{switch}}\right)^{-1/2}\left( \frac{1000~\text{days}}{\tTotal}\right)^{1/2} \left( \frac{17.7~\text{meter}^2}{\Acavity}\right)^{1/2} ~,
\end{align}
% \begin{align}
% \epsilon &\approx \left(\frac{8\pi \m Q}{\tTotal \rhoDM \Acavity N
% _l^2} \right)^{1/2} \\
% &\sim 10^{-19} \left(\frac{40}{N_\text{switch}}\right)^{-1/2}\left( \frac{1000~\text{days}}{\tTotal}\right)^{1/2} \left( \frac{17.7~\text{meter}^2}{\Acavity}\right)^{1/2} ~,
% \end{align}
where $\Acavity=(2\pi R)( 2\sqrt{2} R )= 17.7~\text{m}^2$ for $R=1~\text{m}$~(see Fig.~\ref{fig: BREAD open trap}).
Since the backreaction limit corresponds to a 100\% efficient detector, by comparing it with our projected reach, we can calculate the photon detection efficiency of our electron Penning trap. Comparing with the result in Section \ref{sec: projections}, we get an efficiency of
\begin{align}
\text{efficiency} = \left(\frac{\epsilon_{\text{backreaction}}}{\epsilon_{\text{Penning}}} \right)^2 \sim \left(\frac{ 10^{-18}}{10^{-16}} \right)^2 \sim 10^{-4} ~.
\end{align}

%The only caveat is we need to use the same cyclotron quality factor $Q_c=20 \left( \frac{\wc}{0.1 ~\text{meV}} \right)$ for this comparison.

The calculation for axion is identical as long as we make the substitution
\begin{align}
\epsilon \to \frac{\g \Bext}{m_a} ~.
\end{align}
The backreaction limit for axion is therefore
\begin{align}
\g &\approx \frac{m_a}{\Bext} \left(\frac{8\pi m_a}{\tobs \rhoDM \Acavity N_l^2} \right)^{1/2} \\
&\sim 10^{-16} ~\text{GeV}^{-1} \left( \frac{m_a}{0.1~\text{meV}}\right) \left(\frac{40}{N_\text{switch}}\right)^{-1/2} \left( \frac{10 ~\text{T}}{\Bext} \right) \left( \frac{1000~\text{days}}{\tTotal}\right)^{1/2} \left( \frac{17.7~\text{meter}^2}{\Acavity}\right)^{1/2} ~.
\end{align}
\section{Details of Achieving Large Cyclotron Number\label{sec:large nc appendix}}
This appendix collects various supporting arguments for how to achieve a large cyclotron number $n_c$.
\subsection{{Signal Formation Time} \label{sec: detection time}}
When the cyclotron state makes a transition $\ket{\nc} \to \ket{\nc + 1}$, the axial frequency shifts $\wz \to \wz + \delta$ immediately. However, there exists a minimum time $\tsignal$ for this change to be reflected in the signal. We will show that $\tsignal = \frac{1}{\delta}$.

The electron's classical axial motion $z(t)$ is essentially a driven, damped harmonic oscillator. With a driving term $F_0\cos{\omega_d t}$ of drive frequency $\omega_d$, the equation of motion for the electron axial mode can be written in the form
\begin{align} \label{eq:axialEOM}
    \ddot{z}+\gamma_z\dot{z}+\omega_0^2 z = \frac{F_0}{\me}\cos{\omega_d t},
\end{align}
where the initial natural frequency of the axial motion is the axial frequency $\omega_0 = \wz$, and the drive frequency is chosen to match it $\omega_d = \omega_0$. In practice, the drive can be taken to have zero width. $\gamma_z$ is the axial damping rate. We will consider the limit $\wz \gg \delta \gg \gz$. To get a signal, we monitor the electron's axial current, which is proportional to the rate of change $\dot{z}$ of the electron's axial position $z$ \cite{Review}.

Suppose a cyclotron transition occurs at some time $t=0$. We get the current signal:
\begin{align}
    \dot{z}(t<0)= \frac{F_0}{\me}\wdr(S_{ip}^0\cos{(\wdr t+\alpha)}-S_Q^0\sin{(\wdr t+ \alpha)})\label{eq:current before jump}\\
    \dot{z}(t>0)= \frac{F_0}{\me}\wdr(S_{ip}'\cos{(\wdr t+\alpha)}-S_Q'\sin{(\wdr t+ \alpha)})\nonumber\\
    +e^{-\frac{\gamma_z}{2}t}(-A \sin{(\wdr+\delta) t}+B\cos{(\wdr+\delta) t})\label{eq:current after jump},
\end{align}
where $\alpha$ is an unknown phase, $S_{ip}^0$ and $S_Q^0$ are the initial in-phase and quadrature amplitudes, and $S'_{ip}$ and $S'_Q$ are the corresponding amplitudes after the cyclotron jump, and $A$ and $B$ are some $\mathcal{O}(1)$ constants determined by initial conditions~\cite{drivenOsc}.

If we were directly measuring for changes in $S_{ip}$ or $S_Q$, it takes a time $t\sim \frac{1}{\gz}$ for the transients to get damped. However, we can integrate the signal against another known sinusoid with the same phase $\sin{(\wdr t+ \alpha)}$ over several periods of $\frac{2\pi}{\wdr}$. After some uses of trig identities, it can be shown that the only surviving term is proportional to $\sin{(\delta t)}$. This means a frequency-shift signal can be measured after a time $\tsignal \sim \frac{1}{\delta}$. In other words, since our signal is a small frequency shift, we can measure the beat at the corresponding time scale, which is much faster than the naive damping time.
\subsection{Minimum Averaging Time\label{Minimum Averaging Time}}
\subsubsection{Constraints on $\zmax$}
\Eq{\ref{eq: tSNR last step}} suggests that we can lower $\tSNR$ by decreasing $\Rtrap$ and increasing $\zmax$. However, a larger $\zmax$ corresponds to a larger wave function in the axial direction, and this exposes the electron wave function to the inevitable anharmonicity of the static potential $V_0$ \cite{Review}. The magnitude of this axial anharmonicity\footnote{The corresponding radial anharmonicity merely shifts the axial frequency, which can be easily taken into account and does not set a new experimental constraint \cite{Review}.} is set by the size of the trap $\Rtrap$. In particular, the axial frequency gains a dependence on axial position. For a realistic trap, this constraint has a phenomenological formula \cite{ThesisDUrso}
\begin{align}
\Delta\wz|_{\max}=|\wz(z)-\wz(0)|_{\max}<\gz~,
\end{align}
where
\begin{align}\label{eq: wz anharmonicity}
\wz(z)=\wz\left(1+\frac{3C_4}{4C_2}\left(\frac{2 z}{D_{\text{trap}}}\right)^2+\frac{15C_6}{16C_2}\left(\frac{2 z}{D_{\text{trap}}}\right)^4+\frac{35C_8}{32C_2}\left(\frac{2 z}{D_{\text{trap}}}\right)^6+\frac{5C_{10}}{4C_2}\left(\frac{2 z}{D_{\text{trap}}}\right)^8+...\right)~.
\end{align}
Recalling $D_{\text{trap}}=\sqrt{\frac{1}{2}d^2+\Rtrap^2}~$and $d=2.0478~\Rtrap$ \cite{OpenTrap}, we have
\begin{align}
    \Delta\wz\approx\wz\left|\frac{15C_6}{16C_2}\left(\frac{2\zmax}{\sqrt{3}\Rtrap}\right)^4+\frac{35C_8}{32C_2}\left(\frac{2\zmax}{\sqrt{3}\Rtrap}\right)^6+\frac{5C_{10}}{4C_2}\left(\frac{2\zmax}{\sqrt{3}\Rtrap}\right)^8\right|.
\end{align}
Equation \eqref{eq: wz anharmonicity} only considers the first few terms of a Taylor expansion, which is adequate as long as $C_{12} \ll C_{10}$. These parameters can be found in Table 1 of \cite{OpenTrap}. While $C_4$ can be experimentally set to zero, we take $C_6 \approx 5 \times 10^{-3}$ for a 10 $\mu$m machine tolerance \cite{ThesisDUrso}. (We consider higher-order terms because $C_8 \gg C_6$ and $C_{10} \approx C_8$.) 

\begin{figure}[htpb]
    \hspace*{-0.2in}
    \vspace*{0.5cm}
    \centering
    \includegraphics[width=1\textwidth, bb=0 0 980 550]{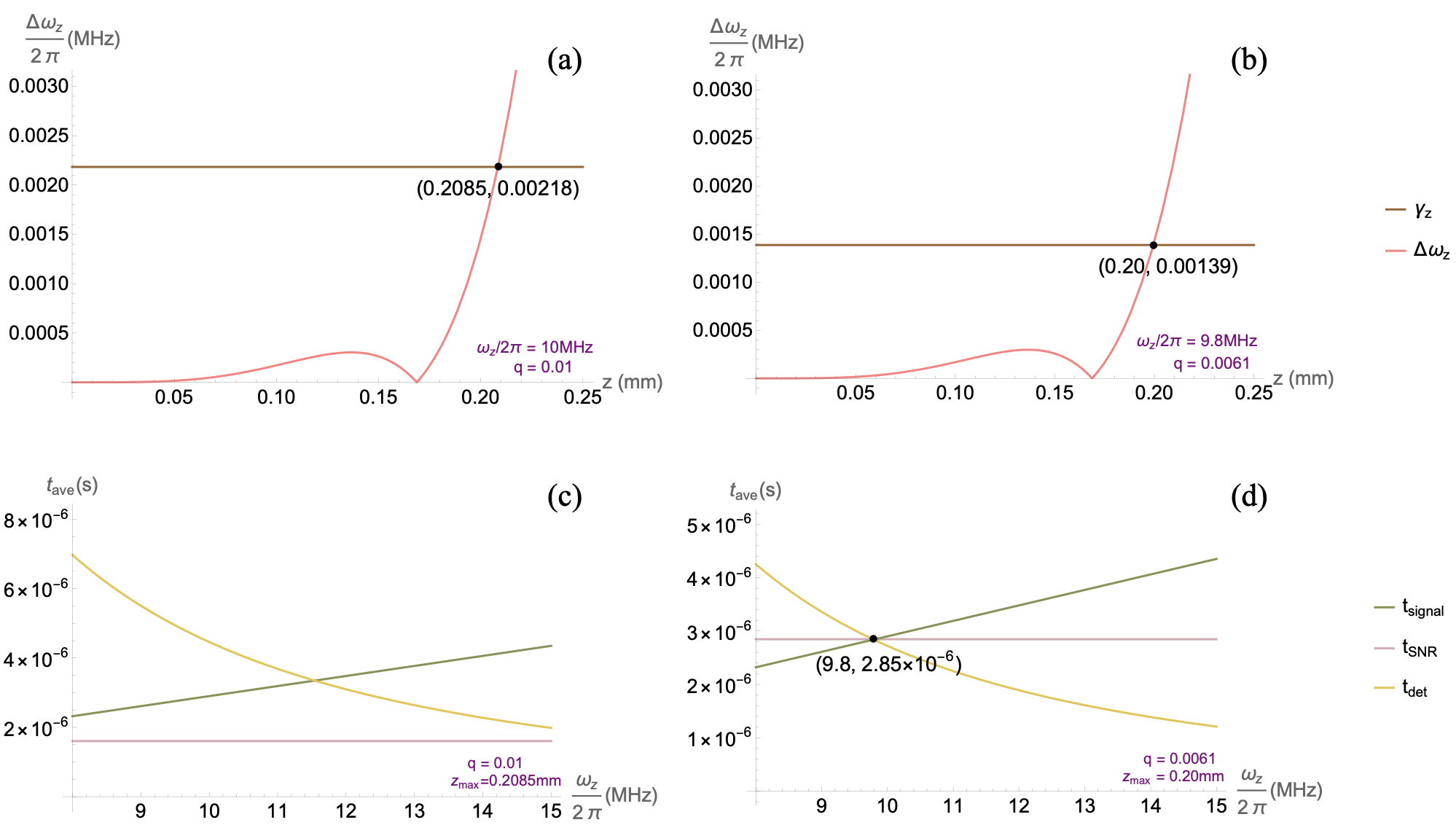}
    \caption{The first row describes the constraint that determines $\zmax$, while the second row incorporates this $\zmax$ into $\tSNR$ to obtain the three times related to $\tave$. The left column uses random parameters, whereas the right column uses optimized parameters. The crossing point in (d) indicates the minimal averaging time. Indeed, the crossing point $\tave = 2.85 \times 10^{-6}$ s is determined with higher precision by the chosen $\wz$ and $q$. Consequently, in our final result (Table \ref{tab: parameters}), the parameters may vary slightly without affecting the outcome due to the precise selection of $\wz$ and $q$.
}
    \label{fig: tave}
\end{figure}

Figures \ref{fig: tave}a and \ref{fig: tave}b show the relationship between $\gamma_z$ and $\Delta \omega_z$ for different sets of $\omega_z$ and $q$. As we can see, before $z = z_{\text{max}}$, the brown line must remain above the red line, indicating that the maximum allowable value of $\zmax$ varies depending on the choice of $\omega_z$ and $q$.

\subsubsection{Reducing \texorpdfstring{$\tave$}{tave}}
As $\tave = \max{\left( \tsignal, \tdet, \tSNR \right)}$, we aim to reduce these three times simultaneously. Indeed, the detailed optimization process is highly nonlinear, but we found the optimal parameters to be $q = 0.0061$, $\frac{\wz}{2 \pi} = 9.8$ MHz, and $\zmax = 0.2$ mm by trial and error. As shown in Figure \ref{fig: tave}d, the three times converge at $\wz = 9.8$ MHz, yielding a minimal averaging time of $\tave \approx 2.85 \times 10^{-6}$ s. 

\subsection{{Consistency Checks} \label{sec: consistency checks}}
We make some consistency checks to show that the large cyclotron number \textcolor{\specialColor}{$\nc \sim 10^6$} does not violate other experimental constraints.

The electron's radial wave function size is given by \eq{\ref{eq: cyclotron radius}}
\begin{align} \label{eq: final cyclotron radius}
r_c \approx \sqrt{n_c \frac{2}{m_e \wc}} \lesssim 0.04 ~\text{mm} ~.
\end{align}
This is still far smaller than the trap size $\Rtrap = 0.5$ mm. 

We should also calculate the fractional relativistic correction to the cyclotron frequency, arising from an effective shift in electron's mass:
\begin{align}
   \left. \Delta \me \right|_{\text{rel}}=\left(\frac{1}{2}+n_c \right)\wc ~.
\end{align} 
Substitute this into
\begin{align}
    \wc=\frac{eB_0}{\me},
\end{align}
the relativistic frequency shift $\left. \Delta \wc \right|_{\text{rel}}$ is given by \cite{Gabrielse85e,Tseng98}
\begin{align}
    \left. \Delta \wc \right|_{\text{rel}}&=\frac{eB_0}{\me+\left. \Delta \me \right|_{\text{rel}}}-\frac{eB_0}{\me}\\
    &\approx -e B_0\frac{\left. \Delta \me \right|_{\text{rel}}}{\me^2}\\
    &\approx -\frac{\wc^2 n_c}{\me}.
\end{align}
For $\wc = 0.1 ~ \text{--} ~ 1$ meV and \textcolor{\specialColor}{$n_c\sim 10^6$}, we get
\begin{align}
    \left|\frac{\left. \Delta \wc \right|_{\text{rel}}}{\wc} \right|\approx  \frac{\wc}{\me} n_c \lesssim 2 \times 10^{-3} ~,
\end{align}
which is negligible.
Despite the fact that our optimization pushes for a much larger $\delta$ than conventional Penning trap experiment, we can see from Table \ref{tab: parameters} that we still respect $\delta \ll \wz$, a requirement for many perturbation expansions.
\section{Cavity Mode Calculation\label{sec: cavity mode appendix}}
\subsection{Spherical Cavity \label{sec: spherical cavity}}
We will calculate $\ksphere^2$ for a spherical cavity of radius $R$. For convenience, we temporarily set up a new set of coordinates, used only in this section. Let the original coordinate system be $\Sigma$, used everywhere else in this paper, in which the $z$-direction aligns with the magnetic field, and the dark photon's component in the $xy$-plane points purely in the $x$-direction (i.e. $A'_y=0$). Now, define a new coordinate system $\Sigma'$, in which the new $z'$-axis aligns with the old $x$-axis; the new $x'$-axis aligns with the old $z$-axis; to maintain a right-handed system, $\hat{y'} = -\hat{y}$, but we still have $A'_{y'}=0$.

We begin with the general equation \eq{\ref{eq: kappa with cylindrical symmetry}},
\begin{align} \label{eq: kappa spherical section}
\ksphere^2 &= \left| \sum_l \frac{\omega^2}{\omega_l^2- \omega^2} \frac{\int d^3 x' \Ebold_{l}^*(\xbold') \cdot \hat{z}'}{\int d^3x' |\mathbf{E_{l}(\xbold')}|^2}  \Ebold_{l} (\mathbf{0}) \cdot \hat{z}'  \right|^2 \\
&\equiv \left| \sum_l \frac{\omega^2}{\omega_l^2- \omega^2} G^{\text{sphere}}_l \right|^2 ~,
\end{align}
where $\omega = \m$ is the driving frequency, $l$ is a generic index running over all cavity modes, and we have omitted the quality factors. Using spherical coordinates ($r$, $\theta'$, $\phi'$) of the $\Sigma'$ system,
\begin{align} \label{eq: z' hat}
\hat{z}'
=\cos \theta' \hat{r} - \sin \theta' \hat{\theta}' ~.
\end{align}

The empty cavity modes of a spherical cavity can be divided into TE and TM modes (transverse electric and magnetic, respectively) with respect to the radial direction. There are three discrete indices $m = 0,1,2,\dots$; $n = 0,1,2, \dots$; and $p = 1,2,3,\dots$, corresponding to the quantization of momentum in the $\phi'$, $\theta'$, and $r$ directions, respectively. The exact expressions of these cavity modes can be found in \cite{DavidHill}. For brevity, instead of explicitly writing them all out, we will start by listing their key properties and using them to argue only a small subset of the cavity modes contribute. Then, we will work directly with the full expressions of those relevant modes.

As we will see shortly, the azimuthal symmetry (in the $\phi'$ direction) collapses the sum over $m$ into $m=0$ modes only and picks out only the TM modes; the polar symmetry (in the $\theta'$ direction) collapses the sum over $n$ into only the $n=1$ modes. So, the summation is only over TM$_{01p}$ modes.

\subsubsection{Only $m=0$ Modes}
All modes of the spherical cavity only depend on the coordinate $\phi'$ via a factor of $\cos(m\phi')$ or $\sin(m \phi')$ \cite{DavidHill}. By \eq{\ref{eq: z' hat}}, $\phi'$ also doesn't appear in $\hat{z}'$. This azimuthal symmetry makes the $\phi'$ integral in $\int d^3 x' \Ebold_{l}^*(\xbold') \cdot \hat{z}'$ trivial, with the result proportional to
\begin{equation}
    \int_0^{2\pi} d\phi' \cos(m\phi') = 2\pi \delta_{m0}
\end{equation}
or
\begin{equation}
    \int_0^{2\pi} d\phi' \sin(m\phi') = 0 ~.
\end{equation}
Hence, only $m=0$ modes contribute.

\subsubsection{Only TM Modes}
Next, we argue that the TE modes do not contribute. First, $\ETE_{r, mnp} = 0$ by definition of ``transverse electric." And $\ETE_{\theta', mnp}$ is proportional to $m$ \cite{DavidHill}, which vanishes when $m=0$. We are then left with $\ETE_{\phi', mnp}$. But it appears in the dot product $\Ebold_l^*(\xbold') \cdot \hat{z}'$ via $\ETE_{\phi', mnp} \hat{z}'_{\phi'}$ and $\hat{z}'_{\phi'}=0$, by \eq{\ref{eq: z' hat}}. Therefore, it suffices to consider only the $m=0$ TM modes.

These modes can, up to some irrelevant constants, be taken to be \cite{DavidHill}
\begin{align} \label{eq: ETMr sphere}
\ETMr = \frac{n(n+1)}{i \omega \kTMnp r^2} \Jhat_n\left(u'_{np}\frac{r}{R}\right) P_n(\cos \theta') ~,
\end{align}
\begin{align} \label{eq: ETM theta}
\ETMt = \frac{1}{i \omega r} \Jhatn'\left(u'_{np}\frac{r}{R}\right) \frac{d}{d\theta'} P_n(\cos \theta') ~,
\end{align}
where $P_n(x)$ are the Legendre polynomials, $\Jhatn(x) \equiv x j_n(x)$ are Harrington's spherical Bessel functions, $ j_n(x)$ are the normal spherical Bessel functions, $u'_{np}$ is the $p$-th zero of the derivative $\Jhatn'(x)$, and the wave number is $\kTMnp = u'_{np}/R$. The last component $\ETM_{\phi', mnp}$ is proportional to $m$ and, hence, can be taken to be zero for our purpose. 

\subsubsection{Only \texorpdfstring{$n=1$}{n=1} Modes}
To show that only $n=1$ modes contribute, consider the numerator in \eq{\ref{eq: kappa spherical section}}:
\begin{equation}
\int d^3 x' \Ebold_l^*(\xbold') \cdot \hat{z}' 
= \int d^3 x' \left( \ETMrs ~ \hat{z}'_r + \ETMts ~ \hat{z}'_{\theta'} \right) ~.
\end{equation}
We will consider these two terms separately.

For the $r$ term, using $\hat{z}'_r=\cos\theta'=P_1\left( \cos\theta' \right)$, the $\theta'$ integral evaluates to 
\begin{equation}
\int d^3x' \ETMrs ~ \hat{z}'_r
\propto  \int_{-1}^{1} d\left( \cos \theta' \right) P_n\left( \cos \theta' \right) P_1\left( \cos \theta' \right)
= \frac{2}{3} \delta_{n1} ~,
\end{equation}
where we used the orthogonality of the Legendre polynomials, $\int_{-1}^{1}dx P_k(x)P_l(x)=\frac{2}{(2l+1)}\delta_{kl}$.

The $\theta'$ term also picks out the $n=1$ modes. We first introduce the associated Legendre polynomials $P^j_n(x)$ given by
\begin{equation} \label{eq: associated Lengendre}
 P^j_n(u) = (-1)^j(1-u^2)^{j/2} \frac{d^j}{du^j} P_n(u)  ~.
\end{equation}
Note that $P^0_n(u)=P_n(u)$. The definition implies two useful identities:
\begin{equation} \label{eq: dp}
\frac{d}{d\theta'} \Pon = P^1_n\left( \cos\theta'\right)
\end{equation}
and
\begin{equation} \label{eq: z' theta identity}
\hat{z}'_{\theta'}=-\sin \theta'=P^1_1(\cos \theta') ~.
\end{equation}
Using \eq{\ref{eq: ETM theta}}, (\ref{eq: dp}), and (\ref{eq: z' theta identity}), the $\theta'$ integral evaluates to
\begin{align}
\int d^3x' \ETMts ~ \hat{z}'_{\theta'} ~ 
&\propto \int_{-1}^{1} d\left(\cos \theta' \right) P^1_n\left( \cos \theta'\right) P^1_1\left(\cos \theta' \right) =  \frac{4}{3} \delta_{n1} ~,
\end{align}
where we used the orthogonality of the associated Legendre polynomials, $\int_{-1}^{1}dx P^j_k(x)P^j_l(x)=\frac{2(l+j)!}{(2l+1)(l-j)!}\delta_{kl}$. We have thus shown that only TM$01p$ modes contribute.

\subsubsection{Final Expression of $\ksphere^2$}
We now compute 
\begin{equation}
\Gsphere = \frac{\int d^3 x' \Ebold^{\text{TM}*}_{01p}(\xbold') \cdot \hat{z}'}
{\int d^3x' |\Ebold^{\text{TM}}_{01p}(\xbold')|^2} \Ebold^{\text{TM}}_{01p}(\mathbf{0}) \cdot \hat{z}' ~.
\end{equation}
We first evaluate $
\Ebold^{\text{TM}}_{01p}(\mathbf{0})\cdot \hat{z}'$, the electric field at the center of the cavity. Approaching the origin from the $\theta' = 0$ direction, the $\theta'$ term vanishes, and the radial direction is equivalent to the $z'$ direction, $\hat{r} = \hat{z}'$. So, the only relevant component is $\ETMr$, given by \eq{\ref{eq: ETMr sphere}}. Substituting $n=1$, $P_1(\cos 0) = 1$, and the Taylor expansion
$\Jhato(x) = \frac{x^2}{3} + O(x^4)$, we have
\begin{align}
\Ebold^{\text{TM}}_{01p}(\mathbf{0}) \cdot \hat{z}'
&= \frac{1}{i \omega} \frac{2}{3} \frac{u'_{1p}}{R} ~.
\end{align}

Using our earlier results, the numerator of the coefficient can be written as
\begin{align}
\int d^3 x' \Ebold^{\text{TM}*}_{01p}(\xbold') \cdot \hat{z}'
=& \int d^3 x' \left[ \ETMros \cos \theta' -  \ETMtos \sin \theta' \right]\\
=& \frac{i}{\omega}(2\pi)  \left[ \left( \frac{2}{3} \right) \int_0^{R} dr \frac{2}{\kTMop} \Jhato \left(u'_{1p}\frac{r}{R}\right) 
+ \left( \frac{4}{3} \right) \int_0^{R} dr ~  r \Jhato'\left(u'_{1p}\frac{r}{R}\right) \right] \\
=& \frac{i}{\omega} (2\pi) \left( \frac{4}{3} \right) \frac{1}{u'_{1p}/R} \int_0^{R} dr \left[  \Jhato \left(u'_{1p}\frac{r}{R}\right) 
+  \left(u'_{1p}\frac{r}{R}\right) \Jhato'\left(u'_{1p}\frac{r}{R}\right) \right] ~.
\end{align}
Integrating by parts, the radial integral can be evaluated trivially
\begin{align}
\int d^3 x' \Ebold^{\text{TM}*}_{01p}(\xbold') \cdot \hat{z}' &= \frac{i}{\omega} (2\pi) \left( \frac{4}{3} \right) \frac{1}{u'_{1p}/R} \int_0^{R} dr \frac{d}{dr}\left[  r\Jhato\left(u'_{1p}\frac{r}{R}\right) \right] \\
=& \frac{i}{\omega} (2\pi) \left( \frac{4}{3} \right) R^2 j_1(u'_{1p})  ~.
\end{align}

Using the orthogonality of the Legendre and associated Legendre polynomials again, the denominator can be written as:
\begin{align}
&\int d^3x |\Ebold^{\text{TM}}_{01p}(\xbold)|^2\\
=&\int d^3x \left[ |\ETM_{r,01p}|^2 + |\ETM_{\theta,01p}|^2
\right] \\
=& \frac{1}{\omega^2} (2\pi)  \int_0^R dr ~ r^2 \left[\left( \frac{2}{\kTMop r^2} \Jhato \left(u'_{1p}\frac{r}{R}\right)\right)^2 \int_{-1}^{1} d(\cos\theta)  \left( P_1(\cos\theta) \right)^2  +\left( \frac{1}{r} \Jhato'\left(u'_{1p}\frac{r}{R}\right) \right)^2 \int_{-1}^{1} d(\cos\theta) \left( P^1_1(\cos \theta) \right)^2  \right] \\
=& \frac{1}{\omega^2} (2\pi) \left[ \left( \frac{2}{3}\right) \frac{4}{(u'_{1p}/R)^2} \int_0^R dr  \frac{1}{r^2}\left(  \Jhato \left(u'_{1p}\frac{r}{R}\right)\right)^2 +\left( \frac{4}{3}\right) \int_0^R dr \left( \Jhato'\left(u'_{1p}\frac{r}{R}\right) \right)^2 \right]\\
=& \frac{1}{\omega^2} (2\pi) \left( \frac{R}{u'_{1p}} \right) \left( \frac{4}{3}\right)  I(\up)
~,
\end{align}
where
\begin{align}
I(\up)= 2 \int_0^{u'_{1p}} dx  \left(  j_1(x) \right)^2 +\int_0^{u'_{1p}} dx \left( \Jhato'(x) \right)^2  ~,
\end{align}
with the change of variables $x=u'_{1p} r/R$. Using the analytic form of the spherical Bessel function
\begin{equation}
j_1(x) = -\frac{\cos(x)}{x}+ \frac{\sin(x)}{x^2} ~,
\end{equation}
the integral can be easily evaluated as
\begin{align}
I(\up)
&=\frac{2 + 2 u'^2_{1p} - 2 u'^4_{1p} + 2 (-1 + u'^2_{1p}) \cos(2 \up) + 
  \up (-4 +u'^2_{1p}) \sin(2 \up)}{-4 u'^3_{1p}} ~. \label{eq:exact Dp}
\end{align}

Putting everything together, we have
\begin{align}
\Gsphere &= \frac{2}{3}j_1(\up) \left[\frac{u'^2_{1p} }{I(\up)} \right]\\
&=- \frac{8}{3}j_1(\up) \left[\frac{u'^5_{1p}}{2 + 2 u'^2_{1p} - 2 u'^4_{1p} + 2 (-1 +u'^2_{1p}) \cos(2 \up) + 
  \up (-4 + u'^2_{1p}) \sin(2 \up)} \right]~,
\end{align}
and, using $\omega_p = \kTMop= \up/R$, we finally arrive at the cavity factor for the sphere
\begin{equation}
\kappa^2_{\text{sphere}}(\omega) =  \left| \sum_{p=1}^{\infty} \frac{(\omega R)^2}{u'^2_{1p} - (\omega R)^2} G^{\text{sphere}}_p \right|^2 ~.
\end{equation}

\subsection{Cylindrical Cavity \label{sec: cylindrical cavity}}
We will calculate $\kcylinder^2$ for a cylindrical cavity of radius $R$ and height $d$. We orient the cylinder's central axis to be parallel to the magnetic field direction $\hat{z}$, as was the case in the proof-of-principle experiment \cite{Fan_2022}. As before, the $\hat{x}$ direction is taken to be the component of the dark photon field in the plane perpendicular to the magnetic field. We will use the cylindrical coordinates $(\rho,\phi,z)$, with the center of the cylinder located at $\mathbf{0'}\equiv\left(0,0,\frac{d}{2} \right)$.

Then we can start with the general equation \eq{\ref{eq: kappa with cylindrical symmetry}},
\begin{align} \label{eq: general kappa equation cylinder}
\kcylinder^2 &= \left| \sum_l \frac{\omega^2}{\omega_l^2- \omega^2} \frac{\int d^3 x' \Ebold_{l}^*(\xbold') \cdot \hat{x}}{\int d^3x' |\mathbf{E_{l}(\xbold')}|^2}  \Ebold_{l} (\mathbf{0'}) \cdot \hat{x}  \right|^2 \\
&\equiv \left| \sum_l \frac{\omega^2}{\omega_l^2- \omega^2} G_l^{\text{cylinder}} \right|^2 ~,
\end{align}
where $\omega = \m$ is the driving frequency, $l$ is a generic index running over all cavity modes, and we have omitted the quality factors. In cylindrical coordinates,
\begin{align}
\hat{x}=\cos(\phi) \hat{\rho} - \sin(\phi)\hat{\phi} ~.
\end{align}

The empty cavity modes of a cylindrical cavity can be divided into TE and TM modes (transverse electric and magnetic, respectively) with respect to the $z$-direction. There are three discrete indices $n = 0,1,2,\dots$; $p = 1,2,3, \dots$; and $q = 0,1,2,\dots$, corresponding to the quantization of momentum in the $\phi$, $\rho$, and $z$ directions, respectively. These modes can, up to some irrelevant constants, be taken to be \cite{DavidHill}:
\begin{align}
E_{z,npq}^{\text{TM}}
&=J_n \left( \frac{x_{np}}{R} \rho \right)
\begin{Bmatrix}
\sin (n\phi)\\
\cos (n\phi)
\end{Bmatrix}
\cos\left(\frac{q\pi}{d}z \right) \\
E_{\rho,npq}^{\text{TM}} 
&=\frac{-1}{\frac{x_{np}}{R}} \frac{q\pi}{d} J_n' \left(\frac{x_{np}}{R} \rho \right)
\begin{Bmatrix}
\sin(n\phi)\\
\cos(n\phi)
\end{Bmatrix}
\sin(\frac{q\pi}{d}z) \\
E_{\phi,npq}^{\text{TM}} 
&=\frac{-1}{(\frac{x_{np}}{R})^2} \frac{nq\pi}{d}\frac{J_n (\frac{x_{np}}{R} \rho)}{\rho}
\begin{Bmatrix}
\cos(n\phi)\\
-\sin(n\phi)
\end{Bmatrix}
\sin(\frac{q\pi}{d}z)\\
E_{z,npq}^{\text{TE}}
&=0\\
E_{\rho,npq}^{\text{TE}} 
&=i \frac{1}{(\frac{x_{np}'}{R})^2} n\frac{J_n (\frac{x_{np}'}{R} \rho)}{\rho} 
\begin{Bmatrix}
\cos(n\phi)\\
-\sin(n\phi)
\end{Bmatrix}
\sin(\frac{q\pi}{d}z) \label{eq: ETE rho cylinder}\\
E_{\phi,npq}^{\text{TE}} 
&=-i\frac{1}{\frac{x_{np}'}{R
}} J'_n \left(\frac{x_{np}'}{R} \rho \right)
\begin{Bmatrix}
\sin(n\phi)\\
\cos(n\phi)
\end{Bmatrix}
\sin(\frac{q\pi}{d}z) ~, 
\end{align}
where $J_n$ is the n-th Bessel function, $x_{np}$ is the $p$-th zero of $J_n(x)$, and $x'_{np}$ is the $p$-th zero of the derivative $J'_n(x)$. As we shall see, only the $\text{TE}_{1pq}$ modes with odd $q$ contribute to $\kappa^2$. The resonant frequencies are
\begin{align}
(\omega^{\text{TM}}_{npq})^2 &= \left(\frac{x_{np}}{R} \right)^2 + \left(\frac{q \pi}{d} \right)^2\\
(\omega^{\text{TE}}_{npq})^2 &= \left(\frac{x'_{np}}{R} \right)^2 + \left(\frac{q \pi}{d} \right)^2 ~.
\end{align}

\subsubsection{Only $n=1$ Modes}
In the numerator of \eq{\ref{eq: general kappa equation cylinder}}, we have
\begin{align}
\int d^3 x' \Ebold_{l}^*(\xbold') \cdot \hat{x} ~ \propto \int_0^{2\pi} d\phi \left( E_{\rho,npq}^{\text{TM}} \cos \phi - E_{\phi,npq}^{\text{TM}} \sin \phi \right) ~.
\end{align}
Unless we choose the matching sinusoidal function with $n=1$, the integral vanishes:
\begin{align}
\int_0^{2\pi} d\phi \sin\phi \sin(n\phi)=\int_0^{2\pi} d\phi \cos\phi \cos(n\phi) = \pi \delta_{n1} ~.
\end{align}
So only $n=1$ modes contribute to $\kappa^2$.

\subsubsection{Only TE Modes}
Consider a term involving a TM $n=1$ mode. The integral in the numerator of \eq{\ref{eq: general kappa equation cylinder}} is proportional to 
\begin{align}
\int d^3 x' \Ebold_{l}^*(\xbold') \cdot \hat{x} ~ \propto& \int_0^R d\rho \rho \left[ J'_1 \left(\frac{x_{1p}}{R} \rho \right) + \left(\frac{x_{1p}}{R}\right)^{-1}\frac{J_1 \left(\frac{x_{1p}}{R} \rho \right)}{\rho}\right]\\
 ~ \propto& \int_0^R d\rho \left[ \rho  \left(\frac{x_{1p}}{R}\right) J'_1 \left(\frac{x_{1p}}{R} \rho \right) + J_1 \left(\frac{x_{1p}}{R} \rho \right)\right]\\
 \propto& ~  \left.  \rho J_1 \left(\frac{x_{1p}}{R} \rho \right) \right|_0^{R}\\
 =& 0 ~,
\end{align}
where we used integration by parts in the third line, and in the last step, we used the property of the Bessel function $J_1(0)=0$ and $J_1(x_{1p})=0$, by definition. Hence, the TM modes do not contribute. We will henceforth only use TE modes and drop the superscript in the resonant frequencies.

\subsubsection{Final Expression of $\kcylinder^2$}
We now compute 
\begin{equation}
\Gcylinder = \frac{\int d^3 x \Ebold^{\text{TE}*}_{1pq}(\xbold') \cdot \hat{x}}
{\int d^3x |\Ebold^{\text{TE}}_{1pq}(\xbold)|^2} \Ebold^{\text{TE}}_{1pq}(\mathbf{0'}) \cdot \hat{x} ~.
\end{equation}
We first evaluate $
\Ebold^{\text{TE}}_{1pq}(\mathbf{0}') \cdot \hat{x}$, the electric field at the center of the cavity. Approaching the origin from the $\phi = 0$ direction, the $\phi$ term vanishes, and the radial direction is equivalent to the $x$ direction, $\hat{\rho} = \hat{x}$. So, the only relevant component is $E^{\text{TM}}_{\rho,1pq}$, given by \eq{\ref{eq: ETE rho cylinder}},
\begin{align}
\Ebold^{\text{TE}}_{1pq}(\mathbf{0'}) \cdot \hat{x}
&= i \left(\frac{x_{1p}'}{R} \right)^{-2} \lim_{\rho \to 0} \frac{J_1 \left(\frac{x_{1p}'}{R} \rho \right)}{\rho} 
\lim_{\phi \to 0} \cos(\phi)
 \lim_{z \to d/2} \sin(\frac{q\pi}{d}z) \\
 &= \frac{i}{2} \left(\frac{x_{1p}'}{R} \right)^{-1} 
\sin(\frac{q\pi}{2})
 ~.
\end{align}
We see that only modes with odd $q$ contribute.

The integral in the numerator is given by
\begin{align}
\int d^3 x \Ebold^{\text{TE}*}_{1pq}(\xbold') \cdot \hat{x} 
&= -i \pi \left(\frac{x_{1p}'}{R} \right)^{-2} \int_0^R d\rho ~ \rho \left[\frac{J_1\left(\frac{x_{1p}'}{R} \rho \right)}{\rho} + \left(\frac{x_{1p}'}{R}\right)J'_1\left(\frac{x_{1p}'}{R} \rho \right) \right] \int_0^d dz ~ \sin(\frac{q\pi}{d} z) \\
&= -i \pi \left(\frac{x_{1p}'}{R} \right)^{-2} \left[\rho J_1\left(\frac{x_{1p}'}{R} \rho \right) \right]_0^R \frac{2d}{q\pi} \\
&= -i \left(\frac{x_{1p}'}{R} \right)^{-2} \left( R  \frac{2d}{q} \right) J_1\left(x_{1p}'\right) ~,
\end{align}
where we integrated parts in the second step. A similar calculation shows that the denominator is
\begin{align}
\int d^3x |\Ebold^{\text{TE}}_{1pq}(\xbold)|^2
= \left(\frac{x_{1p}'}{R} \right)^{-4} \frac{1}{2} \left[x'^2_{1p} J_0^2(x_{1p}')+({x'^2_{1p}}-2)J_1^2(x_{1p}') \right](\pi) \left( \frac{d}{2} \right) ~.
\end{align}

Putting everything together, we have
\begin{align}
\Gcylinder &= \frac{ 4 x_{1p}' J_1\left(x_{1p}'\right)}
{\left[x'^2_{1p} J_0^2(x_{1p}')+({x'^2_{1p}}-2)J_1^2(x_{1p}') \right](q \pi) } \\
&= \frac{4 x'_{1p}}{(x'^2_{1p}-1)J_1(x_{1p}')}\frac{1}{q\pi} \qquad (q \text{ odd})
~,
\end{align}
where, in the second line, we used the identity $x'_{1p}J_0(x'_{1p})=J_1(x'_{1p})$, which can be easily derived from differentiating the relation between the Bessel functions $\int dx ~  x J_0(x) = x J_1(x)$. The cavity factor is then
\begin{align}
\kcylinder^2 = \left| \sum_{p=1}^{\infty} \sum_{q \text{ odd}} \frac{\omega^2}{\left(\frac{x'_{1p}}{R} \right)^2 + \left(\frac{q \pi}{d} \right)^2- \omega^2} \Gcylinder \right|^2 ~.
\end{align}

\section{Linear Focusing of a Cylindrical Cavity\label{sec:cylindrical focusing}}
As derived in Appendix \ref{sec: cylindrical cavity}, the cavity factor for a cylinder of radius $R$ and height $d$ is given by a sum over two discrete indices, $p$ and odd integers $q$:
\begin{align}
\kcylinder^2(\omega) &= \left| \sum_{p=1}^{\infty} \sum_{q \text{ odd}} \frac{\omega^2}{\omega_{1pq}^2- \omega^2} \Gcylinder \right|^2 \\
&= \left| \sum_{p=1}^{\infty} \sum_{q \text{ odd}} \frac{\omega^2}{\left(\frac{x'_{1p}}{R} \right)^2 + \left(\frac{q \pi}{d} \right)^2- \omega^2} \Gcylinder \right|^2 ~,
\end{align}
where
\begin{align}
\Gcylinder = \frac{4 x'_{1p}}{(x'^2_{1p}-1)J_1(x_{1p}')}\frac{1}{q\pi} \qquad (q \text{ odd})~,
\end{align}
$J_n$ is the n-th Bessel function, and $x'_{np}$ is the $p$-th zero of the derivative $J'_n(x)$.

Now, let $\omega_N$ be the $N$-th smallest resonant frequency among all $\wpq$, i.e. we re-index the frequencies using a single number $N$. For a generic $\omega$, the sum is always dominated by the closest resonant mode $\omega_{N_*}$ among all the $\omega_N$: 
\begin{align}
\kcylinder^2(\omega) &\approx \left| \frac{\omega^2}{\wNstar^2 - \omega^2} \GcylinderNstar \right|^2 \\
&\approx \left| \frac{\omega^2}{2\omega (\wNstar-\omega)} \GcylinderNstar \right|^2  ~. \label{eq: first cylinder kappa square}
\end{align}
To understand the term $1/(\wNstar-\omega)$, consider the local density of resonance near a particular $\omega$. Note that, unlike the case of the sphere, which has one index, the density is not constant for the cylinder due to having two indices $p$ and $q$. If the local density gets larger, the typical distance between $\omega$ and $\wNstar$ gets smaller. This quantity is of order the distance between neighbouring resonant frequencies:
\begin{align}
    \wNstar-\omega \lesssim \wNstar - \omega_{N_*-1} ~.
\end{align} 
This, in turn, is of order the ``inverse local density of resonance":
\begin{align} \label{eq: inverse local density of resonance}
\wNstar - \omega_{N_*-1} \sim \frac{\delta \omega}{\delta N} ~,
\end{align}
where there are $\delta N$ resonant frequencies between $\omega$ and $\omega + \delta \omega$.

\begin{figure}[htpb]
    \hspace*{-0.2in}
    \vspace*{0.5cm}
    \centering
    \includegraphics[width=0.7\textwidth, bb=0 0 2250 1250]{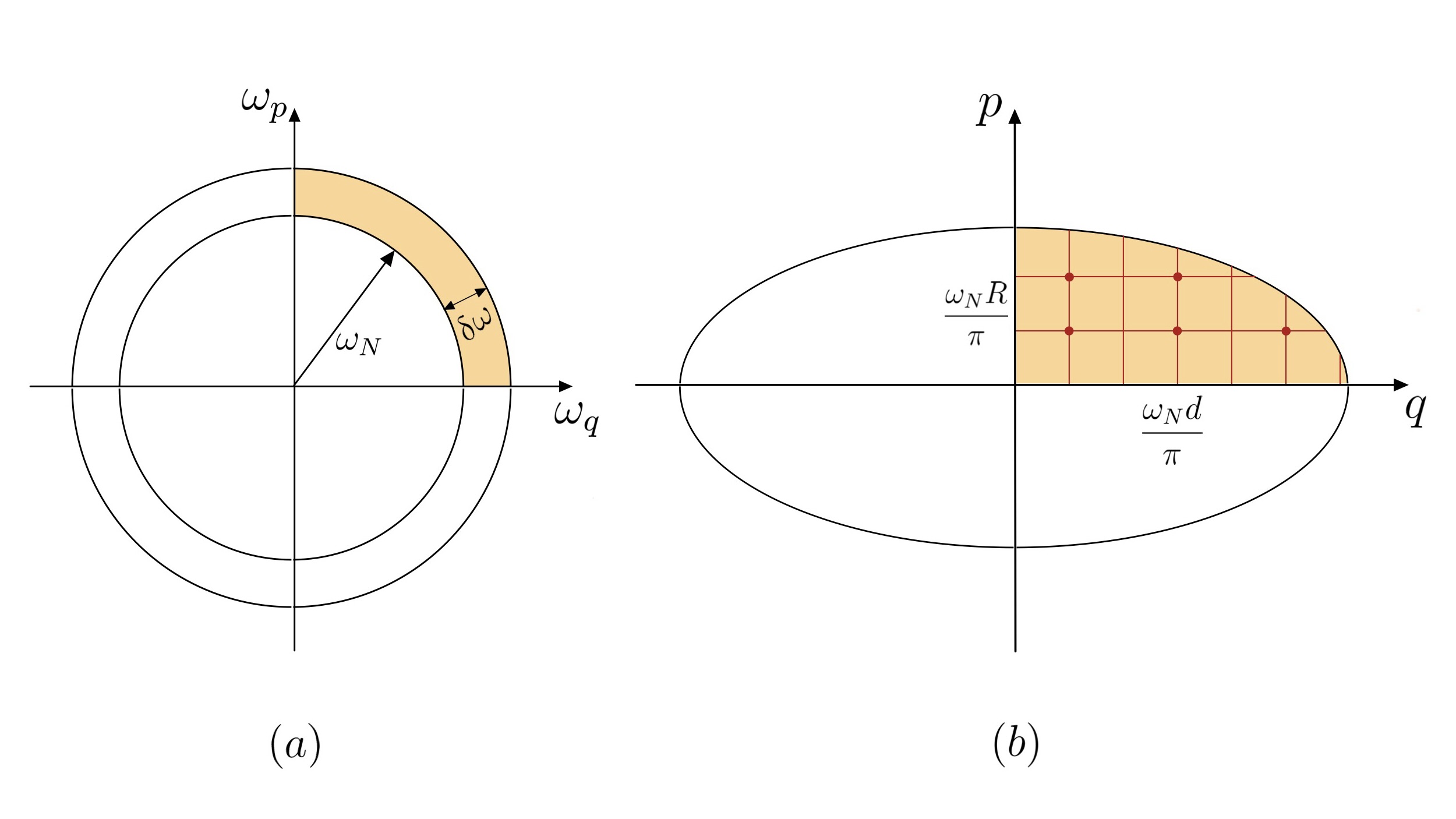}
    \caption{(a) The shaded area of the ring is proportional to the incremental number of resonant frequencies $\delta N$. (b)~The shaded area of the ellipse is approximately equal to the number of resonant frequencies $N$.}
    \label{fig: Cylinder Frequency Density}
\end{figure}

Next, we argue that $\delta N$ is proportional to $\omega \hspace{0.2 em} \delta \omega$ for a two-index family of resonant frequencies. First, consider the following relation between the frequencies:
\begin{align} \label{eq: circle and ellipse}
    \wN^2 = \wpq^2 = \left(\frac{x'_{1p}}{R} \right)^2 + \left(\frac{q \pi}{d} \right)^2 \equiv \omega_p^2 + \omega_q^2  ~.
\end{align}
Notice that this describes a circle of radius $\wN$ in $\omega_p \omega_q$-space. $N$, the number of resonant frequencies less than $\omega_N$, is proportional to (one quarter of) the area of the circle. If we increase the radius of the circle by $\delta \omega$, the additional ring has an area $2\pi \omega_N \delta \omega$, which is proportional to $\delta N$, as shown in Figure~\ref{fig: Cylinder Frequency Density}a. So we have the relation
\begin{align}
    \frac{\delta N }{N} &= \frac{2\pi \wN \delta \omega}{\pi \wN^2} \\
    \frac{\delta \omega}{\delta N } &= \frac{\wN}{2N} ~.
\end{align}
Substituting into \eq{\ref{eq: inverse local density of resonance}} and \eq{\ref{eq: first cylinder kappa square}}, we get
\begin{align}
\kcylinder^2(\omega)
&\approx \left| N \GcylinderNstar \right|^2  ~,
\end{align}
justifying the aforementioned name.

To find the value of $N$, we need to go to $pq$-space. By \eq{\ref{eq: circle and ellipse}}, we have an ellipse in $pq$-space, with semi-major and semi-minor axes $\wN R/ \pi$ and $\wN d/ \pi$:
\begin{align}
1 = \left(\frac{p}{\wN R/ \pi} \right)^2 + \left(\frac{q }{\wN d/ \pi} \right)^2 ~,
\end{align}
where we used $x'_{1p} \approx \pi p$ for large $p$.

Now, consider the integer lattice of $pq$-space. Every point in this lattice with positive $p$ and positive odd $q$ corresponds to a unique resonant frequency, as shown in Fig.~\ref{fig: Cylinder Frequency Density}(b). So, the area of the ellipse is $8 N$:
\begin{align}
8 N = \pi \left(\frac{\wN R}{\pi} \right)\left(\frac{\wN d}{\pi} \right) ~.
\end{align}

Using $\wN \approx \omega$, we now have
\begin{align}
\kcylinder^2(\omega)
&\approx \left| \frac{\omega^2 R d}{8\pi} \Gcylinderstar \right|^2  ~.
\end{align}
Finally, for large $p$, we have the scaling
\begin{align} \label{eq: Gcylinder}
\left|\Gcylinder \right|
&= \left| \frac{4 x'_{1p}}{(x'^2_{1p}-1)J_1(x_{1p}')}\frac{1}{q\pi} \right|
\approx \left| \frac{4}{\sqrt{x'_{1p}}\sqrt{2/\pi}}\frac{1}{q\pi} \right|
~.
\end{align}
Using $\omega_q \approx \omega_p \approx \omega/\sqrt{2}$, this yields
\begin{align}
\kcylinder^2(\omega)
&\approx \left| \sqrt{\frac{1}{8\pi}} \frac{\omega^2 R}{\sqrt{x'_{1p}}}  \frac{d}{q\pi} \right|^2 = \left| \sqrt{\frac{1}{8\pi}} \frac{\omega^2 \sqrt{R}}{\sqrt{\omega_p}}  \frac{1}{\omega_q} \right|^2 \approx  \frac{1}{2\sqrt{2}\pi} \omega R
~,
\end{align}
verifying our prediction of linear focusing for a cylindrical cavity in Section \ref{sec: physical picture}.
\section{Fixed Polarization Scenario\label{sec:fixed polarization}}

Several cosmological generation mechanisms of dark photon dark matter (DPDM) result in a fixed polarization scenario. This has a nontrivial interplay with the rotation of the Earth, modifying our dark photon limits. \cite{DPLimitsReview2021} systematically studied this effect on experiments, but that paper assumes continuous observables. Since our observable is discrete (quantum transitions), we need a different treatment. 

\subsection{Transition Probability}
We first define a coordinate system with the center of Earth as the origin and the Earth rotation axis as $\hat{z}$ (leaving $\hat{x}$ and $\hat{y}$ arbitrary for now). At the location of the experiment, define the unit vector pointing in the direction of the magnetic field of the Penning trap to be $\Bhat(t)$, which is changing due to the rotation of the Earth. For our experiment, this happens to point straight up toward the zenith, making the calculation a bit simpler, although we will show that the result can be easily generalized.

Writing $\Bhat(t)$ in spherical polar coordinate, the polar angle $\theta$ is fixed and is determined by the latitude of the location, while the azimuthal angle is rotating along with the Earth, $\varphi(t) = \ws t + \varphi_0$. Here, $\ws$ is the angular velocity of Earth rotation, given by one cycle per sidereal day: $\ws = \frac{2 \pi}{23.93 \text{ hour}}$. We can write this vector as
\begin{equation}
\Bhat(t) = \begin{pmatrix}
 \sin \theta \cos \varphi(t) \\
 \sin \theta \sin \varphi(t) \\
\cos \theta 
\end{pmatrix} ~.
\end{equation}
Next, let the unit vector along the direction of the fixed DPDM be $\Ahat$. Without loss of generality, we can choose $\hat{x}$ and $\hat{y}$ such that its azimuthal angle $\varphi_A=0$. Then the unknown cosmological polarization of DPDM can be entirely parameterized by the polar angle $\tA$:
\begin{equation}
\Ahat = \begin{pmatrix}
\sin \tA\\
0\\
\cos \tA 
\end{pmatrix} ~.
\end{equation}
Finally, define the angle between $\Bhat(t)$ and $\Ahat$ to be $\alpha(t)$. The quantity we are interested in is their dot product:
\begin{align}
\cos \alpha(t) &= \Ahat \cdot \Bhat(t)\\
&= \sin \theta \sin \tA  \cos(\ws t + \varphi_0) + \cos \theta  \cos \tA ~. \label{eq:cos square alpha}
\end{align}

It is easy to get $\theta$ and $\varphi_0$ for an arbitrary experimental orientation. Suppose the direction of the Penning trap magnetic field $\Bhat$ does not point to the Zenith, but instead has both a North-South tilt $\Delta \theta_{NS}$ and an East-West tilt $\gamma_{EW}$. The North-South tilt can be accounted for by our previous formalism by modifying the definition of polar angle: $\theta \equiv \theta_{\text{location}} + \Delta \theta_{NS}$. Having a $\gamma_{EW}$ amounts to shifting $\varphi \to \varphi + \Delta \varphi_{EW}$. This is degenerate with $\varphi_0$ and hence makes no difference.

Now, consider the probability of observing no cyclotron excitation in the presence of DPDM, which follows a Poisson distribution. Consider the infinitesimal time interval between $t$ and $t+dt$. Let the average number of excitations during this time be 
\begin{equation}
    d\lambda = \Gamma_c dt ~,
\end{equation}
where the cyclotron transition rate $\Gamma_c$ is given by eq. (\ref{eq:Gamma cavity}). We will separate out its angular dependence by
\begin{equation}
    \Gamma_c \equiv \Gamma_0 \sin^2\alpha(t) = \Gamma_0 \left[ 1- \cos^2\alpha(t) \right]~,
\end{equation}
which is always possible for a cavity with azimuthal symmetry (see the discussion around \eq{\ref{eq: kappa with cylindrical symmetry}}). Notice that this quantity depends on $\cos(\tA)$ and $\varphi_0$, which are both drawn from uniform distributions.

The conditional probability of observing no excitation during the time interval $dt$ conditioned on a polarization angle $\tA$ and initial azimuthal phase $\varphi_0$ is:
\begin{equation}
dP_0( \cos \tA, \varphi_0) = e^{-d\lambda(\cos \tA, \varphi_0)} ~,
\end{equation}
where we have made explicit the dependence of $d\lambda$. 
To convert the conditional probability to total probability, we marginalize over the distribution of the two uncorrelated conditions:
\begin{align}
dP_0  &= \int_{-1}^{1} d(\cos \tA) P(\cos \tA) \int_0^{2\pi} d \varphi_0 P(\varphi_0) dP_0( \cos \tA, \varphi_0)  \\
&= \int_{-1}^{1} \frac{d(\cos \tA)}{2} \int_0^{2\pi} \frac{d \varphi_0 }{2\pi} dP_0( \cos \tA, \varphi_0) ~.
\end{align}
Since we have no prior knowledge of the DPDM polarization angle and the initial azimuthal phase, we have taken both to be uniform distributions: $P(\cos \tA)=\frac{1}{2}$ and $P(\varphi_0)=\frac{1}{2\pi}$. Notice that since the area element is proportional to $\sin(\tA) d\tA = d(\cos \tA)$, this is the correct quantity to have a uniform distribution, not $\tA$ itself.

Finally, the total probability of detecting nothing over the entire run time of the experiment at a given frequency bin from $t=0$ to $t=\tobs$ is the successive product of the instantaneous probabilities:
\begin{align}
P_0 &= \prod_{dt} dP_0\\
&= \int_{-1}^{1} \frac{d(\cos \tA)}{2} \int_0^{2\pi} \frac{d \varphi_0 }{2\pi} \prod_{dt}  dP_0( \cos \tA, \varphi_0)\\
&= \int_{-1}^{1} \frac{d(\cos \tA)}{2} \int_0^{2\pi} \frac{d \varphi_0 }{2\pi} \prod_{dt} \exp[-d\lambda]\\
&= \int_{-1}^{1} \frac{d(\cos \tA)}{2} \int_0^{2\pi} \frac{d \varphi_0 }{2\pi} \exp[-\int d\lambda]\\
&= \int_{-1}^{1} \frac{d(\cos \tA)}{2} \int_0^{2\pi} \frac{d \varphi_0 }{2\pi} \exp{-\Gamma_0 \int_0^{\tobs} dt \left[1- \cos^2\alpha(t) \right]} ~.
\end{align}
Substituting \eq{\ref{eq:cos square alpha}} and incorporating the probability of fast decay $\Gamma_c \to \zeta \Gamma_c$, where $\zeta $ is the detection efficiency given by \eq{\ref{eq: efficiency zeta}}, we can finally find the CL=90\% confidence level limit by solving the following equation for $\Gf$:
\begin{align}\label{eq: gamma0 for fixed}
1-CL = P_0(\zeta\tobs,\theta)= \int_{-1}^{1} \frac{d(\cos \tA)}{2} \int_0^{2\pi} \frac{d \varphi_0 }{2\pi} \exp{-\zeta \Gf \int_0^{\tobs} dt \left[1- \left[ \sin \theta \sin \tA  \cos(\ws t + \varphi_0) + \cos \theta  \cos \tA\right]^2 \right]}~.
\end{align}

\subsection{Long Time Limit}

Although our current proposed parameters and the previous proof-of-principle experiment \cite{Fan_2022} both have observation times $\tobs$ of several days, it is easy to adjust our quality factor $Q_c$ to achieve a much shorter $\tobs$. Therefore, we will discuss both the long and short observation time limits (defined relative to a Sidereal day) and show that they, in fact, share the same result.

\Eq{\ref{eq: gamma0 for fixed}} tells us that the solution of $\Gf$ depends on $\tobs$ via an integration limit. Consider the case that
\begin{align}
  T=\tobs\gg1~\text{day}=\frac{2\pi}{\ws}.  
\end{align}
In this limit, we can write the time integral in terms of time average
\begin{equation}
\langle f(t) \rangle_T = \lim_{T \to \infty} \frac{\int_0^T dt f(t)}{T} ~.
\end{equation}
Therefore, in the long time limit, the time integral of \eq{\ref{eq: gamma0 for fixed}} becomes 
\begin{align}\label{eq: average T}
    \int_0^{T} dt \left[1- \left[ \sin \theta \sin \tA  \cos(\ws t + \varphi_0) + \cos \theta  \cos \tA\right]^2 \right]\approx T\times(1- \langle  \left[ \sin \theta \sin \tA  \cos(\ws t + \varphi_0) + \cos \theta  \cos \tA\right]^2 \rangle_T)~.
\end{align}

Using $ \langle \cos(\ws t + \varphi_0) \rangle_T = 0$ and $\langle \cos^2(\ws t + \varphi_0)\rangle_T = 1/2$, we can then write the integral as
\begin{align}
T \times(1- \frac{1}{2} \sin^2 \theta \sin^2 \tA  - \cos^2 \theta  \cos^2 \tA ) ~.  
\end{align}

Now we are able to rewrite \eq{\ref{eq: gamma0 for fixed}} and simplify it,
\begin{align}\label{eq: trans prob fix}
    1-CL=P_0(\zeta\tobs,\theta)&= \int_{-1}^{1} \frac{d(\cos \tA)}{2} \int_0^{2\pi} \frac{d \varphi_0 }{2\pi} \exp{-\zeta \Gf \tobs \times(1- \frac{1}{2} \sin^2 \theta \sin^2 \tA  - \cos^2 \theta  \cos^2 \tA )}\\
    &= \exp{-\zeta\Gf \tobs(1-\frac{1}{2}\sin^2 \theta)} \int_{-1}^{1} \frac{dx}{2} \exp{\zeta\Gf \tobs(1-\frac{3}{2} \sin^2 \theta) x^2} ~.
\end{align}
For comparison, we also recall the random polarization case from Appendix \ref{sec: limit}:
\begin{align}\label{eq: trans prob random}
1-CL=P_0(\zeta\tobs) = \exp{-\zeta\Gamma_{c,\text{random}}\tobs} = \exp{-\zeta\Gr \langle \sin^2 \theta_A \rangle \tobs}= \exp{-\frac{2}{3}\zeta \Gr \tobs} ~.
\end{align}

\subsubsection{Experiment in Chicago}
The $\theta$ in \eq{\ref{eq: trans prob fix}} is a parameter that depends on the location of the experiment. In our previous work \cite{ThesisXingFan}, we did a multi-day proof-of-principle experiment in Chicago, which satisfies the long time limit requirement $\tobs\gg1$ day. Therefore, we can use \eq{\ref{eq: trans prob fix}} to calculate $\Gf$. 

That experiment was located at Evanston, Illinois, which has a latitude of $42.05^{\circ}$N, corresponding to a polar angle of $\theta_{\text{Chicago}} = 90^{\circ}- 42.05^{\circ} =0.837 $. The total run time was $\tobs= 638870 \text{ seconds}$. Substituting these two numbers, we numerically found a limit of
\begin{equation}
\Gamma_{0,\text{fixed}} < 6.55 \times 10^{-6} \text{ s}^{-1} ~.
\end{equation}
 By comparison, the limit from \eq{\ref{eq: trans prob random}} is equivalent to
 \begin{equation}
\Gamma_{0,\text{random}} < \frac{3}{2} \times 4.33 \times 10^{-6} \text{ s}^{-1} = 6.50\times 10^{-6} \text{ s}^{-1} ~.
 \end{equation}
We see that the fixed polarization limit on $\epsilon$ is only a little bit weaker than that of random polarization, and we will see below this is due to $\theta_{\text{Chicago}}=0.837$ being not too far from the optimal angle $\theta_* = 0.955$. 

\subsubsection{Optimal Angle}
\begin{figure}[h]
\hspace{-0.2in}
\vspace{0.5cm}
\centering
  \includegraphics[width=0.5\textwidth, bb=0 0 630 420]{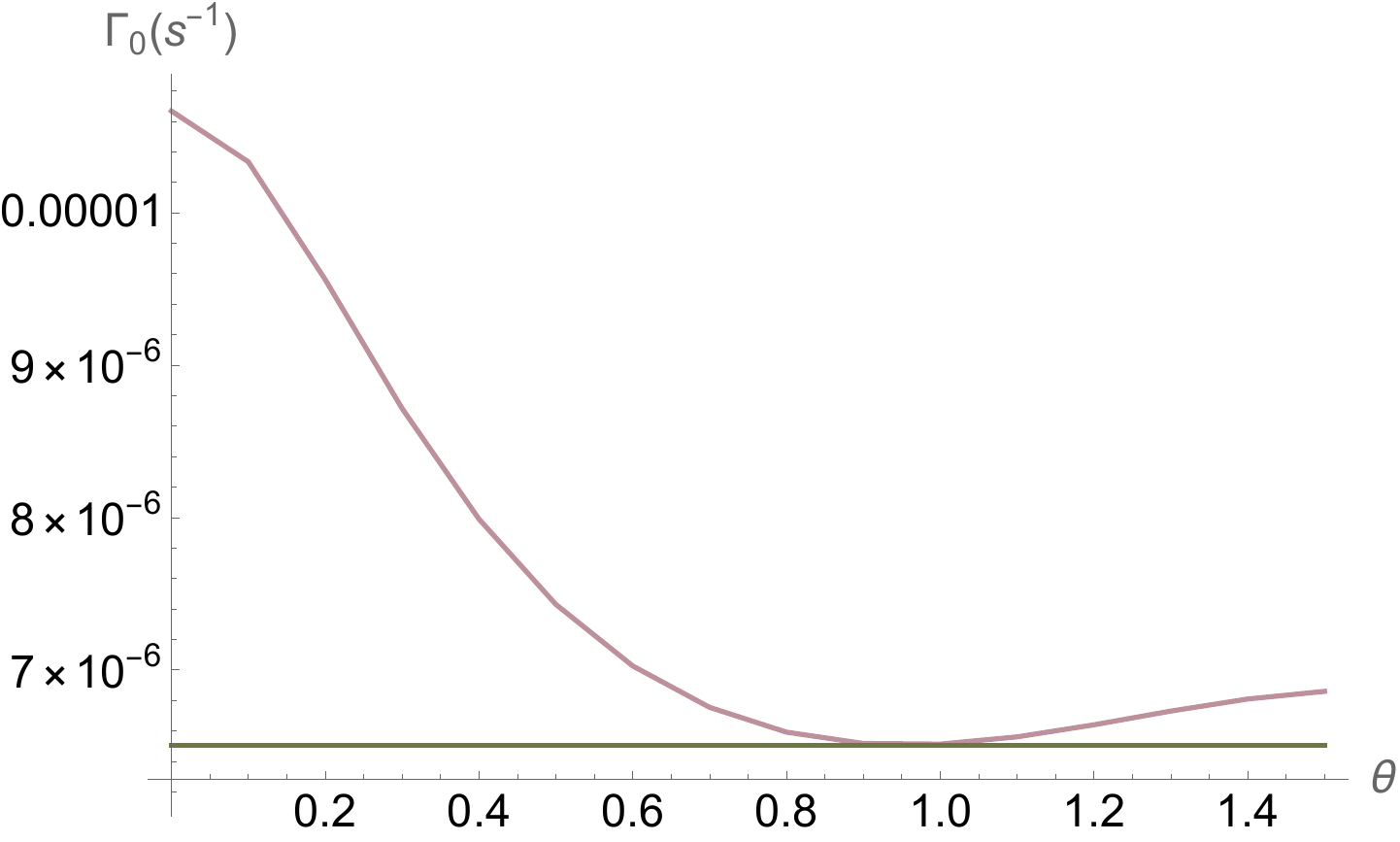}
  \caption{Transition rate vs polar angle of the Penning trap magnetic field. The purple curve corresponds to the fixed polarization case, while the green line is the random polarization case.}
  \label{fig:Transition Rate vs Polar Angle}
\end{figure}

In Figure \ref{fig:Transition Rate vs Polar Angle},  we plot the numerical result from \eq{\ref{eq: trans prob fix}} and \eq{\ref{eq: trans prob random}}, and it shows that the fixed polarization scenario, in general, sets a worse limit than the random polarization case. However, there is an optimal angle where they are equal. By comparing  \eq{\ref{eq: trans prob fix}} and \eq{\ref{eq: trans prob random}}, it is obvious that this optimal polar angle occurs exactly at $\sin^2(\theta_*)= \frac{2}{3}$, with $\theta_* =0.955$.

\subsection{Short Time Limit}
For short observation time where $\tobs\ll1~\text{day}$, \eq{\ref{eq: gamma0 for fixed}} can be simplified as,
\begin{align}
1-CL = P_0(\zeta\tobs,\theta,t_0)&= \int_{-1}^{1} \frac{d(\cos \tA)}{2} \int_0^{2\pi} \frac{d \varphi_0 }{2\pi} \exp{-\zeta \Gf \int_{t_0}^{t_0+\tobs} dt \left[1- \left[ \sin \theta \sin \tA  \cos(\ws t + \varphi_0) + \cos \theta  \cos \tA\right]^2 \right]}\\
&\approx \int_{-1}^{1} \frac{d(\cos \tA)}{2} \int_0^{2\pi} \frac{d \varphi_0 }{2\pi} \exp{-\zeta \Gf \tobs \left[1- \left[ \sin \theta \sin \tA  \cos(\ws t_0 + \varphi_0) + \cos \theta  \cos \tA\right]^2 \right]}~.
\end{align}
Here, $t_0$ represents the initial time of each observation. Since we are scanning in frequency continuously, different values of 
$t_0$ can be regarded as representing an average value over the long observation period. Averaging $P_0(\zeta\tobs,\theta,t_0)$ over one day gives the expected value of $P_0(\zeta\tobs,\theta)$:
\begin{align}
\langle P_0(\zeta\tobs,\theta) \rangle= \int_{-1}^{1} \frac{d(\cos \tA)}{2} \int_0^{2\pi} \frac{d \varphi_0 }{2\pi} \exp{-\zeta \Gf \tobs \langle\left[1- \left[ \sin \theta \sin \tA  \cos(\ws t_0 + \varphi_0) + \cos \theta  \cos \tA\right]^2 \right]\rangle_{T=1\text{day}}}~.
\end{align}
Comparing this expression of $\langle P_0(\zeta\tobs,\theta) \rangle$ with \eq{\ref{eq: average T}}, it turns out that we get the same answer for these two limits.

\newpage
\section{Experimental Parameters\label{sec: experimental parameters}}
\begin{table}[H]
    \centering
\resizebox{\textwidth}{!}{
    \begin{tabular}{|c|c|c|c|c|} \hline  
          \textbf{parameter}&\textbf{symbol}& \textbf{optimal}&  \textbf{lower limit}&\textbf{upper limit}\\ \hline  
  external magnetic field&$\Bext$& 10 T& N/A&BREAD experiment\\ \hline  
          cavity size&$R$&  1 m&  \multicolumn{2}{|c|}{BREAD experiment}\\ \hline  
  bottle ring magnetic field perturbation& $\tilde{B_2}$& 0.27 T& N/A&maximum magnetic field\\ \hline  
          bottle size&$\Rbot$ &  0.5 mm&  fabrication technique&N/A\\ \hline  
 magnetic field gradient& $B_2$& $1.08\times 10^6~\text{T}/\text{m}^2$& \multicolumn{2}{|c|}{$B_2=  \frac{\tilde{B_2}}{\Rbot^2} $}\\ \hline  
 cyclotron radii& $r_c$& $ <3.9 \times 10^{-2} \text{ mm }  $& \multicolumn{2}{|c|}{$r_c=\sqrt{n_c \frac{2}{m_e \wc}}$}\\ \hline  
  effective magnetic field perturbation& $\tilde{B}_\text{2,eff}$& $<1.6\times 10^{-3}\text{ T}$& \multicolumn{2}{|c|}{$\tilde{B}_\text{2,eff}=B_2\times r_c^2$}\\ \hline  
 axial amplitude& $\zmax$& 0.2 mm& N/A&$\Rbot$ (diffraction  \& $B_2$ anharmonicity)\\ \hline  
  trap radius and height&$\Rtrap$& 0.5 mm& axial anharmonicty&$\Rbot$\\ \hline  
 trap height& $d$& 1.02 mm& \multicolumn{2}{|c|}{$d=2.0478\Rtrap$ (axial anharmonicity \cite{OpenTrap})}\\ \hline  
  trapping potential&$|V_0|$& $\approx$225 V$\left(\frac{0.1~\text{meV}}{\wc} \right)^2$&  trapping in axial direction&power supply\\ \hline 
 effective axial frequency& $\frac{\wz}{2\pi}$& 9.8 MHz& \multicolumn{2}{|c|}{$ \wz^2 = \frac{ e V_0}{\me(\Rtrap^2+\frac{1}{2}d^2)}+2 n_c \frac{e B_2}{m_e^2} $}\\ \hline  
 effective axial frequency shift& $\delta$& $56.9$  kHz& \multicolumn{2}{|c|}{$\delta = \frac{e B_2}{m_e^2 \wz}$}\\ \hline  
 magnetron frequency & $\omega_m$& -114 neV $\left(\frac{0.1~\text{meV}}{\wc} \right)^3$& \multicolumn{2}{|c|}{$\wm = \frac{1}{2 \wc}\times\frac{ e V_0}{\me(\Rtrap^2+\frac{1}{2}d^2)}$}\\ \hline  
 signal formation time& $\tsignal$& $2.8 \times 10^{-6}$ s& \multicolumn{2}{|c|}{$\tsignal=\frac{1}{\delta}$}\\ \hline  
 detector quality factor& $\Qdet$& $2.9 \times 10^4$& \multicolumn{2}{|c|}{$\Qdet \leq 2.8 \times 10^5 \left(\frac{1~\text{MHz}}{\wz/2\pi} \right)~$}\\ \hline  
 quality factor efficiency& q& $0.0061$& \multicolumn{2}{|c|}{$q=\frac{\Qdeteff}{\Qdet}$}\\ \hline  
 detector damping rate& $\frac{\gdet}{2\pi}$& $56.2$ kHz& \multicolumn{2}{|c|}{$\gdet=\frac{\wz}{\Qdeteff}$}\\ \hline  
 detection time& $\tdet$& $2.8 \times 10^{-6}$ s& \multicolumn{2}{|c|}{$\tdet=\frac{1}{\gdet}$}\\ \hline  
 circuit effective resistance& $\Reff$& $380\text{k}\Omega$& \multicolumn{2}{|c|}{$ \Reff =q\times 6000 ~\text{M}\Omega \left( \frac{1 ~\text{MHz}}{\wz/2\pi} \right)^2$}\\ \hline  
 image charge parameter& $d_1$& $0.9$& \multicolumn{2}{|c|}{depends on the aspect ratio $\frac{d}{\Rtrap}$ \cite{OpenTrap}}\\ \hline  
 axial damping rate& $\frac{\gamma_z}{2 \pi}$& 1.4 kHz& \multicolumn{2}{|c|}{$\gamma_z=\frac{1}{m_e} \left(\frac{e d_1}{d} \right)^2 \Reff$}\\ \hline  
 circuit temperature& $T_R$& 0.01K& cryogenic technology&N/A\\ \hline  
 axial energy& $E_z$& $0.43$ meV& \multicolumn{2}{|c|}{$E_z=\frac{1}{2}\me \wz^2\zmax^2$}\\ \hline  
 signal to noise ratio& SNR& 5& background&N/A\\ \hline  
SNR time & $t_{\text{SNR}}$& $2.87 \times 10^{-6}$ s& \multicolumn{2}{|c|}{$\tSNR = \frac{\SNR^2 T_R}{2 E_z \gz} $}\\ \hline  
 averaging time& $t_{\text{ave}}$& $2.87 \times 10^{-6}$ s& $ \max\left(\tsignal,\tdet,\tSNR \right)~$&$\tau_c$\\ \hline  
 cyclotron lifetime& $\tau_c$& $2.87\times 10^{-6} \text{ s} \left( \frac{1.2\times 10^6}{n_c} \right) \left( \frac{0.1 \text{ meV}}{\wc}\right)^2 $& \multicolumn{2}{|c|}{$\tau_c= \frac{1}{n_c} \frac{3 m_e} {4\alpha \wc^2} ~$}\\ \hline  
 cyclotron number& $n_c$& $ 1.2 \times 10^6 \left(\frac{0.1~\text{meV}}{\wc} \right)^2$& \multicolumn{2}{|c|}{N/A}\\ \hline  
 cyclotron linewidth& $\Delta \wc$& $5\times 10^{-3}$ meV& \multicolumn{2}{|c|}{$\Delta \wc = \wc \frac{B_2}{B_0} \zmax^2 + \frac{1}{\tau_c}$}\\ \hline  
 cyclotron quality factor& $Q_c$& $20 \sim 200$& \multicolumn{2}{|c|}{$Q_c = \frac{\wc}{\Delta \wc}$}\\ \hline  
 total detection time& $t_{\text{total}}$& 1000 days& N/A&N/A\\ \hline  
 detection time per frequency bin& $\tobs$& 5.5 days& \multicolumn{2}{|c|}{$\tobs=\tTotal \frac{\Delta\wc}{m_{A',\text{max}}-m_{A',\text{min}}}$}\\ \hline 
    \end{tabular}
    }
    \caption{Optimal experimental parameters chosen to maximize cyclotron number $n_c$ and satisfy all other experimental constraints. Independent variables' lower and upper limits are listed, while dependent variables' formulas are given.}
    \label{tab: parameters}
\end{table}

\bibliography{NewRef.bib}

@String { hvd           = {Harvard University} }

@String { ieee          = {IEEE Trans. Instrum. Meas.} }

@String { jetp          = {Sov. Phys. JETP} }

@String { massspec      = {Intl. J. of Mass Spec. and Ion Proc.} }

@String { nature        = {Nature} }

@String { nim           = {Nucl. Instrum. Meth.} }

@String { nima          = {Nucl. Instrum. Methods Phys. Res., Sect. A} }

@String { pr            = {Phys. Rev.} }

@String { pra           = {Phys. Rev. A} }

@String { prd           = {Phys. Rev. D} }

@String { prl           = {Phys. Rev. Lett.} }

@String { rmp           = {Rev. Mod. Phys.} }

@article{Agrawal:2018vin,
    author = "Agrawal, Prateek and Kitajima, Naoya and Reece, Matthew and Sekiguchi, Toyokazu and Takahashi, Fuminobu",
    title = "{Relic Abundance of Dark Photon Dark Matter}",
    eprint = "1810.07188",
    archivePrefix = "arXiv",
    primaryClass = "hep-ph",
    reportNumber = "TU-1074,IPMU18-015,RESCEU-13/18,MIT-CTP/5066, TU-1074, IPMU18-015, RESCEU-13/18, MIT-CTP/5066",
    doi = "10.1016/j.physletb.2019.135136",
    journal = "Phys. Lett. B",
    volume = "801",
    pages = "135136",
    year = "2020"
}

@article{Dror:2018pdh,
    author = "Dror, Jeff A. and Harigaya, Keisuke and Narayan, Vijay",
    title = "{Parametric Resonance Production of Ultralight Vector Dark Matter}",
    eprint = "1810.07195",
    archivePrefix = "arXiv",
    primaryClass = "hep-ph",
    doi = "10.1103/PhysRevD.99.035036",
    journal = "Phys. Rev. D",
    volume = "99",
    number = "3",
    pages = "035036",
    year = "2019"
}

@article{Graham:2015rva,
    author = "Graham, Peter W. and Mardon, Jeremy and Rajendran, Surjeet",
    title = "{Vector Dark Matter from Inflationary Fluctuations}",
    eprint = "1504.02102",
    archivePrefix = "arXiv",
    primaryClass = "hep-ph",
    doi = "10.1103/PhysRevD.93.103520",
    journal = "Phys. Rev. D",
    volume = "93",
    number = "10",
    pages = "103520",
    year = "2016"
}

@article{Ahmed:2020fhc,
    author = "Ahmed, Aqeel and Grzadkowski, Bohdan and Socha, Anna",
    title = "{Gravitational production of vector dark matter}",
    eprint = "2005.01766",
    archivePrefix = "arXiv",
    primaryClass = "hep-ph",
    doi = "10.1007/JHEP08(2020)059",
    journal = "J. High Energy Phys.",
    volume = "08",
    pages = "059",
    year = "2020"
}

@article{Hochberg:2021yud,
    author = "Hochberg, Yonit and Lehmann, Benjamin V. and Charaev, Ilya and Chiles, Jeff and Colangelo, Marco and Nam, Sae Woo and Berggren, Karl K.",
    title = "{New Constraints on Dark Matter from Superconducting Nanowires}",
    journal="arXiv:2110.01586"
}

@article{Chiles:2021gxk,
    author = "Chiles, Jeff and others",
    title = "{New Constraints on Dark Photon Dark Matter with Superconducting Nanowire Detectors in an Optical Haloscope}",
    eprint = "2110.01582",
    archivePrefix = "arXiv",
    primaryClass = "hep-ex",
    doi = "10.1103/PhysRevLett.128.231802",
    journal = "Phys. Rev. Lett.",
    volume = "128",
    number = "23",
    pages = "231802",
    year = "2022"
}

@Article{Review,
  Title                    = {{Geonium Theory: Physics of a Single Electron or Ion in a Penning Trap}},
  Author                   = {L. S. Brown and G. Gabrielse},
  Journal                  = rmp,
  Year                     = {1986},
  Pages                    = {233-311},
  Volume                   = {58},

  Groupsearch              = {1}
}

@PhdThesis{ThesisDUrso,
  Title                    = {{Cooling and Self-Excitation of a One-Electron Oscillator}},
  Author                   = {B. {D'Urso}},
  School                   = hvd,
  Year                     = {2003},
  Note                     = {(thesis advisor: G. Gabrielse)}
}

@Article{DehmeltMeasurementTime,
  Title                    = {''Continuous Stern-Gerlach Effect: Noise and the Measurement Process''},
  Author                   = {H. Dehmelt},
  Journal                  = {Proc. Natl. Acad. Sci.},
  Year                     = {1986},
  Pages                    = {3074-3077},
  Volume                   = {83},

  Groupsearch              = {0}
}

@Article{DehmeltWalls1968,
  Title                    = {"Bolometric" Technique for the rf Spectroscopy of Stored Ions},
  Author                   = {H. Dehmelt and F. Walls},
  Journal                  = prl,
  Year                     = {1968},
  Pages                    = {127},
  Volume                   = {21}
}

@Article{InhibitionLetter,
  Title                    = {``Observation of Inhibited Spontaneous Emission''},
  Author                   = {G. Gabrielse and H. Dehmelt},
  Journal                  = PRL,
  Year                     = {1985},
  Pages                    = {67-70},
  Volume                   = {55},

  Groupsearch              = {1}
}

@Article{Gabrielse85e,
  Title                    = {``Observation of a Relativistic Bistable Hysteresis in the Cyclotron Motion of a Single Electron''},
  Author                   = {G. Gabrielse and H. Dehmelt and W. Kells},
  Journal                  = prl,
  Year                     = {1985},
  Pages                    = {537-540},
  Volume                   = {54},
  Groupsearch              = {1}
}

@Article{OpenTrap,
  Title                    = {{Open-Endcap Penning Traps for High-Precision Experiments}},
  Author                   = {G. Gabrielse and L. Haarsma and S. L. Rolston},
  Journal                  = massspec,
  Year                     = {1989},
  Note                     = {ibid. {93:}, 121 (1989)},
  Pages                    = {319--332},
  Volume                   = {88},

  Groupsearch              = {1}
}

@Article{CylindricalPenningTrap,
  Title                    = {{Cylindrical Penning Traps with Orthogonalized Anharmonicity Compensation}},
  Author                   = {G. Gabrielse and F. Colin MacKintosh},
  Journal                  = {Int. J. Mass Spec. Ion Proc.},
  Year                     = {1984},
  Pages                    = {1-17},
  Volume                   = {57},

  Groupsearch              = {1}
}

@Article{GlauberCoherentState,
  Title                    = {``Coherent and Incoherent States of the Radiation Field''},
  Author                   = {Roy J. Glauber},
  Journal                  = PR,
  Year                     = {1963},
  Pages                    = {2766-2788},
  Volume                   = {131},

  Groupsearch              = {0}
}

@article{HarvardMagneticMoment2011,
  title = {Cavity control of a single-electron quantum cyclotron: Measuring the electron magnetic moment},
  author = {Hanneke, D. and Fogwell Hoogerheide, S. and Gabrielse, G.},
  journal = {Phys. Rev. A},
  volume = {83},
  issue = {5},
  pages = {052122},
  numpages = {26},
  year = {2011},
  month = {May},
  publisher = {American Physical Society},
  doi = {10.1103/PhysRevA.83.052122},
  url = {https://link.aps.org/doi/10.1103/PhysRevA.83.052122}
}

@Article{HarvardMagneticMoment2008,
  Title                    = {{New Measurement of the Electron Magnetic Moment and the Fine Structure Constant}},
  Author                   = {D. Hanneke and S. Fogwell and G. Gabrielse},
  Journal                  = prl,
  Year                     = {2008},
  Pages                    = {120801},
  Volume                   = {100}
}

@Article{QuantumCyclotron,
  Title                    = {{Observing the Quantum Limit of an Electron Cyclotron: {QND} Measurements of Quantum Jumps Between Fock States}},
  Author                   = {S. Peil and G. Gabrielse},
  Journal                  = prl,
  Year                     = {1999},
  Pages                    = {1287-1290},
  Volume                   = {83},

  Groupsearch              = {1}
}

@Article{Penning,
  Author                   = {F. M. Penning},
  Journal                  = {Physica (Utrecht)},
  Year                     = {1936},
  Pages                    = {873},
  Volume                   = {3},

  Groupsearch              = {0}
}

@Article{DehmeltMagneticMoment,
  Title                    = {{New High-Precision Comparison of Electron and Positron $g$ Factors}},
  Author                   = {{R. S. Van Dyck, Jr.} and P. B. Schwinberg and H. G. Dehmelt},
  Journal                  = prl,
  Year                     = {1987},
  Pages                    = { 26-29},
  Volume                   = {59},

  Groupsearch              = {0}
}

@Article{Tseng98,
  Title                    = {``One-Bit Memory Using One Electron: Parametric Oscillations in a Penning Trap''},
  Author                   = {C. H. Tseng and D. Enzer and G. Gabrielse and F. L. Walls},
  Journal                  = PRA,
  Year                     = {1999},
  Pages                    = {2094-2104},
  Volume                   = {59},

  Groupsearch              = {1}
}

@Article{DehmeltMagneticBottle,
  Title                    = {Axial, Magnetron, Cyclotron and Spin-Cyclotron-Beat Frequencies Measured on Single Electron Almost at Rest in Free Space (Geonium)},
  Author                   = {R. {Van Dyck, Jr.} and P. Ekstrom and H. Dehmelt},
  Journal                  = {Nature (London)},
  Year                     = {1976},
  Pages                    = {776},
  Volume                   = {262}
}

@ARTICLE{atomsNewMeasurement2019,
       author = {{Gabrielse}, G. and {Fayer}, S.~E. and {Myers}, T.~G. and {Fan}, X.},
        title = "{Towards an Improved Test of the Standard Model's Most Precise Prediction}",
      journal = {Atoms},
         year = 2019,
        month = apr,
       volume = {7},
       number = {2},
        pages = {45},
          doi = {10.3390/atoms7020045},
       adsurl = {https://ui.adsabs.harvard.edu/abs/2019Atoms...7...45G}
}

@Article{atomsTheoryReview2019,
       author = {{Aoyama}, Tatsumi and {Kinoshita}, Toichiro and {Nio}, Makiko},
        title = "{Theory of the Anomalous Magnetic Moment of the Electron}",
      journal = {Atoms},
         year = 2019,
        month = feb,
       volume = {7},
       number = {1},
        pages = {28},
          doi = {10.3390/atoms7010028},
       adsurl = {https://ui.adsabs.harvard.edu/abs/2019Atoms...7...28A}
}

@Article{FanBackActionPRL2021,
  title = {Circumventing Detector Backaction on a Quantum Cyclotron},
  author = {Fan, X. and Gabrielse, G.},
  journal = {Phys. Rev. Lett.},
  volume = {126},
  issue = {7},
  pages = {070402},
  numpages = {6},
  year = {2021},
  month = {Feb},
  publisher = {American Physical Society},
  doi = {10.1103/PhysRevLett.126.070402},
  url = {https://link.aps.org/doi/10.1103/PhysRevLett.126.070402}
}

@article{DPTheory1986,
title = {Two U(1)'s and $\epsilon$ charge shifts},
journal = {Phys. Lett.},
volume = {166B},
number = {2},
pages = {196-198},
year = {1986},
issn = {0370-2693},
doi = {https://doi.org/10.1016/0370-2693(86)91377-8},
url = {https://www.sciencedirect.com/science/article/pii/0370269386913778},
author = {Bob Holdom}
}

@article{ATheoryOfDarkMatter2009,
  title = {A theory of dark matter},
  author = {Arkani-Hamed, Nima and Finkbeiner, Douglas P. and Slatyer, Tracy R. and Weiner, Neal},
  journal = {Phys. Rev. D},
  volume = {79},
  issue = {1},
  pages = {015014},
  numpages = {16},
  year = {2009},
  month = {Jan},
  publisher = {American Physical Society},
  doi = {10.1103/PhysRevD.79.015014},
  url = {https://link.aps.org/doi/10.1103/PhysRevD.79.015014}
}

@article{Planck2018-1,
	author = {{Planck Collaboration}},
	title = {Planck 2018 results - I. Overview and the cosmological legacy of Planck},
	DOI= "10.1051/0004-6361/201833880",
	url= "https://doi.org/10.1051/0004-6361/201833880",
	journal = {Astron. Astrophys.},
	year = 2020,
	volume = 641,
	pages = "A1",
}

@article{DarkPhotonMisalignmentMechanism2011,
  title = {Dark light, dark matter, and the misalignment mechanism},
  author = {Nelson, Ann E. and Scholtz, Jakub},
  journal = {Phys. Rev. D},
  volume = {84},
  issue = {10},
  pages = {103501},
  numpages = {9},
  year = {2011},
  month = {Nov},
  publisher = {American Physical Society},
  doi = {10.1103/PhysRevD.84.103501},
  url = {https://link.aps.org/doi/10.1103/PhysRevD.84.103501}
}

@article{WISPyDarkMatter2012,
	doi = {10.1088/1475-7516/2012/06/013},
	url = {https://doi.org/10.1088/1475-7516/2012/06/013},
	year = 2012,
	month = {jun},
	publisher = {{IOP} Publishing},
	number = {06},
	pages = {013--013},
	author = {Paola Arias and Davide Cadamuro and Mark Goodsell and Joerg Jaeckel and Javier Redondo and Andreas Ringwald},
	title = {{WISPy} cold dark matter},
	journal = {J. Cosmol. Astropart. Phys.}
}

@article{FuzzyCDM2000,
  title = {Fuzzy Cold Dark Matter: The Wave Properties of Ultralight Particles},
  author = {Hu, Wayne and Barkana, Rennan and Gruzinov, Andrei},
  journal = {Phys. Rev. Lett.},
  volume = {85},
  issue = {6},
  pages = {1158--1161},
  numpages = {0},
  year = {2000},
  month = {Aug},
  publisher = {American Physical Society},
  doi = {10.1103/PhysRevLett.85.1158},
  url = {https://link.aps.org/doi/10.1103/PhysRevLett.85.1158}
}

@article{Okun:1982xi,
    author = "Okun, L. B.",
    title = "{LIMITS OF ELECTRODYNAMICS: PARAPHOTONS?}",
    reportNumber = "ITEP-48-1982",
    journal = "Sov. Phys. JETP",
    volume = "56",
    pages = "502",
    year = "1982"
}

@article{LowEnergyFrontierReview2010,
author = {Jaeckel, Joerg and Ringwald, Andreas},
title = {The Low-Energy Frontier of Particle Physics},
journal = {Annu. Rev. Nucl. Part. Sci.},
volume = {60},
number = {1},
pages = {405-437},
year = {2010},
doi = {10.1146/annurev.nucl.012809.104433},

URL = { 
        https://doi.org/10.1146/annurev.nucl.012809.104433
    
},
eprint = { 
        https://doi.org/10.1146/annurev.nucl.012809.104433
    
}
}

@article{DPLimitsReview2021,
  title = {Dark photon limits: A handbook},
  author = {Caputo, Andrea and Millar, Alexander J. and O'Hare, Ciaran A. J. and Vitagliano, Edoardo},
  journal = {Phys. Rev. D},
  volume = {104},
  issue = {9},
  pages = {095029},
  numpages = {35},
  year = {2021},
  month = {Nov},
  publisher = {American Physical Society},
  doi = {10.1103/PhysRevD.104.095029},
  url = {https://link.aps.org/doi/10.1103/PhysRevD.104.095029}
}

@article{HuntForDarkPhoton2020,
title = {Searching in the dark: the hunt for the dark photon},
journal = {Rev. Phys.},
volume = {5},
pages = {100042},
year = {2020},
issn = {2405-4283},
doi = {https://doi.org/10.1016/j.revip.2020.100042},
url = {https://www.sciencedirect.com/science/article/pii/S2405428320300058},
author = {Alessandra Filippi and M. {De Napoli}}
}

@article{ADMXTechnicalDesignReview2021,
author = {Khatiwada,R.  and Bowring,D.  and Chou,A. S.  and Sonnenschein,A.  and Wester,W.  and Mitchell,D. V.  and Braine,T.  and Bartram,C.  and Cervantes,R.  and Crisosto,N.  and Du,N.  and Rosenberg,L. J.  and Rybka,G.  and Yang,J.  and Will,D.  and Kimes,S.  and Carosi,G.  and Woollett,N.  and Durham,S.  and Duffy,L. D.  and Bradley,R.  and Boutan,C.  and Jones,M.  and LaRoque,B. H.  and Oblath,N. S.  and Taubman,M. S.  and Tedeschi,J.  and Clarke,John  and Dove,A.  and Hashim,A.  and Siddiqi,I.  and Stevenson,N.  and Eddins,A.  and O’Kelley,S. R.  and Nawaz,S.  and Agrawal,A.  and Dixit,A. V.  and Gleason,J. R.  and Jois,S.  and Sikivie,P.  and Sullivan,N. S.  and Tanner,D. B.  and Solomon,J. A.  and Lentz,E.  and Daw,E. J.  and Perry,M. G.  and Buckley,J. H.  and Harrington,P. M.  and Henriksen,E. A.  and Murch,K. W.  and Hilton,G. C. },
title = {Axion Dark Matter Experiment: Detailed design and operations},
journal = {Rev. Sci. Instrum.},
volume = {92},
number = {12},
pages = {124502},
year = {2021},
doi = {10.1063/5.0037857},

URL = { 
        https://doi.org/10.1063/5.0037857
    
},
eprint = { 
        https://doi.org/10.1063/5.0037857
    
}

}

@article{ADMXSideCar2018,
  title = {Piezoelectrically Tuned Multimode Cavity Search for Axion Dark Matter},
  author = {Boutan, C. and Jones, M. and LaRoque, B. H. and Oblath, N. S. and Cervantes, R. and Du, N. and Force, N. and Kimes, S. and Ottens, R. and Rosenberg, L. J. and Rybka, G. and Yang, J. and Carosi, G. and Woollett, N. and Bowring, D. and Chou, A. S. and Khatiwada, R. and Sonnenschein, A. and Wester, W. and Bradley, R. and Daw, E. J. and Agrawal, A. and Dixit, A. V. and Clarke, J. and O'Kelley, S. R. and Crisosto, N. and Gleason, J. R. and Jois, S. and Sikivie, P. and Stern, I. and Sullivan, N. S. and Tanner, D. B. and Harrington, P. M. and Lentz, E.},
  collaboration = {ADMX Collaboration},
  journal = {Phys. Rev. Lett.},
  volume = {121},
  issue = {26},
  pages = {261302},
  numpages = {7},
  year = {2018},
  month = {Dec},
  publisher = {American Physical Society},
  doi = {10.1103/PhysRevLett.121.261302},
  url = {https://link.aps.org/doi/10.1103/PhysRevLett.121.261302}
}

@article{SinglePhotonReview2011,
author = {Eisaman,M. D.  and Fan,J.  and Migdall,A.  and Polyakov,S. V. },
title = {Invited Review Article: Single-photon sources and detectors},
journal = {Rev. Sci. Instrum.},
volume = {82},
number = {7},
pages = {071101},
year = {2011},
doi = {10.1063/1.3610677},

URL = { 
        https://doi.org/10.1063/1.3610677
    
},
eprint = { 
        https://doi.org/10.1063/1.3610677
    
}

}

@article{Xenon1TResultDarkPhoton,
  title = {New limits on dark photons from solar emission and keV scale dark matter},
  author = {An, Haipeng and Pospelov, Maxim and Pradler, Josef and Ritz, Adam},
  journal = {Phys. Rev. D},
  volume = {102},
  issue = {11},
  pages = {115022},
  numpages = {7},
  year = {2020},
  month = {Dec},
  publisher = {American Physical Society},
  doi = {10.1103/PhysRevD.102.115022},
  url = {https://link.aps.org/doi/10.1103/PhysRevD.102.115022}
}

@article{CAPP2021_10ueV,
  title = {First Results from an Axion Haloscope at CAPP around $10.7\text{ }\text{ }\ensuremath{\mu}\mathrm{eV}$},
  author = {Kwon, Ohjoon and Lee, Doyu and Chung, Woohyun and Ahn, Danho and Byun, HeeSu and Caspers, Fritz and Choi, Hyoungsoon and Choi, Jihoon and Chong, Yonuk and Jeong, Hoyong and Jeong, Junu and Kim, Jihn E. and Kim, Jinsu and Kutlu, \ifmmode \mbox{\c{C}}\else \c{C}\fi{}a\ifmmode \breve{g}\else \u{g}\fi{}lar and Lee, Jihnhwan and Lee, MyeongJae and Lee, Soohyung and Matlashov, Andrei and Oh, Seonjeong and Park, Seongtae and Uchaikin, Sergey and Youn, SungWoo and Semertzidis, Yannis K.},
  journal = {Phys. Rev. Lett.},
  volume = {126},
  issue = {19},
  pages = {191802},
  numpages = {6},
  year = {2021},
  month = {May},
  publisher = {American Physical Society},
  doi = {10.1103/PhysRevLett.126.191802},
  url = {https://link.aps.org/doi/10.1103/PhysRevLett.126.191802}
}

@article{CAPP2020_13ueV,
  title = {Search for Invisible Axion Dark Matter with a Multiple-Cell Haloscope},
  author = {Jeong, Junu and Youn, S. W. and Bae, Sungjae and Kim, Jihngeun and Seong, Taehyeon and Kim, Jihn E. and Semertzidis, Yannis K.},
  journal = {Phys. Rev. Lett.},
  volume = {125},
  issue = {22},
  pages = {221302},
  numpages = {6},
  year = {2020},
  month = {Nov},
  publisher = {American Physical Society},
  doi = {10.1103/PhysRevLett.125.221302},
  url = {https://link.aps.org/doi/10.1103/PhysRevLett.125.221302}
}

@article{CAPP2020_7ueV,
  title = {Axion Dark Matter Search around $6.7\text{ }\text{ }\ensuremath{\mu}\mathrm{eV}$},
  author = {Lee, S. and Ahn, S. and Choi, J. and Ko, B. R. and Semertzidis, Y. K.},
  journal = {Phys. Rev. Lett.},
  volume = {124},
  issue = {10},
  pages = {101802},
  numpages = {5},
  year = {2020},
  month = {Mar},
  publisher = {American Physical Society},
  doi = {10.1103/PhysRevLett.124.101802},
  url = {https://link.aps.org/doi/10.1103/PhysRevLett.124.101802}
}

@ARTICLE{DOSUE2022_100ueV,
       author = {{Kotaka}, Shumpei and {Adachi}, Shunsuke and {Fujinaka}, Ryo and {Honda}, Shunsuke and {Nakata}, Hironobu and {Seino}, Yudai and {Sueno}, Yoshinori and {Sumida}, Toshi and {Suzuki}, Junya and {Tajima}, Osamu and {Takeichi}, Soichiro},
        title = "Search for Dark Photon Dark Matter in the Mass Range 74-100 $\mu\text{eV}$ with a Cryogenic Millimeter-Wave Receiver",
        pages = {arXiv:2205.03679},
    journal=""
}

@article{ReviewScalarFieldDarkMatter,
	doi = {10.1088/1742-6596/378/1/012012},
	url = {https://doi.org/10.1088/1742-6596/378/1/012012},
	year = 2012,
	month = {aug},
	publisher = {{IOP} Publishing},
	volume = {378},
	pages = {012012},
	author = {Juan Maga{\~{n}}a and Tonatiuh Matos},
	title = {A brief Review of the Scalar Field Dark Matter model},
	journal = {J. Phys. Conf. Ser.}
}

@article{UltraLightScalarCosmologicalDarkMatter,
  title = {Ultralight scalars as cosmological dark matter},
  author = {Hui, Lam and Ostriker, Jeremiah P. and Tremaine, Scott and Witten, Edward},
  journal = {Phys. Rev. D},
  volume = {95},
  issue = {4},
  pages = {043541},
  numpages = {32},
  year = {2017},
  month = {Feb},
  publisher = {American Physical Society},
  doi = {10.1103/PhysRevD.95.043541},
  url = {https://link.aps.org/doi/10.1103/PhysRevD.95.043541}
}

@article{GravitationalProductionOfDP2021,
	Author = {Kolb, Edward W. and Long, Andrew J.},
	Da = {2021/03/30},
	Doi = {10.1007/JHEP03(2021)283},
	Id = {Kolb2021},
	Isbn = {1029-8479},
	Journal = {J. High Energy Phys.},
	Number = {3},
	Pages = {283},
	Title = {Completely dark photons from gravitational particle production during the inflationary era},
	Ty = {JOUR},
	Url = {https://doi.org/10.1007/JHEP03(2021)283},
	Volume = {03},
	Year = {2021}}

@ARTICLE{DMDiscovery1933,
       author = {{Zwicky}, F.},
        title = "{Die Rotverschiebung von extragalaktischen Nebeln}",
      journal = {Helv. Phys. Acta},
         year = 1933,
        month = jan,
       volume = {6},
        pages = {110-127},
       adsurl = {https://ui.adsabs.harvard.edu/abs/1933AcHPh...6..110Z}
}

@ARTICLE{DMRotationCurve1980,
       author = {{Rubin}, V.~C. and {Ford Jr.}, W.~K. and {Thonnard}, N.},
        title = "{Rotational properties of 21 SC galaxies with a large range of luminosities and radii, from NGC 4605 (R=4kpc) to UGC 2885 (R=122kpc).}",
      journal = {Astrophys. J.},
         year = 1980,
        month = jun,
       volume = {238},
        pages = {471-487},
          doi = {10.1086/158003},
       adsurl = {https://ui.adsabs.harvard.edu/abs/1980ApJ...238..471R}
}

@article{DMGravitationalLensing2006,
	doi = {10.1086/508162},
	url = {https://doi.org/10.1086/508162},
	year = 2006,
	month = {aug},
	publisher = {American Astronomical Society},
	volume = {648},
	number = {2},
	pages = {L109--L113},
	author = {Douglas Clowe and Maru{\v{s}}a Brada{\v{c}} and Anthony H. Gonzalez and Maxim Markevitch and Scott W. Randall and Christine Jones and Dennis Zaritsky},
	title = {A Direct Empirical Proof of the Existence of Dark Matter},
	journal = {Astrophys. J.}
}

@article{DMBulletCollision2000,
  title = {Observational Evidence for Self-Interacting Cold Dark Matter},
  author = {Spergel, David N. and Steinhardt, Paul J.},
  journal = {Phys. Rev. Lett.},
  volume = {84},
  issue = {17},
  pages = {3760--3763},
  numpages = {0},
  year = {2000},
  month = {Apr},
  publisher = {American Physical Society},
  doi = {10.1103/PhysRevLett.84.3760},
  url = {https://link.aps.org/doi/10.1103/PhysRevLett.84.3760}
}

@article{DM_WMAPGalaxyCenter2007,
  title = {Possible evidence for dark matter annihilations from the excess microwave emission around the center of the Galaxy seen by the Wilkinson Microwave Anisotropy Probe},
  author = {Hooper, Dan and Finkbeiner, Douglas P. and Dobler, Gregory},
  journal = {Phys. Rev. D},
  volume = {76},
  issue = {8},
  pages = {083012},
  numpages = {6},
  year = {2007},
  month = {Oct},
  publisher = {American Physical Society},
  doi = {10.1103/PhysRevD.76.083012},
  url = {https://link.aps.org/doi/10.1103/PhysRevD.76.083012}
}

@article{DishAntennaProposal-2,
	doi = {10.1088/1475-7516/2013/04/016},
	url = {https://doi.org/10.1088/1475-7516/2013/04/016},
	publisher = {{IOP} Publishing},
	pages = {04 (2013) 016},
	author = {Dieter Horns and Joerg Jaeckel and Axel Lindner and Andrei Lobanov and Javier Redondo and Andreas Ringwald},
	title = {Searching for {WISPy} cold dark matter with a dish antenna},
	journal = {J. Cosmol. Astropart. Phys.}
}

@article{Co_2019,
	doi = {10.1103/physrevd.99.075002},
	year = 2019,
	month = {apr},
  
	publisher = {American Physical Society ({APS})},
  
	volume = {99},
  
	number = {7},
    pages = {075002},
	author = {Raymond T. Co and Aaron Pierce and Zhengkang Zhang and Yue Zhao},
  
	title = {Dark photon dark matter produced by axion oscillations},
  
	journal = prd
}

@article{Co_2021,
	doi = {10.1007/jhep12(2021)099},
  
	url = {https://doi.org/10.1007%2Fjhep12%282021%29099},
  
	year = 2021,
	month = {dec},
  
	publisher = {Springer Science and Business Media {LLC}
},
  
	volume = {12},
    pages = 099,
	author = {Raymond T. Co and Keisuke Harigaya and Aaron Pierce},
  
	title = {Gravitational waves and dark photon dark matter from axion rotations},
  
	journal = {J. High Energy Phys.}
}

@article{LeoParticlePhysicsTextBook,
title = {Techniques for nuclear and particle physics experiments},
author = {Leo, W R},
abstractNote = {This book is based on a laboratory course in nuclear physics and treats the experimental techniques and instrumentation most often used in nuclear and particle physics experiments as well as in various other experimental sciences. It provides results, formulae, and informative details on interaction of radiation in matter; radioprotection and radioactive sources; statistics for the interpretation and analysis of data; principles and operation of the main types of detectors (ionization, scintillation and semiconductor detectors); nuclear electronics instrumentation (NIM, CAMAC); and, various systems and techniques for experiments.},
doi = {},
url = {https://www.osti.gov/biblio/5711102}, journal = {},
place = {United States},
year = {1987},
month = {1}
}

@article{ItanoWineland1982,
  title = {Laser cooling of ions stored in harmonic and Penning traps},
  author = {Itano, Wayne M. and Wineland, D. J.},
  journal = {Phys. Rev. A},
  volume = {25},
  issue = {1},
  pages = {35--54},
  numpages = {0},
  year = {1982},
  month = {Jan},
  publisher = {American Physical Society},
  doi = {10.1103/PhysRevA.25.35},
  url = {https://link.aps.org/doi/10.1103/PhysRevA.25.35}
}

@article{DMRadio,
	doi = {10.1103/physrevd.92.075012},
  
	url = {https://doi.org/10.1103%2Fphysrevd.92.075012},
  
	year = 2015,
	month = {oct},
  
	publisher = {American Physical Society ({APS})},
  
	volume = {92},
  
	number = {7},
  
	author = {Saptarshi Chaudhuri and Peter W. Graham and Kent Irwin and Jeremy Mardon and Surjeet Rajendran and Yue Zhao},
  
	title = {Radio for hidden-photon dark matter detection},
  
	journal = {Physical Review D}
}

@book{QMCT,
      author        = "Cohen-Tannoudji, Claude and Diu, Bernard and Laloë,
                       Franck",
      title         = "{Quantum mechanics; 1st ed.}",
      publisher     = "Wiley",
      address       = "New York, NY",
      year          = "1977",
      url           = "https://cds.cern.ch/record/101367",
      note          = "Trans. of : Mécanique quantique. Paris : Hermann, 1973",
}

@book{DavidHill,
      author        = "Hill, David A.",
      title         = "{Electromagnetic fields in cavities: Deterministic and statistical theories.}",
      publisher     = "Wiley",
      year          = "2009",
}

@article{Fan_2022,
   title={One-Electron Quantum Cyclotron as a Milli-eV Dark-Photon Detector},
   volume={129},
   ISSN={1079-7114},
   url={http://dx.doi.org/10.1103/PhysRevLett.129.261801},
   DOI={10.1103/physrevlett.129.261801},
   number={26},
   journal={Physical Review Letters},
   publisher={American Physical Society (APS)},
   author={Fan, Xing and Gabrielse, Gerald and Graham, Peter W. and Harnik, Roni and Myers, Thomas G. and Ramani, Harikrishnan and Sukra, Benedict A.~D. and Wong, Samuel S.~Y. and Xiao, Yawen},
   year={2022},
   month=dec }

@article{BREAD,
   title={Broadband Solenoidal Haloscope for Terahertz Axion Detection},
   volume={128},
   ISSN={1079-7114},
   url={http://dx.doi.org/10.1103/PhysRevLett.128.131801},
   DOI={10.1103/physrevlett.128.131801},
   number={13},
   journal={Physical Review Letters},
   publisher={American Physical Society (APS)},
   author={Liu, Jesse and Dona, Kristin and Hoshino, Gabe and Knirck, Stefan and Kurinsky, Noah and Malaker, Matthew and Miller, David W. and Sonnenschein, Andrew and Awida, Mohamed H. and Barry, Peter S. and Berggren, Karl K. and Bowring, Daniel and Carosi, Gianpaolo and Chang, Clarence and Chou, Aaron and Khatiwada, Rakshya and Lewis, Samantha and Li, Juliang and Nam, Sae Woo and Noroozian, Omid and Zhou, Tony X.},
   year={2022},
   month=mar }

@article{MillichargeIonTrap,
    author = "Budker, Dmitry and Graham, Peter W. and Ramani, Harikrishnan and Schmidt-Kaler, Ferdinand and Smorra, Christian and Ulmer, Stefan",
    title = "{Millicharged Dark Matter Detection with Ion Traps}",
    eprint = "2108.05283",
    archivePrefix = "arXiv",
    primaryClass = "hep-ph",
    doi = "10.1103/PRXQuantum.3.010330",
    journal = "PRX Quantum",
    volume = "3",
    number = "1",
    pages = "010330",
    year = "2022"
}

@inproceedings{FocusingLimit,
    author = "Jaeckel, Joerg and Knirck, Stefan",
    title = "{Dish Antenna Searches for WISPy Dark Matter: Directional Resolution Small Mass Limitations}",
    booktitle = "{12th Patras Workshop on Axions, WIMPs and WISPs}",
    eprint = "1702.04381",
    archivePrefix = "arXiv",
    primaryClass = "hep-ph",
    doi = "10.3204/DESY-PROC-2009-03/Knirck_Stefan",
    pages = "78--81",
    year = "2017"
}

@article{BetheDiffractionSmallHole,
  title = {Theory of Diffraction by Small Holes},
  author = {Bethe, H. A.},
  journal = {Phys. Rev.},
  volume = {66},
  issue = {7-8},
  pages = {163--182},
  numpages = {0},
  year = {1944},
  month = {Oct},
  publisher = {American Physical Society},
  doi = {10.1103/PhysRev.66.163},
  url = {https://link.aps.org/doi/10.1103/PhysRev.66.163}
}

@PhdThesis{ThesisXingFan,
  Title                    = {An Improved Measurement of the Electron Magnetic Moment},
  Author                   = {Xing Fan},
  School                   = hvd,
  Year                     = {2022}
}

@article{PostInflationAxionMass,
   title={Dark matter from axion strings with adaptive mesh refinement},
   volume={13},
   ISSN={2041-1723},
   url={http://dx.doi.org/10.1038/s41467-022-28669-y},
   DOI={10.1038/s41467-022-28669-y},
   number={1},
   journal={Nature Communications},
   publisher={Springer Science and Business Media LLC},
   author={Buschmann, Malte and Foster, Joshua W. and Hook, Anson and Peterson, Adam and Willcox, Don E. and Zhang, Weiqun and Safdi, Benjamin R.},
   year={2022},
   month=feb }

@article{MultilayerOpticalHaloscopes,
  title = {Axion and hidden photon dark matter detection with multilayer optical haloscopes},
  author = {Baryakhtar, Masha and Huang, Junwu and Lasenby, Robert},
  journal = {Phys. Rev. D},
  volume = {98},
  issue = {3},
  pages = {035006},
  numpages = {21},
  year = {2018},
  month = {Aug},
  publisher = {American Physical Society},
  doi = {10.1103/PhysRevD.98.035006},
  url = {https://link.aps.org/doi/10.1103/PhysRevD.98.035006}
}

@ARTICLE{SubwavelengthAperture,
  author={Polycarpou, Anastasis C. and Christou, Marios A.},
  journal={IEEE Transactions on Antennas and Propagation}, 
  title={Closed-Form Expressions for the On-Axis Scattered Fields by a Subwavelength Circular Aperture in an Infinite Conducting Plane}, 
  year={2017},
  volume={65},
  number={2},
  pages={978-982},
  doi={10.1109/TAP.2016.2634278}}

@article{StrongCP1,
    author = "Peccei, R. D. and Quinn, Helen R.",
    title = "{CP Conservation in the Presence of Instantons}",
    reportNumber = "ITP-568-STANFORD",
    doi = "10.1103/PhysRevLett.38.1440",
    journal = "Phys. Rev. Lett.",
    volume = "38",
    pages = "1440--1443",
    year = "1977"
}

@article{StrongCP2,
    author = "Peccei, R. D. and Quinn, Helen R.",
    title = "{Constraints Imposed by CP Conservation in the Presence of Instantons}",
    reportNumber = "ITP-572-STANFORD",
    doi = "10.1103/PhysRevD.16.1791",
    journal = "Phys. Rev. D",
    volume = "16",
    pages = "1791--1797",
    year = "1977"
}

@article{DMRadioGUT,
    author = "Brouwer, L. and others",
    collaboration = "DMRadio",
    title = "{Proposal for a definitive search for GUT-scale QCD axions}",
    eprint = "2203.11246",
    archivePrefix = "arXiv",
    primaryClass = "hep-ex",
    doi = "10.1103/PhysRevD.106.112003",
    journal = "Phys. Rev. D",
    volume = "106",
    number = "11",
    pages = "112003",
    year = "2022"
}

@article{AxionSearchExp,
    author = "Graham, Peter W. and Irastorza, Igor G. and Lamoreaux, Steven K. and Lindner, Axel and van Bibber, Karl A.",
    title = "{Experimental Searches for the Axion and Axion-Like Particles}",
    eprint = "1602.00039",
    archivePrefix = "arXiv",
    primaryClass = "hep-ex",
    doi = "10.1146/annurev-nucl-102014-022120",
    journal = "Ann. Rev. Nucl. Part. Sci.",
    volume = "65",
    pages = "485--514",
    year = "2015"
}

@article{AxionThyReview,
    author = "Kim, Jihn E. and Carosi, Gianpaolo",
    title = "{Axions and the Strong CP Problem}",
    eprint = "0807.3125",
    archivePrefix = "arXiv",
    primaryClass = "hep-ph",
    doi = "10.1103/RevModPhys.82.557",
    journal = "Rev. Mod. Phys.",
    volume = "82",
    pages = "557--602",
    year = "2010",
    note = "[Erratum: Rev.Mod.Phys. 91, 049902 (2019)]"
}

@article{WeinbergAxion,
    author = "Weinberg, Steven",
    title = "{A New Light Boson?}",
    reportNumber = "HUTP-77/A074",
    doi = "10.1103/PhysRevLett.40.223",
    journal = "Phys. Rev. Lett.",
    volume = "40",
    pages = "223--226",
    year = "1978"
}

@article{HarmlessAxion,
    author = "Dine, Michael and Fischler, Willy and Srednicki, Mark",
    title = "{A Simple Solution to the Strong CP Problem with a Harmless Axion}",
    reportNumber = "Print-81-0320 (IAS,PRINCETON)",
    doi = "10.1016/0370-2693(81)90590-6",
    journal = "Phys. Lett. B",
    volume = "104",
    pages = "199--202",
    year = "1981"
}

@article{DFSZ2,
    author = "Zhitnitsky, A. R.",
    title = "{On Possible Suppression of the Axion Hadron Interactions. (In Russian)}",
    journal = "Sov. J. Nucl. Phys.",
    volume = "31",
    pages = "260",
    year = "1980"
}

@article{KSVZ1,
    author = "Kim, Jihn E.",
    title = "{Weak Interaction Singlet and Strong CP Invariance}",
    reportNumber = "UPR-0120T",
    doi = "10.1103/PhysRevLett.43.103",
    journal = "Phys. Rev. Lett.",
    volume = "43",
    pages = "103",
    year = "1979"
}

@article{KSVZ2,
    author = "Shifman, Mikhail A. and Vainshtein, A. I. and Zakharov, Valentin I.",
    title = "{Can Confinement Ensure Natural CP Invariance of Strong Interactions?}",
    reportNumber = "ITEP-64-1979",
    doi = "10.1016/0550-3213(80)90209-6",
    journal = "Nucl. Phys. B",
    volume = "166",
    pages = "493--506",
    year = "1980"
}

@article{ALPObs,
    author = "Graham, Peter W. and Rajendran, Surjeet",
    title = "{New Observables for Direct Detection of Axion Dark Matter}",
    eprint = "1306.6088",
    archivePrefix = "arXiv",
    primaryClass = "hep-ph",
    doi = "10.1103/PhysRevD.88.035023",
    journal = "Phys. Rev. D",
    volume = "88",
    pages = "035023",
    year = "2013"
}

@article{InflationaryAxion3,
title = {The not-so-harmless axion},
journal = {Physics Letters B},
volume = {120},
number = {1},
pages = {137-141},
year = {1983},
issn = {0370-2693},
doi = {https://doi.org/10.1016/0370-2693(83)90639-1},
url = {https://www.sciencedirect.com/science/article/pii/0370269383906391},
author = {Michael Dine and Willy Fischler}
}

@article{InflationaryAxion2,
title = {A cosmological bound on the invisible axion},
journal = {Physics Letters B},
volume = {120},
number = {1},
pages = {133-136},
year = {1983},
issn = {0370-2693},
doi = {https://doi.org/10.1016/0370-2693(83)90638-X},
url = {https://www.sciencedirect.com/science/article/pii/037026938390638X},
author = {L.F. Abbott and P. Sikivie}
}

@article{InflationaryAxion1,
title = {Cosmology of the invisible axion},
journal = {Physics Letters B},
volume = {120},
number = {1},
pages = {127-132},
year = {1983},
issn = {0370-2693},
doi = {https://doi.org/10.1016/0370-2693(83)90637-8},
url = {https://www.sciencedirect.com/science/article/pii/0370269383906378},
author = {John Preskill and Mark B. Wise and Frank Wilczek}
}

@article{ElectronMagneticMomentPRL2022,
  title = {Measurement of the Electron Magnetic Moment},
  author = {Fan, X. and Myers, T. G. and Sukra, B. A. D. and Gabrielse, G.},
  journal = {Phys. Rev. Lett.},
  volume = {130},
  issue = {7},
  pages = {071801},
  numpages = {6},
  year = {2023},
  month = {Feb},
  publisher = {American Physical Society},
  doi = {10.1103/PhysRevLett.130.071801},
  url = {https://link.aps.org/doi/10.1103/PhysRevLett.130.071801}
}

@article{Graham:2013gfa,
    author = "Graham, Peter W. and Rajendran, Surjeet",
    title = "{New Observables for Direct Detection of Axion Dark Matter}",
    eprint = "1306.6088",
    archivePrefix = "arXiv",
    primaryClass = "hep-ph",
    reportNumber = "SLAC-PUB-15483",
    doi = "10.1103/PhysRevD.88.035023",
    journal = "Phys. Rev. D",
    volume = "88",
    pages = "035023",
    year = "2013"
}

@article{Hewett:2012ns,
    author = "Hewett, JoAnne L. and Weerts, Harry and Brock, Raymond and Butler, Joel N. and Casey, Brendan C. K. and Collar, Juan and de Gouvêa, André and Essig, Rouven and etc.",
    title = "{Fundamental Physics at the Intensity Frontier}",
    eprint = "1205.2671",
    archivePrefix = "arXiv",
    primaryClass = "hep-ex",
    reportNumber = "FERMILAB-CONF-12-879-PPD-T",
    year = "2012",
    journal=""
}

@article{Ringwald:2012cu,
    author = "Ringwald, Andreas",
    title = "{Searching for axions and ALPs from string theory}",
    eprint = "1209.2299",
    archivePrefix = "arXiv",
    primaryClass = "hep-ph",
    year = "2012",
    journal=""
}

@article{Ringwald:2012hr,
    author = "Ringwald, Andreas",
    title = "{Exploring the Role of Axions and Other WISPs in the Dark Universe}",
    eprint = "1210.5081",
    archivePrefix = "arXiv",
    primaryClass = "hep-ph",
    year = "2012",
    journal=""
}

@article{Baker:2011na,
    author = "Baker, O. K. and Betz, M. and Caspers, F. and Jaeckel, J. and Lindner, A. and Ringwald, A. and Semertzidis, Y. K. and Sikivie, P. and etc.",
    title = "{Prospects for Searching Axion-like Particle Dark Matter with Dipole, Toroidal and Wiggler Magnets}",
    eprint = "1110.2180",
    archivePrefix = "arXiv",
    primaryClass = "physics.ins-det",
    doi = "10.1103/PhysRevD.85.035018",
    journal = "Phys. Rev. D",
    volume = "85",
    pages = "035018",
    year = "2012"
}

@article{Arias:2010bh,
    author = "Arias, Paola and Jaeckel, Joerg and Redondo, Javier and Ringwald, Andreas",
    title = "{Optimizing Light-Shining-through-a-Wall Experiments for Axion and other WISP Searches}",
    eprint = "1009.4875",
    archivePrefix = "arXiv",
    primaryClass = "hep-ph",
    doi = "10.1103/PhysRevD.82.115018",
    journal = "Phys. Rev. D",
    volume = "82",
    pages = "115018",
    year = "2010"
}

@article{Jaeckel:2010ni,
    author = "Jaeckel, Joerg and Ringwald, Andreas",
    title = "{The Low-Energy Frontier of Particle Physics}",
    eprint = "1002.0329",
    archivePrefix = "arXiv",
    primaryClass = "hep-ph",
    doi = "10.1146/annurev.nucl.012809.104433",
    journal = "Ann. Rev. Nucl. Part. Sci.",
    volume = "60",
    pages = "405-437",
    year = "2010"
}

@article{Ehret:2010mh,
    author = "Ehret, Klaus and Frede, Marcus and Ghazaryan, Samvel and Hildebrandt, Martin and Knabbe, Ernst-Axel and Kracht, Dietmar and Lindner, Axel and List, Jörn and etc.",
    title = "{New ALPS Results on Hidden-Sector Lightweights}",
    eprint = "1004.1313",
    archivePrefix = "arXiv",
    primaryClass = "hep-ex",
    doi = "10.1016/j.physletb.2010.04.066",
    journal = "Phys. Lett. B",
    volume = "689",
    pages = "149-155",
    year = "2010"
}

@inproceedings{OSQAR:2011xex,
    author = "Schott, Matthias and others",
    collaboration = "OSQAR",
    title = "{First Results of the Full-Scale OSQAR Photon Regeneration Experiment}",
    booktitle = "{International Conference on the Structure and Interactions of the Photon and 19th International Workshop on Photon-Photon Collisions}",
    eprint = "1110.0774",
    archivePrefix = "arXiv",
    primaryClass = "hep-ex",
    month = "10",
    year = "2011"
}

@article{Battesti:2010dm,
    author = "Battesti, Rémy and Fouché, Mathieu and Detlefs, Carsten and Roth, Thierry and Berceau, Paul and Duc, François and Frings, Pierre and Rikken, Geert L. J. A. and etc.",
    title = "{A Photon Regeneration Experiment for Axionlike Particle Search using X-rays}",
    eprint = "1008.2672",
    archivePrefix = "arXiv",
    primaryClass = "hep-ex",
    doi = "10.1103/PhysRevLett.105.250405",
    journal = "Phys. Rev. Lett.",
    volume = "105",
    pages = "250405",
    year = "2010"
}

@article{Conlon:2006tq,
    author = "Conlon, Joseph P.",
    title = "{The QCD axion and moduli stabilisation}",
    eprint = "hep-th/0602233",
    archivePrefix = "arXiv",
    reportNumber = "OUTP-06-07P",
    doi = "10.1088/1126-6708/2006/05/078",
    journal = "JHEP",
    volume = "0605",
    pages = "078",
    year = "2006"
}

@article{Arvanitaki:2009fg,
    author = "Arvanitaki, Asimina and Dimopoulos, Savas and Dubovsky, Sergei and Kaloper, Nemanja and March-Russell, John",
    title = "{String Axiverse}",
    eprint = "0905.4720",
    archivePrefix = "arXiv",
    primaryClass = "hep-th",
    doi = "10.1103/PhysRevD.81.123530",
    journal = "Phys. Rev. D",
    volume = "81",
    pages = "123530",
    year = "2010"
}

@article{Acharya:2010zx,
    author = "Acharya, Bobby Samir and Bobkov, Konstantin and Kumar, Piyush",
    title = "{An M Theory Solution to the Strong CP Problem and Constraints on the Axiverse}",
    eprint = "1004.5138",
    archivePrefix = "arXiv",
    primaryClass = "hep-th",
    doi = "10.1007/JHEP11(2010)105",
    journal = "JHEP",
    volume = "1011",
    pages = "105",
    year = "2010"
}

@article{Pospelov:2012mt,
    author = "Pospelov, Maxim and Pustelny, Szymon and Ledbetter, Matthew P. and Jackson Kimball, Derek F. and Gawlik, Wojciech and Budker, Dmitry",
    title = "{Detecting Domain Walls of Axionlike Models Using Terrestrial Experiments}",
    eprint = "1205.6260",
    archivePrefix = "arXiv",
    primaryClass = "hep-ph",
    doi = "10.1103/PhysRevLett.110.021803",
    journal = "Phys. Rev. Lett.",
    volume = "110",
    pages = "021803",
    year = "2013"
}

@article{globularCluster,
    author = "Dolan, Matthew J. and Hiskens, Frederick J. and Volkas, Raymond R.",
    title = "{Advancing globular cluster constraints on the axion-photon coupling}",
    eprint = "2207.03102",
    archivePrefix = "arXiv",
    primaryClass = "hep-ph",
    doi = "10.1088/1475-7516/2022/10/096",
    journal = "JCAP",
    volume = "10",
    pages = "096",
    year = "2022"
}

@article{ORGAN,
    author = "Quiskamp, Aaron and McAllister, Ben T. and Altin, Paul and Ivanov, Eugene N. and Goryachev, Maxim and Tobar, Michael E.",
    title = "{Exclusion of Axionlike-Particle Cogenesis Dark Matter in a Mass Window above 100\,\,\ensuremath{\mu}eV}",
    eprint = "2310.00904",
    archivePrefix = "arXiv",
    primaryClass = "hep-ex",
    doi = "10.1103/PhysRevLett.132.031601",
    journal = "Phys. Rev. Lett.",
    volume = "132",
    number = "3",
    pages = "031601",
    year = "2024"
}

@article{PriceTinyKineticMixing,
    author = {Gherghetta, Tony and Kersten, J\"orn and Olive, Keith and Pospelov, Maxim},
    title = "{Evaluating the price of tiny kinetic mixing}",
    eprint = "1909.00696",
    archivePrefix = "arXiv",
    primaryClass = "hep-ph",
    reportNumber = "FTPI-MINN-19/23, UMN-TH-3832/19",
    doi = "10.1103/PhysRevD.100.095001",
    journal = "Phys. Rev. D",
    volume = "100",
    number = "9",
    pages = "095001",
    year = "2019"
}

@article{195pagesDMRadioReview,
    author = "Chaudhuri, Saptarshi and Irwin, Kent and Graham, Peter W. and Mardon, Jeremy",
    title = "{Optimal Impedance Matching and Quantum Limits of Electromagnetic Axion and Hidden-Photon Dark Matter Searches}",
    eprint = "1803.01627",
    archivePrefix = "arXiv",
    primaryClass = "hep-ph",
    month = "3",
    year = "2018",
    journal=""
}

@article{TransmonQubit,
  title = {Detecting Hidden Photon Dark Matter Using the Direct Excitation of Transmon Qubits},
  author = {Chen, Shion and Fukuda, Hajime and Inada, Toshiaki and Moroi, Takeo and Nitta, Tatsumi and Sichanugrist, Thanaporn},
  journal = {Phys. Rev. Lett.},
  volume = {131},
  issue = {21},
  pages = {211001},
  numpages = {8},
  year = {2023},
  month = {Nov},
  publisher = {American Physical Society},
  doi = {10.1103/PhysRevLett.131.211001},
  url = {https://link.aps.org/doi/10.1103/PhysRevLett.131.211001}
}

@article{transmonAxion,
    author = "Chen, Shion and Fukuda, Hajime and Inada, Toshiaki and Moroi, Takeo and Nitta, Tatsumi and Sichanugrist, Thanaporn",
    title = "{Search for QCD axion dark matter with transmon qubits and quantum circuit}",
    eprint = "2407.19755",
    archivePrefix = "arXiv",
    primaryClass = "hep-ph",
    month = "7",
    year = "2024",
    journal=""
}

@misc{drivenOsc,
  author       = {Matthew Schwartz},
  title        = {Lecture 2: Driven Oscillators},
  year         = {2016},
  note         = {Lecture notes, The Physics of Waves},
  howpublished = {\url{https://scholar.harvard.edu/files/schwartz/files/lecture2-driven-oscillators.pdf}},
}

@article{MADMAX2024,
    author = "Egge, J. and others",
    collaboration = "MADMAX",
    title = "{First search for dark photon dark matter with a MADMAX prototype}",
    eprint = "2408.02368",
    archivePrefix = "arXiv",
    primaryClass = "hep-ex",
    reportNumber = "FERMILAB-PUB-24-0466-PPD",
    month = "8",
    year = "2024",
    journal=""
}

@article{PhysRevD.88.115002,
  title = {Resonant to broadband searches for cold dark matter consisting of weakly interacting slim particles},
  author = {Jaeckel, Joerg and Redondo, Javier},
  journal = {Phys. Rev. D},
  volume = {88},
  issue = {11},
  pages = {115002},
  numpages = {11},
  year = {2013},
  month = {Dec},
  publisher = {American Physical Society},
  doi = {10.1103/PhysRevD.88.115002},
  url = {https://link.aps.org/doi/10.1103/PhysRevD.88.115002}
}

@article{PhysRevLett.118.091801,
  title = {Dielectric Haloscopes: A New Way to Detect Axion Dark Matter},
  author = {Caldwell, Allen and Dvali, Gia and Majorovits, B\'ela and Millar, Alexander and Raffelt, Georg and Redondo, Javier and Reimann, Olaf and Simon, Frank and Steffen, Frank},
  collaboration = {MADMAX Working Group},
  journal = {Phys. Rev. Lett.},
  volume = {118},
  issue = {9},
  pages = {091801},
  numpages = {6},
  year = {2017},
  month = {Mar},
  publisher = {American Physical Society},
  doi = {10.1103/PhysRevLett.118.091801},
  url = {https://link.aps.org/doi/10.1103/PhysRevLett.118.091801}
}

@article{Millar_2017,
doi = {10.1088/1475-7516/2017/01/061},
url = {https://dx.doi.org/10.1088/1475-7516/2017/01/061},
year = {2017},
month = {jan},
publisher = {},
volume = {2017},
number = {01},
pages = {061},
author = {Alexander J. Millar and Georg G. Raffelt and Javier Redondo and Frank D. Steffen},
title = {Dielectric haloscopes to search for axion dark matter: theoretical foundations},
journal = {Journal of Cosmology and Astroparticle Physics}
}

@article{garcia2024first,
  title={First search for axion dark matter with a Madmax prototype},
  author={Garcia, B and Bergermann, D and Caldwell, A and Dabhi, V and Diaconu, C and Diehl, J and Dvali, G and Egge, J and Garutti, E and Heyminck, S and others},
  journal={arXiv preprint arXiv:2409.11777},
  year={2024}
}

@article{PhysRevLett.132.131004,
  title = {First Results from a Broadband Search for Dark Photon Dark Matter in the 44 to $52\text{ }\text{ }\mathrm{\ensuremath{\mu}}\mathrm{eV}$ Range with a Coaxial Dish Antenna},
  author = {Knirck, Stefan and Hoshino, Gabe and Awida, Mohamed H. and Cancelo, Gustavo I. and Di Federico, Martin and Knepper, Benjamin and Lapuente, Alex and Littmann, Mira and Miller, David W. and Mitchell, Donald V. and Rodriguez, Derrick and Ruschman, Mark K. and Sawtell, Matthew A. and Stefanazzi, Leandro and Sonnenschein, Andrew and Teafoe, Gary W. and Bowring, Daniel and Carosi, G. and Chou, Aaron and Chang, Clarence L. and Dona, Kristin and Khatiwada, Rakshya and Kurinsky, Noah A. and Liu, Jesse and Pena, Cristi\'an and Salemi, Chiara P. and Wang, Christina W. and Yu, Jialin},
  collaboration = {BREAD Collaboration},
  journal = {Phys. Rev. Lett.},
  volume = {132},
  issue = {13},
  pages = {131004},
  numpages = {6},
  year = {2024},
  month = {Mar},
  publisher = {American Physical Society},
  doi = {10.1103/PhysRevLett.132.131004},
  url = {https://link.aps.org/doi/10.1103/PhysRevLett.132.131004}
}

@article{PhysRevD.105.052010,
  title = {Search for dark photons using a multilayer dielectric haloscope equipped with a single-photon avalanche diode},
  author = {Manenti, Laura and Mishra, Umang and Bruno, Gianmarco and Roberts, Henry and Oikonomou, Panos and Pasricha, Renu and Sarnoff, Isaac and Weston, James and Arneodo, Francesco and Di Giovanni, Adriano and Millar, Alexander John and Mora, Knut Dundas},
  journal = {Phys. Rev. D},
  volume = {105},
  issue = {5},
  pages = {052010},
  numpages = {19},
  year = {2022},
  month = {Mar},
  publisher = {American Physical Society},
  doi = {10.1103/PhysRevD.105.052010},
  url = {https://link.aps.org/doi/10.1103/PhysRevD.105.052010}
}

@article{PhysRevLett.128.231802,
  title = {New Constraints on Dark Photon Dark Matter with Superconducting Nanowire Detectors in an Optical Haloscope},
  author = {Chiles, Jeff and Charaev, Ilya and Lasenby, Robert and Baryakhtar, Masha and Huang, Junwu and Roshko, Alexana and Burton, George and Colangelo, Marco and Van Tilburg, Ken and Arvanitaki, Asimina and Nam, Sae Woo and Berggren, Karl K.},
  journal = {Phys. Rev. Lett.},
  volume = {128},
  issue = {23},
  pages = {231802},
  numpages = {7},
  year = {2022},
  month = {Jun},
  publisher = {American Physical Society},
  doi = {10.1103/PhysRevLett.128.231802},
  url = {https://link.aps.org/doi/10.1103/PhysRevLett.128.231802}
}
\end{document}